\documentclass[useAMS,usenatbib]{mn2e}
\usepackage{times,graphicx,fancyhdr,amsmath,amssymb}
\usepackage{psfig,natbib,longtable,lscape}

\def\hmpc{~h^{-1}{\rm Mpc}}
\def\kps{~{\rm kms}^{-1}}
\def\deg2{{\rm deg}^2}
\def\degm2{deg$^{-2}$}
\def\arc2{~{\rm arcmin}^2}
\def\lya{Ly$\alpha$}
\def\xibar{\overset{\_}{\xi}}
\title[The VLT LBG Redshift Survey I]{The VLT LBG Redshift Survey I: Clustering and Dynamics of $\approx1000$ Galaxies at $z\approx3$
\thanks{Based on data obtained with the NOAO Mayall 4m Telescope at Kitt Peak National Observatory, USA (programme ID: 06A-0133), the NOAO Blanco 4m Telescope at Cerro Tololo Inter-American Observatory, Chile (programme IDs: 03B-0162, 04B-0022) and the ESO VLT, Chile (programme IDs: 075.A-0683, 077.A-0612, 079.A-0442).}
}
\author[R. Bielby et al]{R. M. Bielby$^{1,2}$\thanks{E-mail:rmbielby@gmail.com (RMB)}, T. Shanks$^{1}$, P. M. Weilbacher$^{3}$, L. Infante$^{4}$, N. H. M. Crighton$^{1}$,  
\newauthor C. Bornancini$^{5,6}$, N. Bouch\'{e}$^{7}$, P. H\'{e}raudeau$^{8,9}$, D. G. Lambas$^{5,6}$, J. Lowenthal$^{10}$, 
\newauthor D. Minniti$^{4,11}$, N. Padilla$^{4}$, P. Petitjean$^{2}$, T. Theuns$^{1,12}$  \\
$^{1}$Department of Physics, Durham University, South Road, Durham, DH1 3LE, UK\\
$^{2}$Institut d'Astrophysique de Paris, UMR7095 CNRS, Univerit\'{e} Pierre et Marie Curie, 98 bis Bld Arago, 75014, Paris, France\\
$^{3}$Astrophysikalisches Institut Potsdam, An der Sternwarte 16, D-14482 Potsdam, Germany\\
$^{4}$Departamento de Astronom\'\i a y Astrof\'isica, Pontificia Universidad Catolica de Chile, Casilla 306, Santiago 22, Chile \\
$^{5}$Instituto de Astronom\'i a Te\'orica y Experimental, Observatorio Astron\'omico, Universidad Nacional de C\'ordoba, Laprida 854, \\
X5000BGR C\'ordoba, Argentina\\
$^{6}$Consejo Nacional de Investigaciones Cient\'ificas y T\'ecnicas, Avenida Rivadavia 1917, C1033AAJ, Buenos Aires, Argentina\\
$^{7}$University of Santa Barbara, Broida, Santa Barbara, 93106, USA\\
$^{8}$Argelander Institut f\"ur Astronomie, Auf dem H\"ugel 71, 53121 Bonn, Germany\\
$^{9}$Kapteyn Astronomical Institute, University of Groningen, PO Box 800, 9700 AV, Groningen, The Netherlands\\
$^{10}$Department of Astronomy, Smith College, Northampton, MA 01063, USA\\
$^{11}$Vatican Observatory, V00120 Vatican City State, Italy\\
$^{12}$Universiteit Antwerpen, Campus Groenenborger, Groenenborgerlaan 171, B-2020 Antwerpen, Belgium\\
}
\begin{document}
\date{}
\pagerange{\pageref{firstpage}--\pageref{lastpage}} \pubyear{2010}
\maketitle
\label{firstpage}
\begin{abstract}
We present the initial imaging and spectroscopic data acquired as part
of the VLT VIMOS Lyman-break galaxy Survey. $UBR$ (or $UBVI$) imaging 
covers five $\approx36\arcmin\times36\arcmin$ fields centred on bright
$z>3$ QSOs, allowing $\approx21,000$ $2<z<3.5$ galaxy candidates to be
selected using the Lyman-break technique. We performed spectroscopic
follow-up using VLT VIMOS, measuring redshifts for 1020 $z>2$
Lyman-break galaxies and 10 $z>2$ QSOs from a total of 19 VIMOS
pointings. From the galaxy spectra, we observe a $625\pm510\kps$
velocity offset between the interstellar absorption and Ly$\alpha$
emission line redshifts, consistent with previous results. Using the
photometric and spectroscopic catalogues, we have analysed the galaxy
clustering at $z\approx3$. The angular correlation function,
$w(\theta)$, is well fit by a double power-law with clustering
scale-length, $r_0=3.19^{+0.32}_{-0.54}\hmpc$ and slope $\gamma=2.45$
for $r<1\hmpc$ and $r_0=4.37^{+0.43}_{-0.55}\hmpc$ with
$\gamma=1.61\pm0.15$ at larger scales. Using the redshift sample we
estimate the semi-projected correlation function, $w_p(\sigma)$ and, for
a $\gamma=1.8$ power-law, find $r_0=3.67^{+0.23}_{-0.24}\hmpc$ for the
VLT sample and $r_0=3.98^{+0.14}_{-0.15}\hmpc$ for a combined VLT$+$Keck
sample. From $\xi(s)$ and
$\xi(\sigma,\pi)$, and assuming the above $\xi(r)$ models, we find that
the combined VLT and Keck surveys require a galaxy pairwise velocity dispersion
of $\approx700\kps$, higher than the $\approx400\kps$ assumed by
previous authors. We also measure a value for the gravitational growth
rate parameter of $\beta(z=3)=0.48\pm0.17$, again
higher than previously found and implying a low value for the bias of
$b=2.06^{+1.1}_{-0.5}$. This value is consistent with the galaxy
clustering amplitude which gives $b=2.22\pm0.16$, assuming the standard
cosmology, implying that the evolution of the gravitational growth rate
is also consistent with Einstein gravity. Finally, we have compared our
Lyman-break galaxy  clustering amplitudes with lower redshift
measurements and find that the clustering strength is not inconsistent
with that of low-redshift $L^*$ spirals for  simple `long-lived' galaxy
models.
\end{abstract}


\section{Introduction}

Observations of the $z\sim3$ galaxy population present a valuable tool
for studying cosmology and galaxy formation and evolution. For
cosmology, the interest is in measuring the galaxy clustering
amplitudes and redshift space distortions at high redshift. They both
lead to virtually independent estimates of the bias whose consistency
leads to a test of the standard cosmological model. For theories of
galaxy formation and evolution, this is a key period in the history of
the Universe in which significant levels of star formation shape both
galaxies and the inter-galactic medium (IGM) around them. An
especially vital direction of study is the effect of galactic winds at
this epoch. Such winds have been directly observed at low
\citep{Heckman90, lehnert99, Martin05, Martin06} and high
\citep{Pettini01, adelberger03, wilman05, adelberger05} redshift and
are invoked to explain a range of astrophysical phenomena.

A basic item of cosmological interest is the spatial clustering of the
$z\approx3$ galaxy population itself. In $\Lambda$CDM, structure in
the Universe is known to grow hierarchically through gravitational
instability (e.g. \citealt{mowhite96,1998ApJ...499...20J,2006Natur.440.1137S}) and
testing this model requires the measurement of the clustering of
matter in the Universe across cosmic time (e.g.
\citealt{2005Natur.435..629S,2008MNRAS.391.1589O,2009MNRAS.400.1527K}). Surveys
of matter at $z\approx3$ currently focus on two main populations, LBGs
and Lyman-$\alpha$ emitters (LAEs). A number of measurements of galaxy
clustering are available at $z\approx3$. For example,
\citet{adelberger03} and \citet{daangela05b} use the Keck LBG sample
with spectroscopic redshfits of \citet{steidel03} to measure LBG
clustering clustering lengths of $r_0=3.96\pm0.29\hmpc$ and
$r_0=4.48^{+0.17}_{-0.18}\hmpc$ respectively. Further surveys of LBGs
at $z\approx3$ have produced a range of results with, for example,
\citet{foucaud03} measuring a clustering length for a photometric
sample selected from the CFHT Legacy Survey of $r_0=5.9\pm0.5\hmpc$,
\citet{2005ApJ...619..697A} measured $r_0 = 4.0 \pm 0.6 \hmpc$ at
$\langle z \rangle = 2.9$ using a different photometric sample whilst
\citet{hildebrandt07} measured a value of $r_0=4.8\pm0.3\hmpc$ from an
LBG sample taken from GaBoDS data.

\citet{daangela05b} go on to use the Keck LBG sample to investigate,
via redshift space distortions, the gravitational growth rate of the
galaxy population at $z\approx3$, measuring an infall parameter of
$\beta(z=3)=0.25^{+0.05}_{-0.06}$. The infall parameter, $\beta$,
quantifies the large-scale infall towards density inhomogeneities
\citep{hamilton92,hawkins03} and is defined as
$\beta(z)=\Omega_m(z)^{0.6}/b(z)$, where $\Omega_m(z)$ is the matter
density and $b(z)$ is the bias of the galaxy population. The 2dF
Galaxy Redshift Survey (2dFGRS) measurement of the infall parameter
nearer the present epoch gave $\beta(z\approx0.1)=0.49\pm0.09$
\citep{hawkins03}, similar to values obtained by previous local
measurements (e.g. \citealt{1998MNRAS.296..191R}). There have also
been dynamical measurements of $\beta$ at intermediate redshifts using
Luminous Red Galaxies where \citet{ross07} found
$\beta(z=0.55)=0.4\pm0.05$. \citet{daangela05a} used the combined 2dF
and 2SLAQ QSO redshift surveys to find
$\beta(z=1.5)=0.60\pm0.14$. Finally, \citet{2008Natur.451..541G} used
the VVDS galaxy redshift survey to measure
$\beta(z=0.77)=0.70\pm0.26$. As emphasised by
\citet{2008Natur.451..541G}, if there are independent estimates of
$b(z)$ for each redshift sample, then the standard model prediction
for the evolution with redshift of the gravitational growth rate of
$f=\Omega_m(z)^{0.6}$ can be tested against alternative gravity
models.  Here, we shall follow
\citet{daangela05b,daangela05a,2002MNRAS.329..336H} in making their
version of the redshift-space distortion cosmological test which also
incorporates the \citet{1979Natur.281..358A} geometric cosmological
test.

From redshift-space distortions, we can also determine the small-scale
dynamics of the galaxy population which are usually simply modelled as
a Gaussian velocity dispersion, measured from the length of the
`fingers-of-God' \citep{1972MNRAS.156P...1J,1987MNRAS.227....1K} in
redshift space clustering. This velocity dispersion will generally
also include the effects of velocity measurement error. Although
\citet{daangela05b} had to assume a fixed value of
$<w_z^2>^{1/2}=400\kps$ for the mean pairwise velocity dispersion when
making their LBG measurement of $\beta(z=3)$, in bigger surveys it is
possble to fit for $<w_z^2>^{1/2}$ and $\beta$ simultaneously.  Thus
in 2dFGRS at $z\approx0.1$, \cite{hawkins03} measured a pairwise
velocity dispersion of $<w_z^2>^{1/2}\approx500\kps$. As well as being
of interest cosmologically, the intrinsic galaxy-galaxy velocity
dispersion is interesting in terms of establishing the group
environment for galaxy formation.  Furthermore, these random peculiar
velocities dominate at the smallest spatial scales, significantly
affecting clustering measurements on scales $r\lesssim5\hmpc$. They
influence both the observed galaxy-galaxy clustering and the observed
correlation between galaxy positions and nearby Ly$\alpha$ forest
absorption from the IGM (as measured in
\citealt{adelberger03,adelberger05} and \citealt{2010arXiv1006.4385C}).  To
interpret galaxy-IGM clustering results we shall see that measurements
of the small scale dynamical velocity dispersion of the galaxy
population are very important.

Galactic winds powered by supernovae are a crucial ingredient in
models of galaxy formation \citep{Dekel86, White91}.  Such negative
feedback is required to quench the formation of small galaxies and
make the observed faint-end of the galaxy luminosity function much
flatter than the low-mass end of the dark-matter mass function, see
for example the semi-analytical model of \citet{Cole00}. Simulations
without such strong feedback tend to produce galaxies with too massive
a bulge, which consequently do not lie on the observed Tully-Fisher
relation \citep{Steinmetz99, Governato10}. Such winds can also remove
a significant fraction of baryons from the forming galaxy, thereby
explaining why galaxies are missing most of their baryons
\citep{2009astro2010S..25B}, and hence are much fainter in X-ray emission than
expected \citep{2010MNRAS.407.1403C}. In addition, observations of the IGM as
probed with QSO sightlines reveal the presence of metals even in the
low density regions producing \lya\ forest absorption
\citep{songailacowie96, pettini03, aguirre04,
  2004A&A...419..811A}. Other than enrichment from galactic scale
winds, it is difficult to see from where these metals originate and
this is confirmed by simulations (e.g. \citealt{2009MNRAS.399..574W}).

Direct evidence for outflows in high redshift galaxies came from the
Keck LBG survey spectra analysed by \citet{adelberger03} and
\citet{shapley03} who found evidence for offsets in the positions of
ISM absorption lines, Ly$\alpha$ emission and rest-frame optical
emission lines (see also
\citealt{2000ApJ...528...96P,2002ApJ...569..742P}). \citet{shapley03}
present a model in which the optical emission lines arise in nebular
star-forming HII regions, giving the intrinsic galaxy redshift, whilst
the ISM absorption lines originate from outflowing material
surrounding the stellar/nebular component. Ly$\alpha$ emission arises
in the stellar component, but outflowing neutral material scatters and
absorbs the blue Ly$\alpha$ wing, leaving a peak redshifted with
respect to the intrisic galaxy redshift \citep[e.g.][]{2010ApJ...717..289S}. One
of our prime aims here is to test the observations underpinning this
model in an independent sample of LBGs.

In this paper, we present the first instalment of data of a $z\sim3$
survey of LBGs within wide ($\approx30'$) fields centred on bright
$z\sim3$ QSOs. We discuss the imaging and spectroscopic observations,
the latter including a search for redshift offsets in the LBG spectra,
followed by an analysis of the clustering and dynamics of the LBG
galaxy populations in our fields. In a further paper
\citep{2010arXiv1006.4385C}, we present the analysis of the relationship
between LBGs and the surrounding IGM via QSO sight-lines, with the
intent of further investigating the extent and impact of galactic
winds on the IGM.

The structure of this paper is  as follows. We provide the details of our imaging
survey in section~\ref{sec:imaging}, covering observations and data reduction.
In section~\ref{sec:spec}, we present VLT VIMOS spectroscopic observations,
describing the data reduction and object identification processes.
Section~\ref{sec:clustering} presents a clustering analysis of the
photometrically and spectroscopically identified objects and we finish with our
conclusions and summary in section~\ref{sec:conclusions}. Unless stated
otherwise, we use an $\Omega_m=0.3$, $\Omega_\Lambda=0.7$, $H_0=100h\kps\rm{Mpc}^{-1}$ flat $\Lambda$CDM cosmology, whilst all magnitudes
are quoted in the Vega system.

\section{Imaging} \label{sec:imaging}

\subsection{Target fields}
\label{sec:targfields}

The full VLT survey comprises 45 VIMOS pointings across nine quasar fields. In this paper we analyse an initial sample of 19 pointings across 5 fields, where we have reduced and identified LBG spectra. The remaining LBG observations will be presented in a future paper. High-resolution optical spectra are available for all of the QSOs, which are at declinations appropriate for observations from the VLT at Cerro Paranal. The selected quasars for this paper are Q0042-2627 (z=3.29), SDSS J0124+0044 (z=3.84), HE0940-1050 (z=3.05), SDSS J1201+0116 (z=3.23) and PKS2126-158 (z=3.28). Q0042-2627 has been observed by \citet{williger96} using the Argus multifibre spectrograph on the Blanco 4m telescope at Cerro Tololo Inter-American Observatory (CTIO) and as part of the Large Bright QSO Survey (LBQS) using Keck/HIRES \citep{hewett95}. \citet{pichon03} observed HE0940-1050 and PKS2126-158 using the Ultraviolet and Visual Echelle Spectrograph (UVES) on the VLT and SDSS J0124+0044 has been observed by \citet{peroux05} also using UVES. Finally, SDSS J1201+0116 has been observed by the SDSS team using the SLOAN spectrograph and by \citet{omeara07} using the Magellan Inamori Kyocera Echelle (MIKE) high resolution spectrograph on the Magellan 6.5m telescope at Las
Campanas Observatory.

\subsection{Observations}
\label{lbgimaging}

The imaging for our 5 selected fields was obtained using a combination of the MOSAIC Imager on the Mayall 4-m telescope at KPNO, the MOSAIC-II Imager on the Blanco 4-m at CTIO and VLT VIMOS in imaging mode. Q0042-2627, HE0940-1050 and PKS2126-158 were all observed at CTIO between January 2004 and April 2005. J0124+0044 and J1201+0116 were observed at KPNO in September 2001 and April 2006 respectively. All of these fields were observed with the broadband Johnson $U$ (c6001) filter and the Harris $B$ and $R$ filters, except for J0124+0044, which was observed with the Harris $B$, $V$ and $I$ broadband filters but not the Harris $R$. A full description of the observations is given in Table~\ref{photometry}.

We note that during the observations of the HE0940-1050 field, there was a malfunction of one of the 8 CCDs leaving a gap of $\approx8\arcmin\times18\arcmin$ in the field of view. The remaining CCDs provided unaffected data however, which we use here.

\begin{table*}
\centering
\caption[{\small Details of the imaging data acquired in each of our five target
fields.}]{\small Details of the imaging data acquired in each of our five target
fields. Coordinates are given for the imaging centre, which is not necessarily
the same as the position of the bright corresponding QSO.}
\label{photometry}
\begin{tabular}{@{}lcclccccc@{}}
\hline
\hline
Field & $\alpha$ & $\delta$ &Facility& Band & Exp time &Seeing& \multicolumn{2}{c}{Depth} \\
      & \multicolumn{2}{c}{(J2000)}  &        &      &(s)       &      & 50\% comp.&  3$\sigma$\\
\hline
\hline
Q0042-2627  & 00:46:45 & -25:42:35 & CTIO/MOSAIC2 & $U$ & 12,600 & 1.8$\arcsec$ & 24.09 & 26.16 \\
                         &                  &                    &        & $B$ & 3,300   & 1.8$\arcsec$ & 25.15 & 26.93  \\
                         &                  &                    & VLT/VIMOS        & $R$ & 235      & 1.1$\arcsec$ & 24.72 & 25.79  \\
\hline
J0124+0044  & 01:24:03     & +00:44:32    & KPNO/MOSAIC & $U$ & 13,400 & $1.5\arcsec$ &  ...      & 25.60  \\
                         &                      &                       &                               & $B$ & 2,800   & $1.5\arcsec$ &  ...      & 26.44 \\
                         &                      &                       &                               & $V$ & 3,100   & $1.4\arcsec$ &  ...      & 26.14   \\
                         &                      &                       &                               & $I$ & 7,500     & $1.1\arcsec$ & 24.48& 25.75  \\
\hline
HE0940-1050 & 09:42:53     & -11:04:25    & CTIO/MOSAIC2 & $U$ & 29,000 & 1.3$\arcsec$ & 25.69& 26.75  \\
                           &                      &                      &                               & $B$ & 4,800   & 1.3$\arcsec$ & 25.62 & 26.66  \\
                           &                      &                      &                               & $R$ & 2,250   & 1.0$\arcsec$ & 25.44 & 26.24  \\
\hline
J1201+0116  & 12:01:43 & +01:16:05 & KPNO/MOSAIC  & $U$ & 9,900  & 1.6$\arcsec$ & 24.50 & 26.11  \\
                         &                  &                    &                               & $B$ & 6,000   & 2.4$\arcsec$ &24.43 & 26.56 \\
                         &                  &                    & VLT/VIMOS         & $R$ & 235      & 0.7$\arcsec$ & 25.47& 26.24  \\
\hline
PKS2126-158 & 21:29:12     &-15:38:42     & CTIO/MOSAIC2 & $U$ & 26,400 & 1.3$\arcsec$ &25.08& 26.97  \\
                          &                       &                      &                              & $B$ & 7,800  & 1.6$\arcsec$ &24.94& 27.49 \\
                          &                       &                      &                              & $R$ & 6,400  & 1.5$\arcsec$ &24.65& 26.79  \\
\hline
\hline
\end{tabular}
\end{table*}

The MOSAIC Imagers each have a field of view of $36\arcmin\times36\arcmin$, covered by 8 $2048\times4092$ CCDs. Adjacent chips are separated by a gap of up
to $12\arcsec$ and we have therefore performed a dithered observing strategy for the acquisition of all our imaging data. For all observations we took bias
frames, sky flats (during twilight periods), dome flats and also observed \citet{landolt92a} standard-star fields with each filter on each night of observation for the calibration process.

In the Q0042-2627 and J1201+0116 fields, we also use imaging from the VLT VIMOS
instrument with the broadband R filter. VIMOS consists of 4 CCDs each covering
an area of $7\arcmin\times8\arcmin$, with gaps of $2\arcmin$ between adjacent
chips. The fields were observed with 4 separate pointings, with $<1\arcmin$
overlap between adjacent pointings.

\subsection{Data Reduction}

All data taken using the MOSAIC Imagers were reduced using the {\small MSCRED} package within {\small IRAF}, in accordance with the NOAO Deep Wide-Field Survey guidelines of \citet{januzziweb}. Bias images were created using {\small ZEROCOMBINE} and dome and sky-flats were processed using {\small CCDPROC}. Removal of the ``pupil-ghost'' artifact was performed for the $U$-band calibration and science images using {\small MSCPUPIL}.

The science images were processed using {\small CCDPROC}. Cosmic ray rejection was performed with {\small CRAVERAGE} in the early data-reductions (HE0940-1050 and PS2126-158), whilst in the later reductions, {\small CRREJECT} was used. The {\small FIXPIX} task was used to remove marked bad-pixels and cosmic-rays from the images, using the interpolation setting.

Deprojection of the images was performed using the {\small MSCIMAGE} task, with optimization of the astrometry conducted using {\small MSCCMATCH}. Large-scale sky-variations were removed from science images using {\small MSCSKYSUB} and the resultant final images were combined using {\small MSCIMATCH} and {\small MSCSTACK}.

For the HE0940-1050 and PKS 2126-158, short exposure imaging was obtained. These were used in the selection of QSO candidates (at brighter magnitudes than the LBG candidates) in these fields and were reduced and combined in the same way as the long exposure images described above. As there are typically only one or two short exposures per filter, the gaps between the CCDs still exist in the final short images, and no extra effort was made to remove blemishes 
by hand.

The data reduction for the $R$-band imaging from VLT VIMOS was performed using the VIMOS pipeline. Again bias frames were subtracted and the images were flat fielded using dome flats acquired on the night of observation. Individual exposures were then deprojected and stacked using the {\small SWARP} software \citep{swarp}.

\subsection{Photometry}

We performed object extraction using {\small SEXTRACTOR}, with a detection threshold of 1.2$\sigma$ and a minimum object size of 5 pixels. Object detection was performed on the $R$-band images and fluxes were calculated in all bands using Kron, fixed-width (with a diameter of twice the image seeing FWHM) and isophotal width apertures. Zeropoints for each of the observations were calculated from the Landolt standard-star field observations made during the observing runs and we correct the photometry for galactic extinction using the dust maps of \citet{1998ApJ...500..525S}. Each of the standard-star field images were processed using the same method as for the science frames. The depths reached in the $U$, $B$ and $R$ bands for each field are given in table~\ref{photometry}. We quote the $3\sigma$ depths, which give the limit for detecting an object 5 pixels in size with a signal of $3\times$ the background RMS detection, and the $50\%$ completeness level. The $50\%$ completeness levels are calculated by systematically placing simulated point-source objects in the final stacked images at different magnitudes. The $50\%$ level is then the magnitude at which we are able to recover $50\%$ of simulated sources.

\begin{figure}
\centering
\includegraphics[width=80.mm]{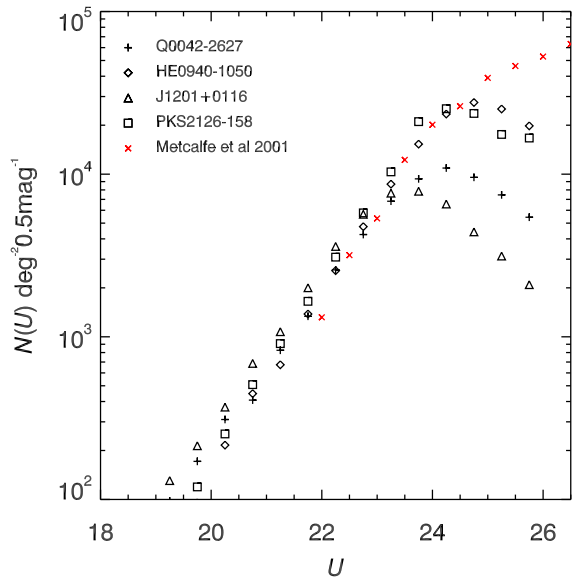}
\caption[{\small $U$-band number counts in the Q0042-2627, HE0940-1050, J1201+0116 and PKS2126-158 QSO fields.}]{{\small $U$-band number counts from the four fields Q0042-2627 (black crosses), HE0940-1050 (diamonds), J1201+0116 (triangles) and PKS2126-158 (squares). The counts of \citet{metcalfe01} from the William Herschel Deep Field are shown for comparison (red crosses).}}
\label{numcountsU}
\end{figure}

\begin{figure}
\centering
\includegraphics[width=80.mm]{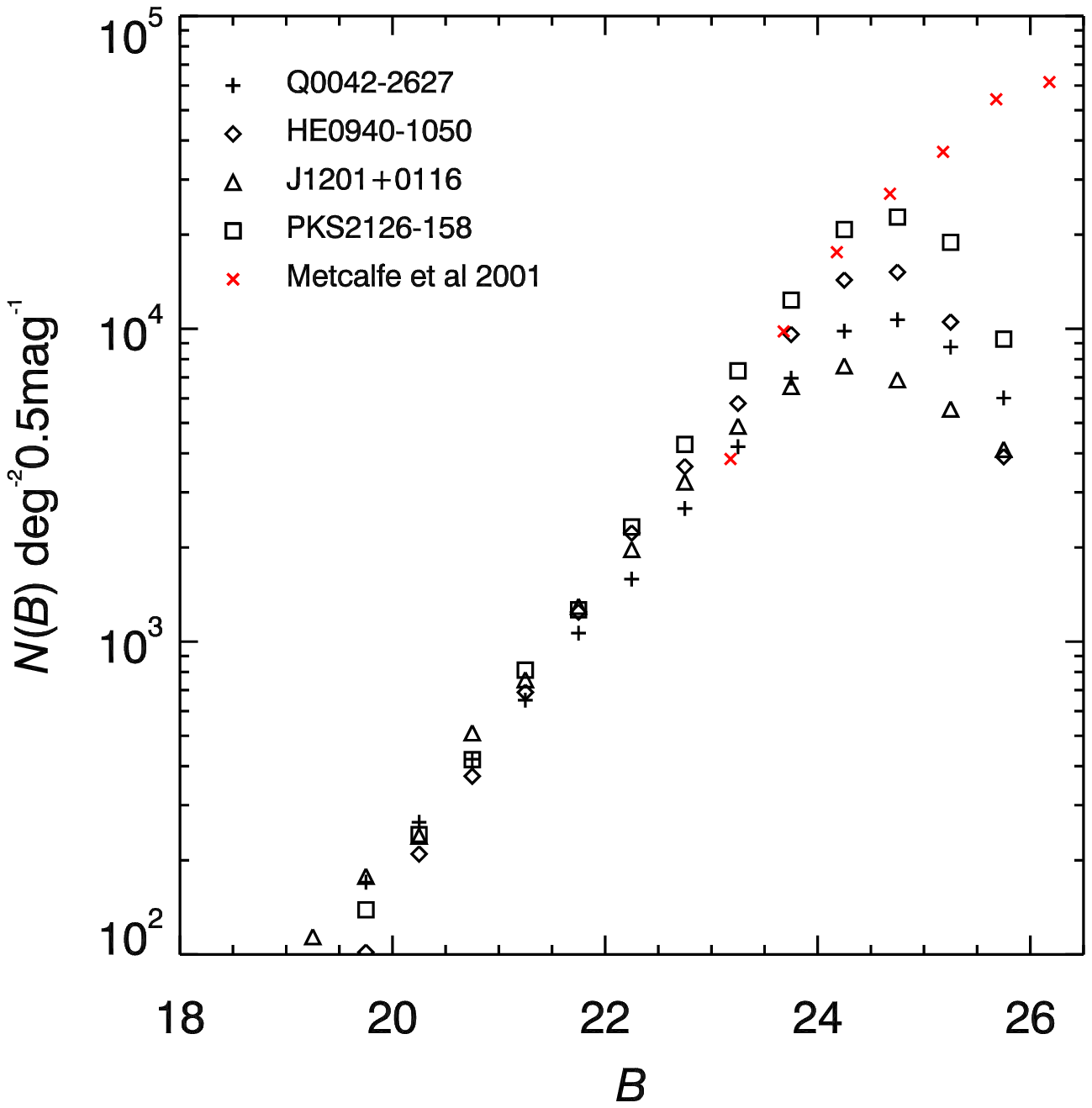}
\caption[{\small $B$-band number counts in the Q0042-2627, HE0940-1050, J1201+0116 and PKS2126-158 QSO fields.}]{{\small $B$-band number counts from the four fields Q0042-2627 (black crosses), HE0940-1050 (diamonds), J1201+0116 (triangles) and PKS2126-158 (squares). The counts of \citet{metcalfe01} from the William Herschel Deep Field are shown for comparison (red crosses).}}
\label{numcountsB}
\end{figure}

\begin{figure}
\centering
\includegraphics[width=80.mm]{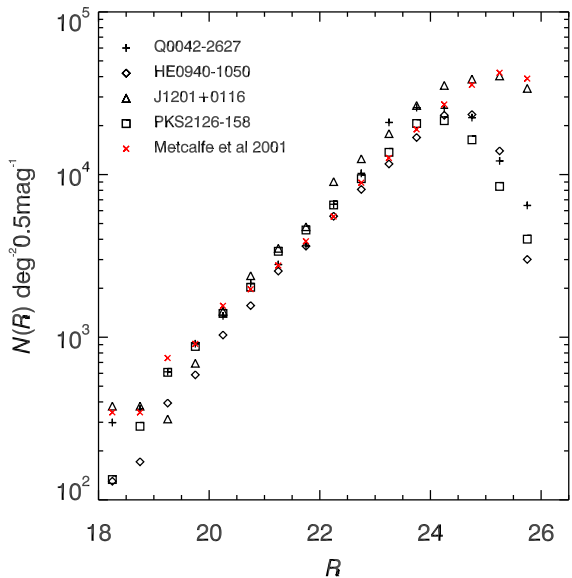}
\caption[{\small $R$-band number counts in the Q0042-2627, HE0940-1050, J1201+0116 and PKS2126-158 QSO fields.}]{{\small $R$-band number counts from the four fields Q0042-2627 (black crosses), HE0940-1050 (diamonds), J1201+0116 (triangles) and PKS2126-158 (squares). The counts of \citet{metcalfe01} from the William Herschel Deep Field are shown for comparison (red crosses).}}
\label{numcountsR}
\end{figure}

The $U$, $B$ and $R$ number counts from the 4 fields are plotted in Figs.~\ref{numcountsU} to \ref{numcountsR}. In general the counts turnover at $\sim0.5\mbox{mag}$ brighter than the 50\% completeness limits, consistent with the counts being dominated by extended sources (whilst the completeness limits are estimated using simulated point-sources). We plot for comparison the number counts of \citet{metcalfe01}. All counts are from our MOSAIC data except for the R band counts of Q0042-2627 and J1201+0116, which are from the VLT VIMOS. The imaging in the J1201+0116 field was taken during relatively poor seeing conditions during observations at CTIO and so reaches shallower depths than the other fields. For these plots, stars have been removed using the {\small SEXTRACTOR} {\small CLASS\_STAR} estimator with a limit of {\small CLASS\_STAR} $<0.8$.

\subsection{Selection Criteria}
\label{sec:sel}

We perform a photometric selection based on that of
\citet{steidel96,steidel03}, but applied to the $U$, $B$ and $R$ band
imaging available from our imaging survey. As in \citet{steidel03} our
selection takes advantage of the Lyman-Break at 912\AA\ and the
Ly$\alpha$-forest passing through the $U$-band and into the $B$-band in
the redshift range $2.0<z<3.5$. To establish the selection in the Vega
$UBR$ system, we convert from the \citet{steidel03} selections using the
photometric transformations of \citet{1993AJ....105.2017S}, moving from
the $U_nG{\cal R}$ AB system to the Johnson-Morgan/Kron-Cousins Vega
photometry. The approximate transformations \citep{1993AJ....105.2017S} are as follows: $U_n=U+0.75$, $G=B-0.17$ and ${\cal R}=R+0.14$ and transform the \citet{steidel03} selection to $(B-R)\leq1.51$ and $(U-B)\geq(B-R)-0.23$.

\begin{figure*}
\centering
\includegraphics[width=80.mm]{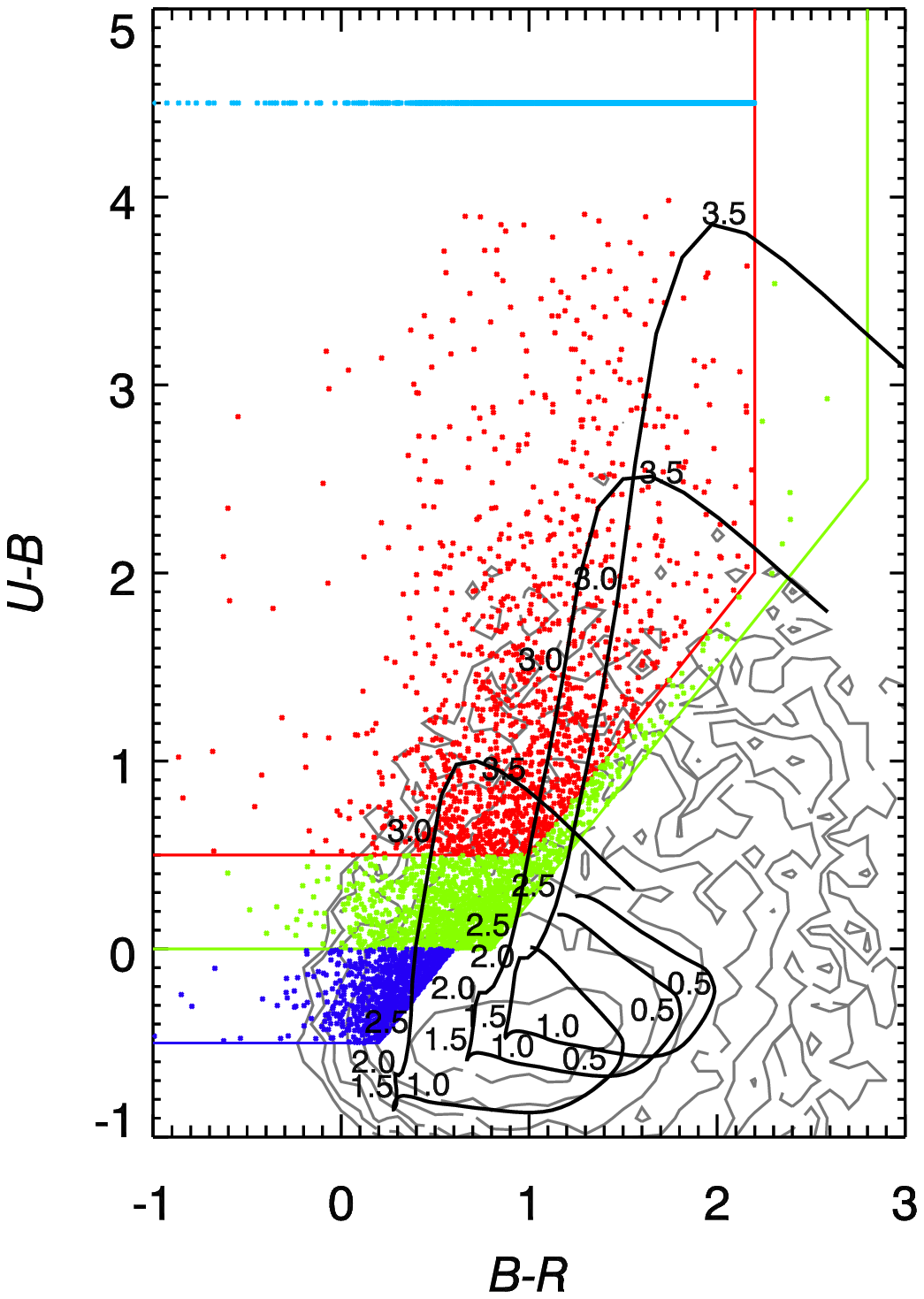}\includegraphics[width=80.mm]{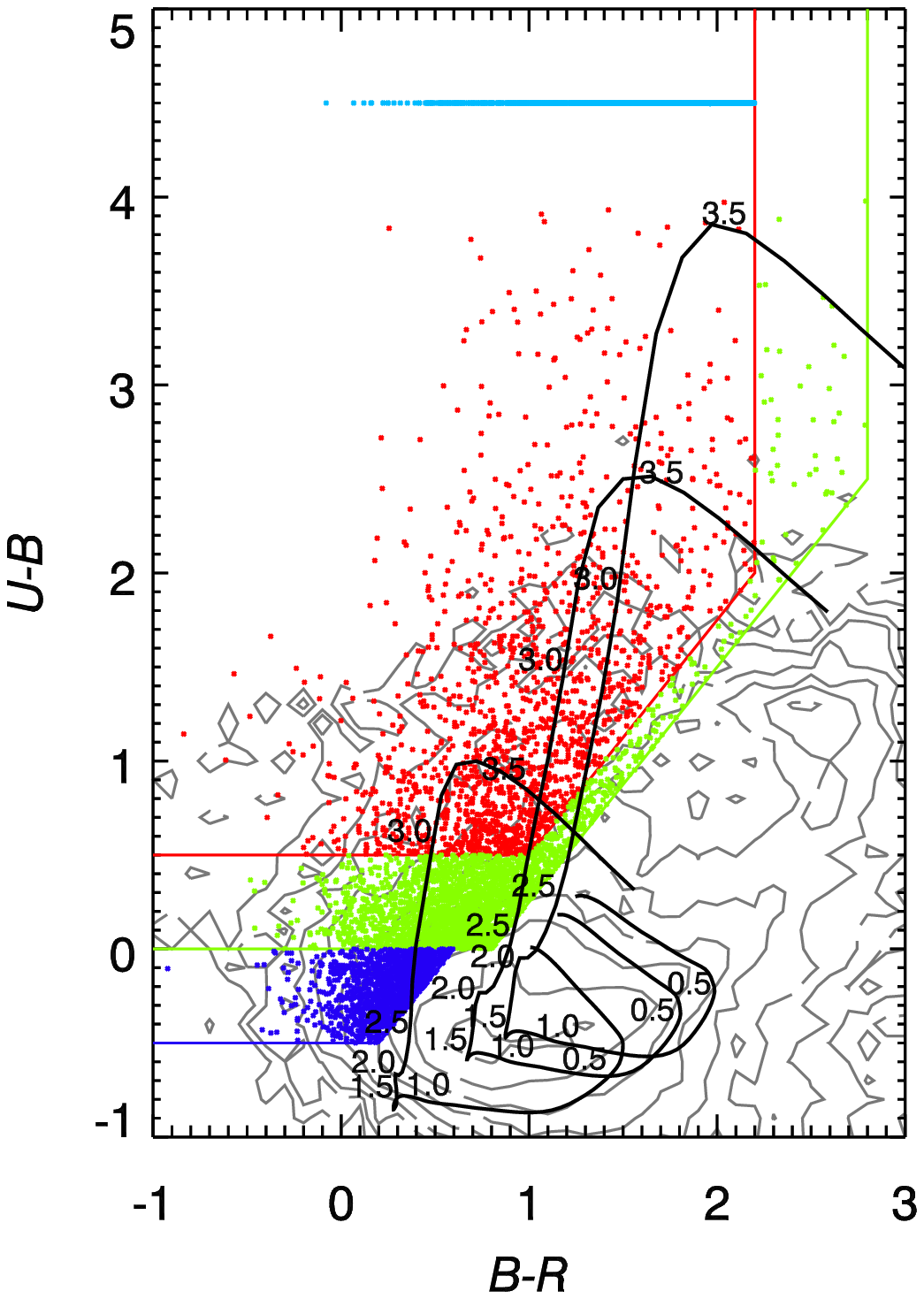}
\caption[{\small Our selection criteria in $UBR$ colour space shown for the Q0042-2627 and HE0940-1050.}]{{\small Our selection criteria in $UBR$ colour space shown for the Q0042-2627 (left) and HE0940-1050 (right). The red line and points show the LBG\_PRI1 selection, the green line and points show the LBG\_PRI1 selection, the blue line and points show the LBG\_PRI3 selection and the cyan line at $U-B=4.5$ shows the LBG\_DROP selection. The grey contours show the entire galaxy population in the fields. The black lines show the galaxy evolution model for a galaxy with a $\tau=9$Gyr exponential SFR formed at $z=6.2$ and are labelled with values of observed redshift from $z=3.83$ to $z=0$.}}
\label{0009selection}
\end{figure*}

\begin{figure*}
\centering
\includegraphics[width=80.mm]{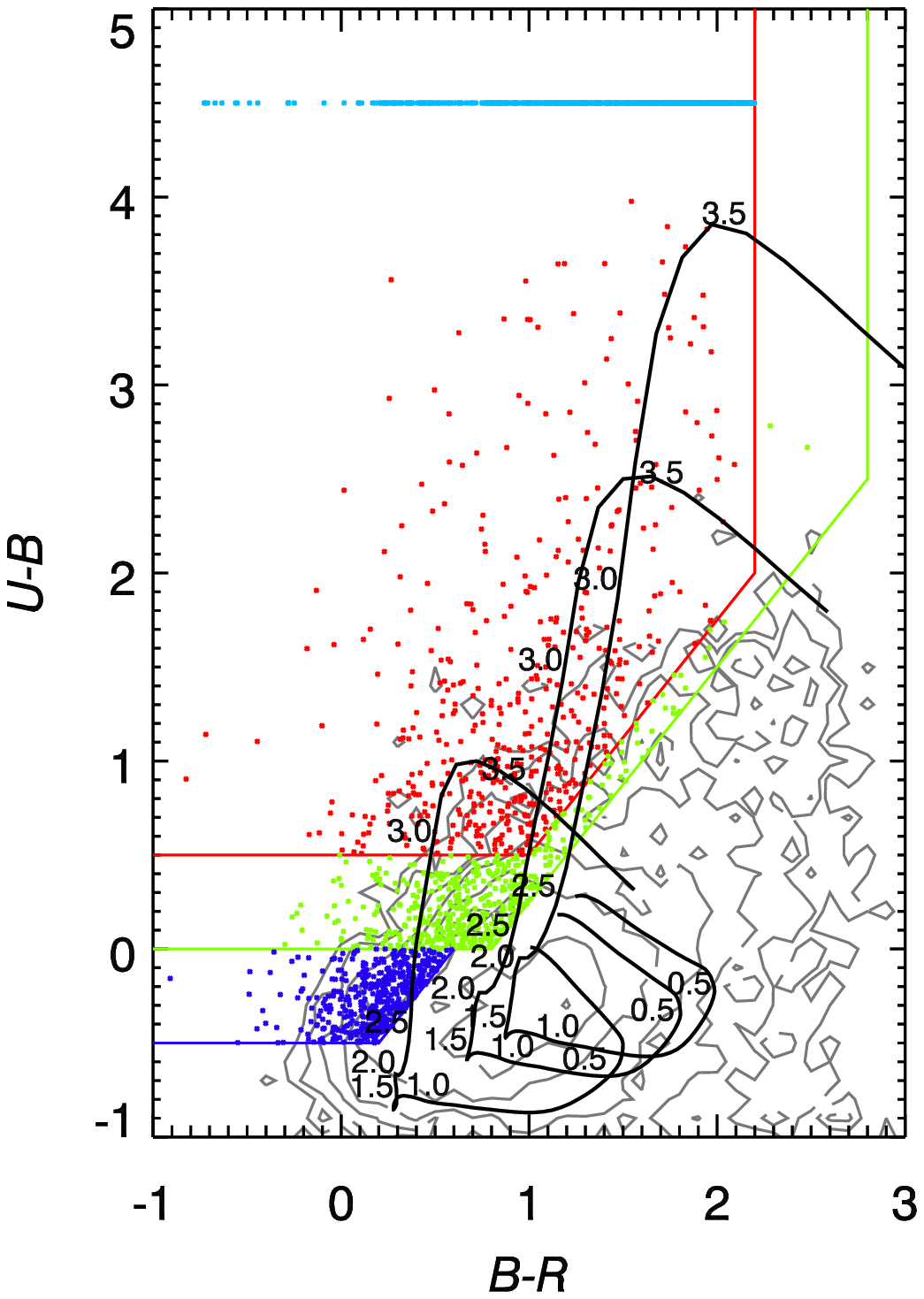}\includegraphics[width=80.mm]{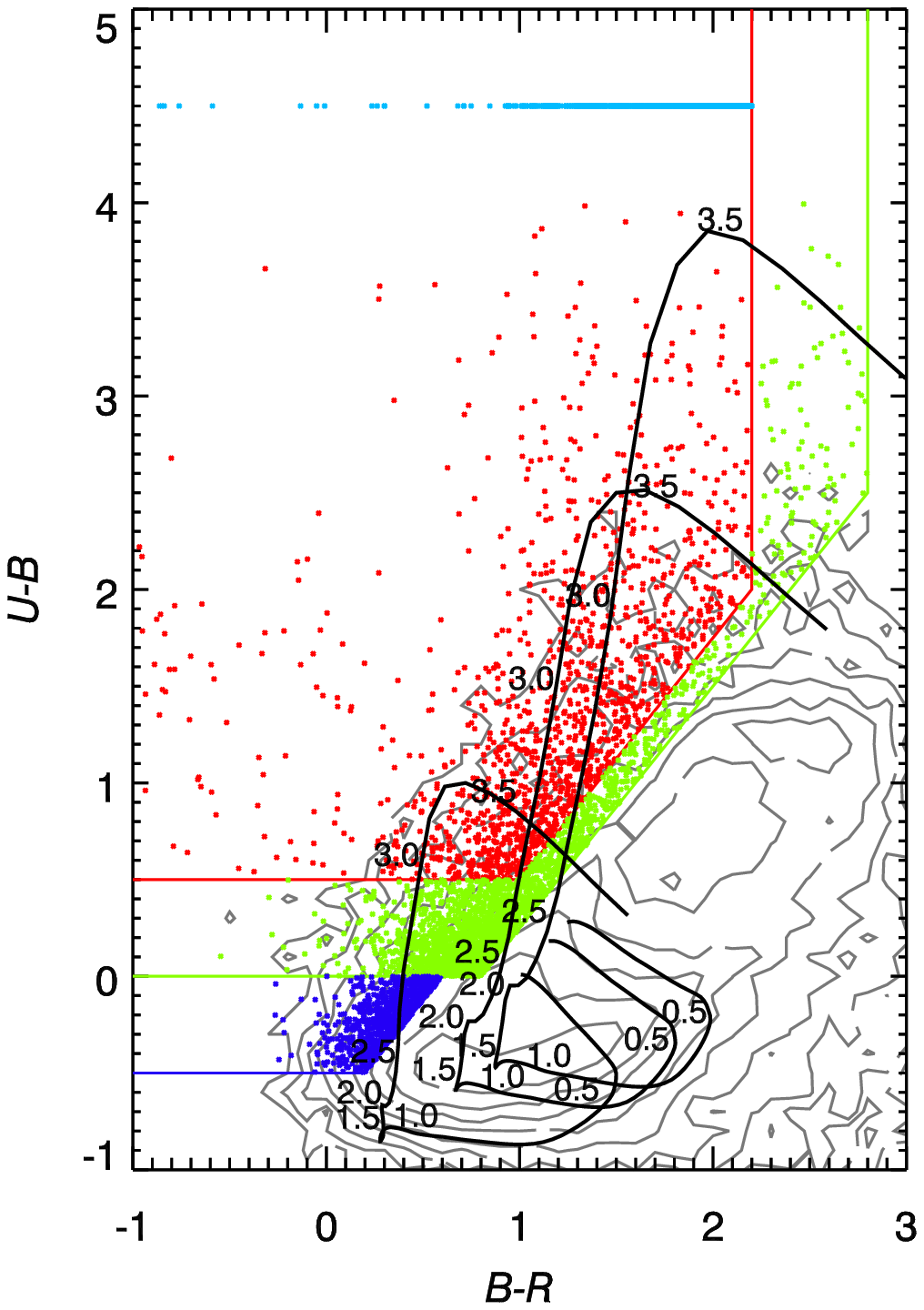}
\caption[{\small Our selection criteria in $UBR$ colour space shown for the J1201+0116 and PKS2126-158 fields.}]{{\small As in Fig.~\ref{0009selection} but for the J1201+0116 and PKS2126-158 fields (left to right).}}
\label{1221selection}
\end{figure*}

We also take into account model colour tracks calculated using GALAXEV \citep{bruzualcharlot03}. The tracks are shown in Figs.~\ref{0009selection} and \ref{1221selection} (solid black curves). We use a Salpeter initial mass-function, assuming solar metallicity with a galaxy formed at $z=6.2$ (i.e. with an age of 12.6 Gyr at $z=0$) and a $\tau=9$ Gyr exponential SFR. The three different curves show the effect of dust extinction with a model given by (left to right) $\tau_\nu=0.5$, $\tau_\nu=1.0$ and $\tau_\nu=2.0$, where $\tau_\nu=2.0$, where $\tau_\nu$ is the effective absorption \citep{2000ApJ...539..718C}. The models agree well with the transformation of the \citet{steidel03} selection criteria, although the dustier models do suggest a greater extension of the $z>3$ population to higher values of $(B-R)$ than the \citet{steidel03} criteria.

Based on the models and the \citet{steidel03} criteria, we develop a number of selection criteria in the $UBR$ system. The key modifications that we make from our initial colour-cut estimates based on the \citet{steidel03} cuts are to extend the selection further redwards in $(B-R)$ and to align the $(U-B)-(B-R)$ axis with the stellar locus in the $UBR$ plane, which has a slope of $(U-B)\sim1.25(B-R)$. We note that the first of these modifications risks increasing the number of contaminants in the form of M-stars \citep{1993AJ....105.2017S} and the second increases the risk of contaminants in the form of lower redshift galaxies. However, given the large number of slits available to us with the VLT VIMOS spectrograph, we deem the risk of increased levels of contamination acceptable, whilst extending the colour-cuts can allow the observation of dusty $z>3$ objects as well as $z\approx3$ galaxies which may be scattered out of the primary selection area due to photometric errors on these faint objects. As such we use four selection criteria with different priorities for spectroscopic observation (taking advantage of the object priority system in arranging the VIMOS slit masks). These selection criteria are as follows:

{\small
\begin{itemize}
\item{LBG\_PRI1}
\begin{enumerate}
\item{$23<R<25.5$}
\item{$U-B>0.5$}
\item{$B-R<0.8(U-B)+0.6$}
\item{$B-R<2.2$}
\end{enumerate}
\item{LBG\_PRI2}
\begin{enumerate}
\item{$23<R<25.5$}
\item{$U-B>0.0$}
\item{$B-R<0.8(U-B)+0.8$}
\item{$B-R<2.8$}
\end{enumerate}
\item{LBG\_PRI3}
\begin{enumerate}
\item{$23<R<25.5$}
\item{$-0.5<U-B<0.0$}
\item{$B-R<0.8(U-B)+0.6$}
\end{enumerate}
\item{LBG\_DROP}
\begin{enumerate}
\item{$23<R<25.5$}
\item{No U detection}
\item{$B-R<2.2$}
\end{enumerate}
\end{itemize}
}

LBG\_PRI1 is our primary sample and selects candidates that are expected to be the most likely $2.5<z<3.0$ galaxies. The LBG\_PRI2 sample targets objects with colours closer to the main sequence of low-redshift galaxies than the LBG\_PRI1 objects. This sample is therefore expected to include a greater level of contamination from low redshift galaxies. In addition, based on the path of the evolution tracks in Figs.~\ref{0009selection} and \ref{1221selection}, we also expect the $z>2.5$ population that this selection samples to have, on average, a lower redshift than the LBG\_PRI1 sample. The next selection sample, LBG\_PRI3, takes this further and is intended to target a $2.0<z<3.0$ galaxy redshift based on the evolution tracks. Finally, we select a sample of $U$-dropout objects (LBG\_DROP) with detections in only our $B$ and $R$ band data.

In none of the above samples do we attempt to remove stellar-like objects due to the risk of losing good LBG candidates. The half-light radius of $z\approx3$ LBGs has been shown to be on average $0.4\arcsec$ and so will not be resolved in our data, which is mostly taken under conditions of $>0.8\arcsec$ seeing.

We apply these selection criteria to four of our QSO fields: Q0042-2627, HE0940-1050, J1201+0116 and PKS2126-158. The candidate selection for the J0124+0044 field was performed separately and is discussed in \citet{bouche04}. Figs.~\ref{0009selection} and ~\ref{1221selection} show the four selection criteria applied to these four fields. The selection boundaries are shown by the red, green and blue lines for the LBG\_PRI1, LBG\_PRI2 and LBG\_PRI3 selections respectively. Objects selected as candidates by each criteria set are shown by red, green, blue and cyan points for the LBG\_PRI1, LBG\_PRI2, LBG\_PRI3 and  LBG\_DROP selections respectively. The grey contours in each plot show the extent of the complete galaxy population in each of the fields.

Returning to the depths of our fields, we now compare these to those of
previous studies in the selection of LBGs. We note that
\citet{steidel03} used photometry with mean 1$\sigma$ depths of
$\left<\sigma(U_n)\right>=28.3$, $\left<\sigma(G)\right>=28.6$ and
$\left<\sigma({\cal R})\right>=28.0$, whilst their imposed ${\cal R}$
band limit was ${\cal R}=25.5$. Using the transformations of
\citet{1993AJ....105.2017S}, the \citet{steidel03} 1$\sigma$ limits
correspond to $U=27.55$, $B=28.77$ and $R=27.86$ in the Vega system.
Comparing this to the average depths in our own fields, we have mean
$3\sigma$ depths of $U=26.2$, $B=26.8$ and $R=26.3$, which equate to
$1\sigma$ depths of $U=27.4$, $B=28.0$ and $R=27.5$, largely comparable
to the \citet{steidel03} imaging data.

The numbers of objects selected by each selection for each field are
given in Table~\ref{sel_tab}. These candidate selections were used as
the basis for the spectroscopic work which is described in the following
sections.

\begin{table*}
\caption[{\small Number of candidate high redshift objects in each of the selected fields.}]{{\small Number of candidate high redshift objects in each of the selected fields. Note that candidates in the J0124+0044 were selected as described in \citet{bouche04} and not using the four selection criteria sets described in this paper.}}
\label{sel_tab}
\centering
{\small
\begin{tabular}{@{}lccccc@{}}
\hline
Field & LBG\_PRI1 & LBG\_PRI2 & LBG\_PRI3 & LBG\_DROP & Total \\
\hline
Q0042-2627  & 1,366 & 1,381 & 650  & 1,390 & 4,787\\
J0124+0044  &  &  &   &  & 3,679 \\
HE0940-1050 & 1,646 & 2,249 & 741  & 1,042 & 5,678\\
J1201+0116  & 477  & 487  & 469  & 606  & 2,029\\
PKS2126-158 & 1,380 & 2,119 & 713  & 667  & 4,879\\
\hline 
Total       & 4,869 & 6,236 & 2,573 & 3,705 & 21,062\\
Observed spectroscopically   & 730 & 569 & 256 & 999 & 2,554\\
\hline
\end{tabular}
}
\end{table*}

\subsection{QSO Candidate Selection}

At redshifts of $z\approx3$, the observed optical spectra of QSOs and galaxies exhibit similar shapes, both being heavily influenced by the Lyman break feature. We therefore add to our targets a number of QSO candidates in each field (except J0124+0044) using the following selection, which is closely based on our high-priority $z\approx3$ LBG selection:

{\small
\begin{enumerate}
\item{CLASS\_STAR $>0.8$}
\item{$U - B > 0.5$}
\item{$B-R < 0.8 (U - B) + 0.8$}
\item{$0.0 < B - R < 2.2$}
\end{enumerate}
}

The magnitude limits used with this selection were $20<R<23$ in the Q0042-2627 and J1201+0116 fields and $18<R<22$ in the HE0940-1050 and PKS2126-158 fields for which we had obtained shallow imaging and could therefore select brighter objects more reliably.

As with the LBGs, QSOs at $z>2$ may be selected by the passage of the Lyman-break through the $U$-band (e.g. \citealt{2009ApJS..180...67R}). This selection is therefore based on the LBG selection, but constrained to brighter magnitudes and stellar-like objects only. This selection gives 71, 39, 15 and 38 QSO candidates in the Q0042-2627, HE0940-1050, J1201+0116 and PKS2126-158 fields respectively. Note that only a small number of these have actually been observed spectroscopically as the LBG candidates remained the higher priority.

\section{Spectroscopy}
\label{sec:spec}

\subsection{Observations}

We observed our LBG candidates using the VIMOS instrument on the VLT UT3 (Melipal) between September 2005 and March 2007. As described earlier, the VIMOS camera consists of four CCDs, each with a field of view of $7\arcmin\times8\arcmin$, arranged in a square configuration, with $2\arcmin$ gaps between the field-of-views of adjacent chips. Each observation therefore covers a field of view of $16\arcmin\times18\arcmin$ with $224\arc2$ being covered by the CCDs. The instrument was set up with the low-resolution blue grating (LR\_Blue) in conjunction with the OS\_Blue filter, giving a wavelength coverage of 3700\AA\ to 6700\AA\ and a resolution of 180 with $1\arcsec$ slits, corresponding to 28\AA\ FWHM at 5000\AA. The dispersion with this setting is 5.3\AA\ per pixel. We note that this configuration also projects the zeroth diffraction order onto the CCDs.

Given the size of our imaging fields ($36\arcmin\times36\arcmin$) it was possible to target 4 distinct sub-fields with the VIMOS field of view. We have therefore observed a total of 19 sub-fields across our 5 fields, i.e. 4 sub-fields in each field except for HE0940-1050 in which only 3 sub-fields were achievable due to the CCD malfunction during the imaging observations. Each sub-field was observed with $10\times1,000$s exposures, apart from sub-field three of the PKS2126-158, which was observed with only $4\times1,000$s due to time constraints in the VIMOS schedule. All observations were performed during dark time, with $<0.8\arcsec$ seeing and $<1.3$ air mass.

Slit masks for each quadrant of each sub-field were designed using the standard VIMOS mask software, VMMPS. We used minimum slit lengths of $8\arcsec$, which equates to 40 pixels given the pixel scale of $0.205\arcsec/\textrm{pixel}$. With the effectively point-like nature of our sources and our maximum seeing constraint of $0.8\arcsec$ this allows us a minimum of $\approx7\arcsec$ for sky spectra per slit (with which to perform the sky-subtraction when extracting the spectra). Using the VMMPS software with the LR\_Blue grism we were able to target up to $\approx60-70$ objects per quadrant (i.e. $\approx250$ objects per sub-field), depending on the sky density of the candidate objects. For the spectroscopic observations, we predominantly used the selections as given in section~\ref{sec:sel}, however to optimize the spectroscopic observations some flexibility was employed in including small numbers of objects outside the selection criteria. However, we note that the LBG\_PRI3 selection was not employed in the spectroscopic observations in the first observations (i.e. the observations of HE0940-1050 and PKS2126-158), whilst the magnitude limit used for selecting objects to observe for later fields was reduced from $R=25.5$ to $R=25$. The total number of spectroscopically observed objects was 3,562.
  
\begin{table*}
\centering
\caption[{\small Details of the spectroscopic data acquired in each of our five target fields.}]{{\small Details of the spectroscopic data acquired in each of our five target fields. Coordinates are given for the targeting centre of each sub-field.}}
\label{tab:lbgspectroscopy}
{\small
\begin{tabular}{lccclcc}
\hline
\hline
Field & Sub-field &$\alpha$  & $\delta$  & Dates & Exp time &Seeing \\
      &           &(J2000)   & (J2000)   &       & (s)      &       \\
\hline
\hline
Q0042-2627  & f1 & 00:45:11.14 & -26:04:22.0 & 8-10,15/08/2007             & $10,000$ & $0.6-1.0\arcsec$ \\
Q0042-2627  & f2 & 00:43:57.30 & -26:04:22.0 & 18-19/08/2007 \& 5-6/09/2007& $10,000$ & $0.9-1.0\arcsec$\\
Q0042-2627  & f3 & 00:45:10.35 & -26:19:06.9 & 11-12/09/2007               & $10,000$ & $0.9-1.0\arcsec$ \\
Q0042-2627  & f4 & 00:43:55.97 & -26:19:16.1 & 7,10/09/2007                & $10,000$ & $0.9-1.0\arcsec$ \\
J0124+0044  & f1 & 01:24:41.82 & +00:52:18.8 & 1-2,4/11/2005               & $10,000$ & $0.8-0.9\arcsec$\\
J0124+0044  & f2 & 01:23:32.06 & +00:52:13.1 & 5,29,31/10/2005             & $10,000$ & $0.6-1.0\arcsec$\\
J0124+0044  & f3 & 01:23:31.29 & +00:37:02.0 & 19-20/09/2007               & $10,000$ & $0.8-1.0\arcsec$\\
J0124+0044  & f4 & 01:24:41.86 & +00:36:51.4 & 4/12/2005 \& 22/08/2006     & $10,000$ & $0.8-0.9\arcsec$ \\
HE0940-1050 & f1 & 09:42:08.02 & -11:08:14.2 & 26-27,29/01/2006            & $10,000$ & $0.5-0.8\arcsec$\\
HE0940-1050 & f2 & 09:43:21.53 & -11:08:35.0 & 30-31/01/2006, 1,25/02/2006 \& 1/03/2006& $10,000$ & $0.5-1.0\arcsec$\\
HE0940-1050 & f3 & 09:43:21.58 & -10:54:31.8 & 14,19/12/2007 \& 31/01/2008 & $10,000$ & $0.6-1.0\arcsec$\\
J1201+0116  & f1 & 12:02:14.01 & +01:09:09.9 & 13-15/04/2007 \& 17/04/2007 & $10,000$ & $0.6-1.0\arcsec$ \\
J1201+0116  & f2 & 12:01:10.01 & +01:09:09.9 & 23/04/2007 \& 8,11,14/05/2007 & $10,000$ & $0.4-0.9\arcsec$ \\
J1201+0116  & f3 & 12:01:10.04 & +01:24:09.8 & 16-17/05/2007               & $10,000$ & $0.5-0.9\arcsec$ \\
J1201+0116  & f4 & 12:02:14.07 & +01:24:08.0 & 18/05/2007 \& 6,8,10/02/2008& $10,000$ & $0.6-0.7\arcsec$ \\
PKS2126-158 & f1 & 21:29:59.57 & -15:31:30.2 & 17/08/2006 \& 1,21-26/09/2006 & $10,000$ & $0.7-1.0\arcsec$ \\
PKS2126-158 & f2 & 21:28:46.20 & -15:31:29.9 & 5-6/08/2005                 & $10,000$ & $0.6-1.0\arcsec$\\
PKS2126-158 & f3 & 21:30:00.41 & -15:47:18.3 & 27/09/2006                  & $4,000$  & $0.8-1.0\arcsec$ \\
PKS2126-158 & f4 & 21:28:46.27 & -15:47:11.9 & 9-11,25,29/08/2005          & $10,000$ & $0.7-0.9\arcsec$\\
\hline
\hline
\end{tabular}
}
\end{table*}

%\end{landscape}

\subsection{Data reduction}

Bias frames were obtained by the VLT service observers at the beginning of each night of observations. Lamp-flats were also taken with each of the masks with the observation setup in place (i.e. the OS\_Blue filter and LR\_Blue grism). These were also taken by the service observers at the beginning of each night's observation. Arc frames were taken during the night with each of the masks with the LR\_Blue grism and OS\_Blue filter.

Data reduction was performed using the VIMOS pipeline software, ESOREX. Firstly the bias frames were combined to form a master bias using VMBIAS. The flat frames were then processed and combined using the VMSPFLAT recipe. VMSPCALDISP was then used to process (bias subtract and flat-field) the arc lamp exposure and to determine the spectral distortions of the instrument. We measured a mean RMS on the inverse dispersion solution (IDS) of $2.3\pm0.6$ \AA. With the bias, flat and arc exposures all processed, the object frames were reduced and combined using the VMMOSOBSSTARE recipe to produce the reduced 2-D spectra. The spectra have not been fully flux calibrated, however we have applied the master response curves for the LR\_Blue grism to correct for the effects of the grism as a function of wavelength.

We extract the 1-D spectra using purpose-written IDL routines. For each spectrum, we first fit the shape of the spectrum across the slit. This is implemented by binning the 2-D aperture along the dispersion axis and then fitting a Gaussian profile to each bin to find the centre of the object signal in each bin. We then fit the resultant spread in the central pixel with a 4th order polynomial function. We then lay an object aperture with a width of $n_{ap}$ pixels over the object and a sky aperture covering all of the usable sky region in the slit. The object and sky spectra are then taken as being the mean over the widths of their respective apertures. Finally, we subtract the sky spectrum from the object spectrum to produce the final object spectrum. The dominant remaining sky-contamination after sky-subtraction were the strong sky emission lines [OI]5577 \AA\, [NaI]5890 \AA\ and [OI]6300 \AA.

\begin{figure}
\centering
\includegraphics[width=80.mm]{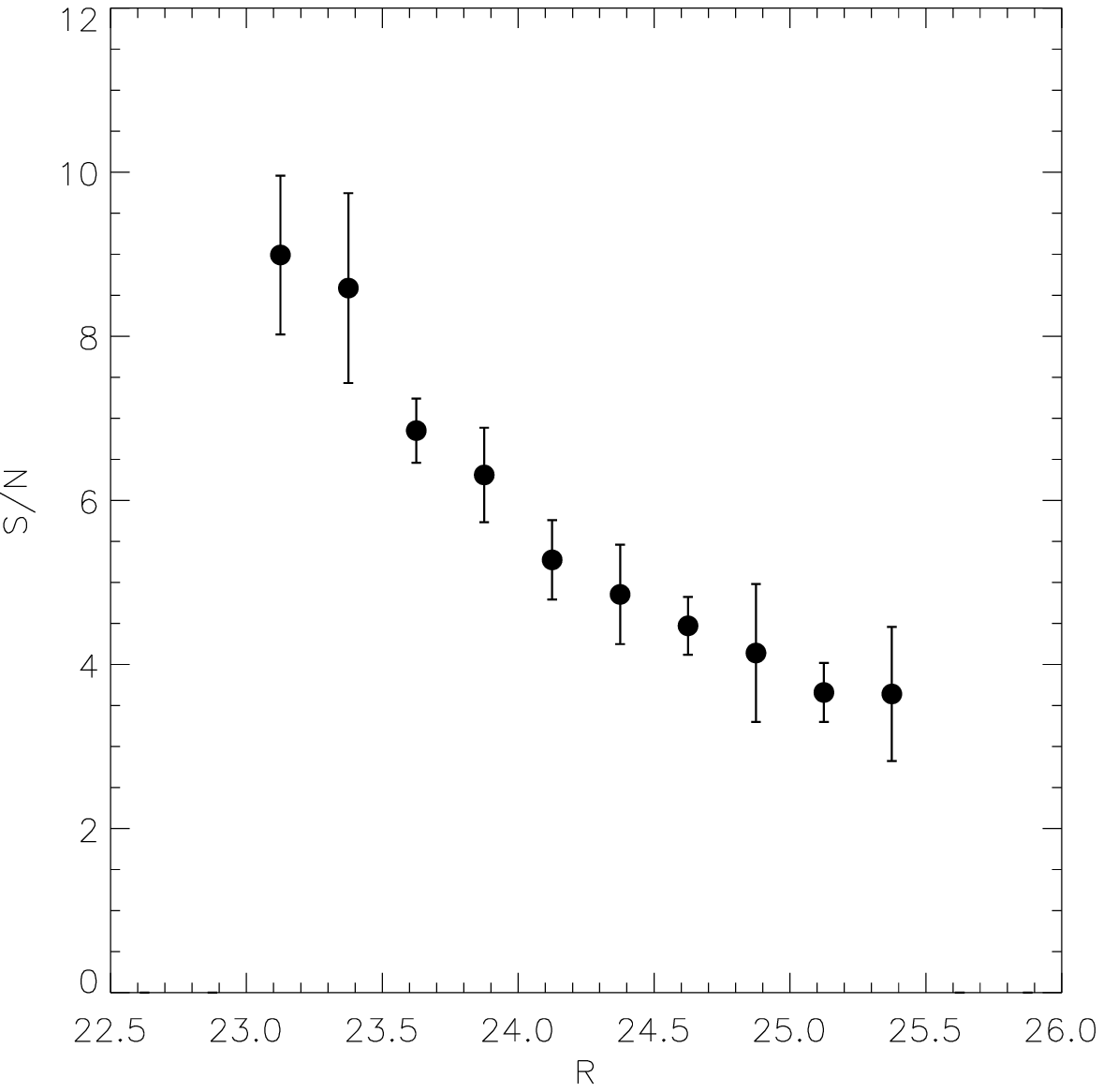}
\caption[{\small Mean continuum signal-to-noise per resolution element obtained from the spectroscopic data.}]{\small Mean signal-to-noise per resolution element (28\AA) in the wavelength range 4100\AA$<\lambda<$5300\AA\ as a function of $R$-band magnitude in our VLT VIMOS spectra with integration times of 10,000s.}
\label{snr}
\end{figure}

We estimate the signal-to-noise by taking the RMS of the sky aperture in each wavelength bin and dividing by $\sqrt{n_{ap}}$, where $n_{ap}$ is the width of the aperture used to extract the 1-D spectrum of a given object. Fig.~\ref{snr} shows the mean signal-to-noise per resolution element (i.e. 28\AA) in the wavelength range 4100\AA$<\lambda<$5300\AA\ in our sky-subtracted spectra as a function of source $R$-band magnitude. The selected range covers many of the key emission and absorption lines exhibited in LBGs in the redshift range $2.5<z<3.5$, whilst excluding the strong sky lines. The points in Fig.~\ref{snr} show the mean spectrum SNR per resolution element, whilst the error bars show the standard deviation within each bin. In the faintest bin ($25.25<R<25.5$), we achieve a mean continuum signal-to-noise of $\approx3.5$. This rises to a continuum signal-to-noise $\approx9$ for our brightest objects ($23<R<23.25$).

\subsection{Object Identification}
\label{sec:lbgredshifting}
We perform the object identification for each slit individually by eye. Given the wavelength range covered by the LR\_Blue grism combined with the redshift range of our targets, $2<z<3.5$, there are several key spectral features that facilitate the identification of those targets. These are primarily:

\begin{itemize}
\item Lyman limit, 912\AA;
\item Ly$\beta$ emission/absorption, 1026\AA\;
\item OVI 1032\AA, 1038\AA;
\item Ly$\alpha$ forest, $<$1215.67\AA;
\item Ly$\alpha$ emission/absorption, 1215.67\AA;
\item Inter-stellar medium (ISM) absorption lines:
\begin{itemize}
\item SiII 1260.4\AA;
\item OI$+$SiII 1303\AA;
\item CII 1334\AA;
\item SiIV doublet 1393\AA\ \& 1403\AA;
\item SiII 1527\AA;
\item FeII 1608\AA;
\item AlII 1670\AA;
\end{itemize}
\item CIV doublet absorption/emission, 1548-1550\AA.
\end{itemize}

The most prominent of these features is most frequently the Ly$\alpha$ emission/absorption feature at 1215\AA. However, as discussed by \citet{shapley03}, the observed optical (rest-frame UV) absorption and emission features are thought to originate from an outflowing shell of material surrounding the core nebular region of the galaxy. These features do not therefore represent the redshift of the rest-frame of the galaxy but in fact of these outflows.

For each confirmed LBG we measure independently the redshift of the Ly$\alpha$ emission/absorption feature and the redshift of the ISM absorption features. In order to measure the Ly$\alpha$ redshift, we fit the feature with a Gaussian function allowing the amplitude, central wavelength and width to be free parameters. From these we determine the redshift and line-width of the feature. We note that absorption blue-wards of the emission wavelength produces an asymmetry in the observed emission line, however given the modest resolution of our observations the Gaussian fit is preferred to any more complex asymmetric fitting to the emission line.

We have performed an estimate of the accuracy of our redshift results by repeating the spectral line fitting method with mock spectra. Each mock spectrum consists of a single Gaussian emission line (i.e. $f=Ae^{-(\lambda-\lambda_\circ)^2/2\sigma^2}$) at a random redshift in the range $2.5<z<3.5$ and a FWHM of $1680\kps$ corresponding to a Gaussian width of $\sigma=850\kps$ (equivalent to the resolution of the instrument). Gaussian random noise was then added to the basic emission line shape to give the required signal-to-noise. For each mock spectrum, we then performed the Gaussian fitting, iteratively performing the process for a total of $10^4$ mock spectra at a given signal-to-noise. The difference between the input redshift and the Gaussian line fitting redshift was then measured for each of the iterations and the error estimated from the distribution of this difference in input and measurement. The process was repeated, increasing the emission line peak flux from 1 to 20$\times$ the Gaussian noise width.

\begin{figure}
\centering
 \includegraphics[width=80.mm]{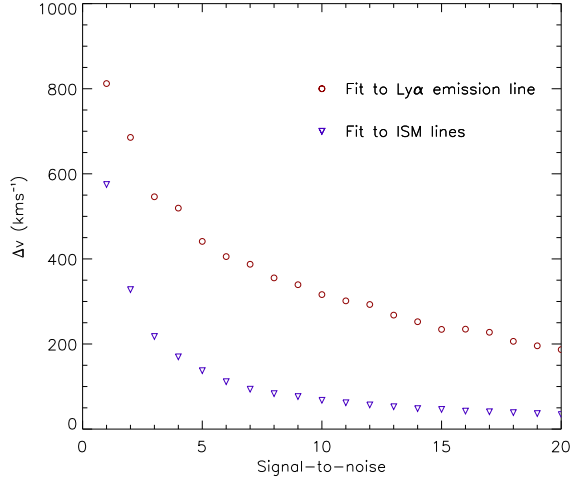}
\caption[{\small Estimate of the accuracy of the velocity measurements as a
function of spectral feature signal-to-noise.}]{\small Estimate of the accuracy
of the Gaussian line-fitting based on iteratively fitting mock spectra with
Gaussian random noise. The open circles show the results of applying the fitting
method to a single emission line spectrum with a range of signal-to-noise (where
the signal-to-noise is defined as the ratio between the peak signal and the
width of the Gaussian noise). The blue triangles show the result of the same
method applied to a simple absorption line spectrum including the ISM lines:
SiII (1260\AA), OI+SiII (1303\AA), CII (1336\AA) and SiIV (1393\AA, 1402\AA).}
\label{lbg:linefitting}
\end{figure}

\begin{figure}
\centering
 \includegraphics[width=80.mm]{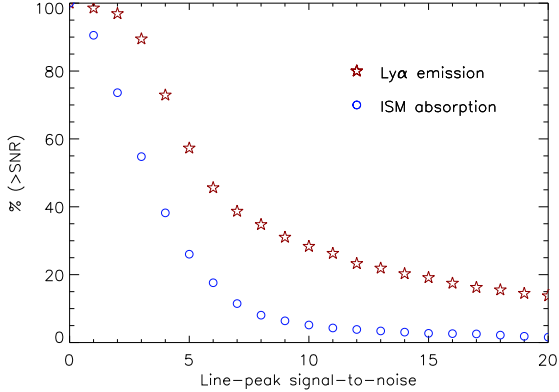}
\caption[{\small Percentage of LBGs with given peak Ly$\alpha$ and ISM
signal-to-noise estimates.}]{\small The distribution of Ly$\alpha$ emission line
(red stars) and ISM absorption line (blue circles) signal-to-noise measurements
in our LBG sample. The calculated signal-to-noise is the ratio between the
emission/absorption line peak (after subtracting the continuum) and the measured
noise. The final ISM signal-to-noise value is taken as the median of the
calculated values for the ISM lines used. See Fig.~\ref{lbg:linefitting} for
the estimated velocity errors based on the feature signal-to-noise.}
\label{lbg:data-linesnr}
\end{figure}

The results are given in Fig.~\ref{lbg:linefitting}, where the measured
accuracy is plotted as a function of the calculated signal-to-noise (red
circles). Further to this, we measure the distribution of Ly$\alpha$ emission
peak signal-to-noise in our galaxy sample, which is shown in
Fig.~\ref{lbg:data-linesnr} as a percentage of the total number of LBGs
exhibiting Ly$\alpha$ emission. If we now compare these two plots, we see that
$\approx90\%$ of our emission line LBGs have an emission line signal-to-noise of
$>3$, which suggests that $90\%$ of the Ly$\alpha$ emission line redshifts have
velocity errors of less than $\approx550\kps$. Further, the median Ly$\alpha$
emission line signal-to-noise is $\approx5.5$ which gives a velocity error of
$\approx400\kps$. Our higher quality spectra (i.e. the top $20\%$) however, are
estimated to achieve velocity errors on the Ly$\alpha$ emission line redshifts
as small as $\approx200\kps$.

Where feasible, we also attempt to measure the redshift of the ISM absorption
lines based on the SiII, OI+SiII, CII and SiIV doublet (despite being a mixture of high and low ionization lines we note that they are all measured to have comparable velocity offsets in \citealt{shapley03}, at least within the resolution constraints afforded by our observations). We primarily use absorption lines between $1215\mbox{\AA}\lesssim\lambda_{rest}\lesssim1500\mbox{\AA}$ as these remain within the wavelength coverage of the low-resolution blue grism over the full redshift range (i.e. $2\lesssim z\lesssim3.5$) of our survey. Measuring the individual absorption lines in most of our spectra is difficult given the SNR of the
absorption features in our spectra, however our ability to estimate the redshift
of the ISM lines can be greatly improved by attempting to determine the mean ISM
redshift by fitting the five lines simultaneously.

To evaluate this method we repeat the iterative error analysis performed for the
Ly$\alpha$ emission line fitting, but fitting five absorption lines (with
$\sigma_{ISM}=850\kps$) simultaneously. Again we measure the offset between the
input redshift and the output redshift measured from the Gaussian line fitting.
The result is again plotted in Fig.~\ref{lbg:linefitting} (blue triangles),
whilst the distribution of ISM signal-to-noise measurements in the data is again
given in Fig.~\ref{lbg:data-linesnr}. This suggests that we may reasonably
expect a significant improvement in the estimated redshift compared to measuring
just a single line. We now predict an accuracy of $\approx200\kps$ at a
signal-to-noise of $\approx3$, which based on Fig.~\ref{lbg:data-linesnr}
accounts for $55\%$ of our sample.

With the Ly$\alpha$ and ISM redshifts determined, we estimated the intrinsic
redshifts, $z_{\mathrm{int}}$, of our LBG sample using the relations of
\citet{adelberger05}. These relations were derived from a sample of 138 LBGs
observed spectroscopically in both the optical and the near infrared and are
based on the offsets found between the Ly$\alpha$ plus ISM lines and the nebular
emission lines, [OII]3727\AA, H$\beta$, [OIII]5007\AA\ and H$\alpha$. These
lines are all associated with the central star-forming regions of LBGs as
opposed to the outflowing material and are thus expected to be more
representative of the intrinsic redshift of a given LBG. The relations of
\citet{adelberger05} that we use here are as follows:

For LBGs with only a redshift from the Ly$\alpha$ emission line we used:
\begin{equation}
\label{eq:ad1}
z_{\mathrm{int}} = z_{\mathrm{Ly}\alpha} - 0.0033 - 0.0050(z_{\mathrm{Ly}\alpha}-2.7)
\end{equation}
For objects with Ly$\alpha$ absorption and a measurement of $z_{\mathrm{ISM}}$ we used:
\begin{equation}
\label{eq:ad2}
z_{\mathrm{int}} = z_{\mathrm{ISM}} + 0.0022 + 0.0015(z_{\mathrm{ISM}}-2.7)
\end{equation}
And for objects with redshifts measured from both the Ly$\alpha$ emission line and the ISM absorption lines we used:
\begin{equation}
\label{eq:ad3}
z_{\mathrm{int}} = \overset{\_}{z} + 0.070\Delta z - 0.0017 - 0.0010(\overset{\_}{z}-2.7)
\end{equation}
where $\overset{\_}{z}$ is the mean of the Ly$\alpha$ redshift ($z_{\mathrm{Ly}\alpha}$) and the ISM absorption line redshift ($z_{\mathrm{ISM}}$) and $\Delta z\equiv z_{\mathrm{Ly}\alpha}-z_{\mathrm{ISM}}$. \citet{adelberger05} quote rms scatters of $\sigma_z=0.0027$ ($200\kps$), $0.0033$ ($250\kps$) and $0.0024$ ($180\kps$) respectively for each of the above relations based on their application to their optical and IR spectroscopic sample of LBGs. 

\begin{figure*}
\centering
\includegraphics[width=160.mm]{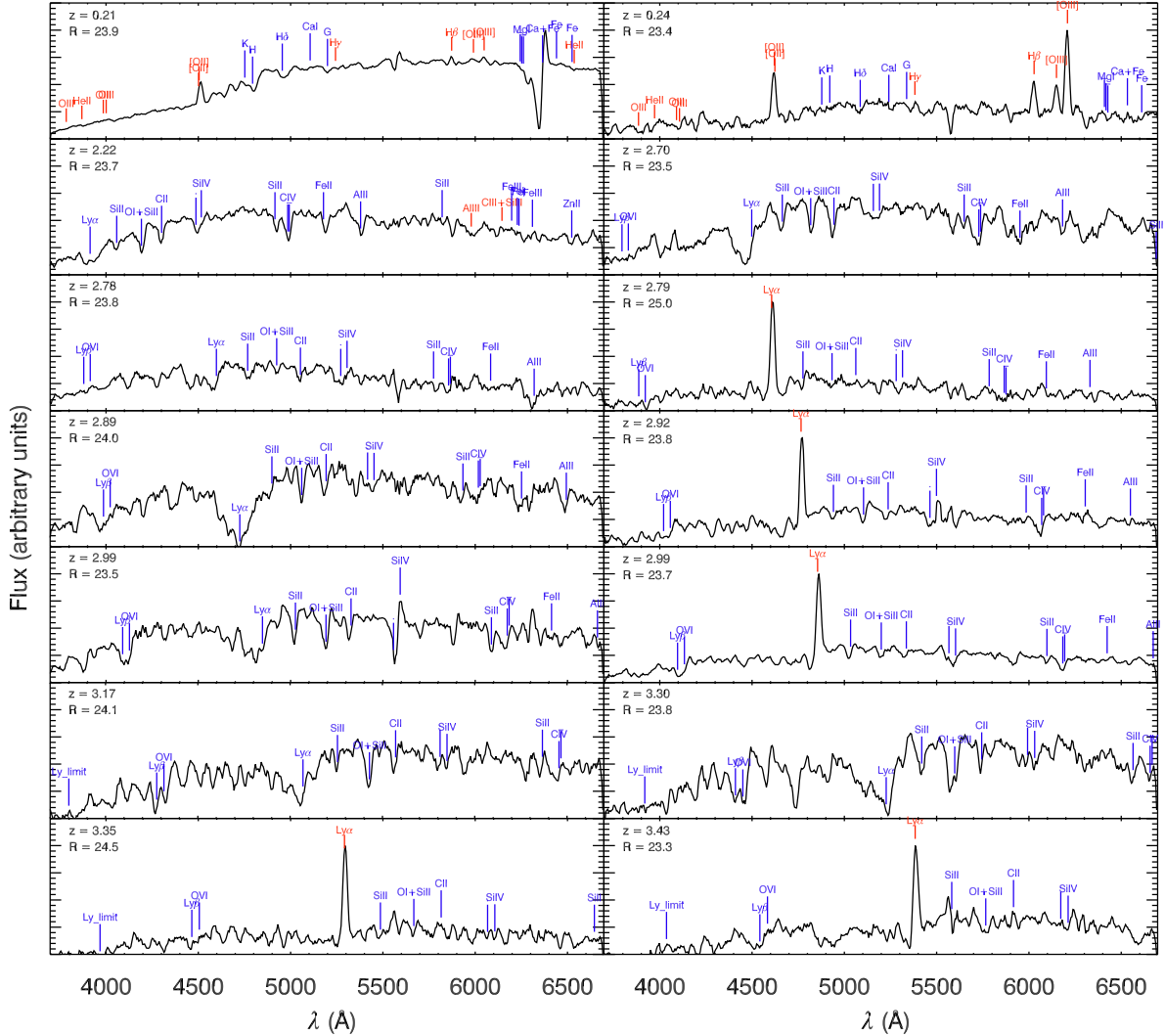}
\caption[{\small Example spectra taken using 10,000s integration time with the
LR\_Blue grism on the VLT VIMOS instrument.}]{{\small Example spectra taken
using 10,000s integration time with the LR\_Blue grism on the VLT VIMOS
instrument. The top two spectra are examples of contaminating low-redshift
galaxies. The remaining 12 panels show LBG spectra exhibiting both Ly$\alpha$
emission and absorption over the redshift range $2<z<3.5$. ISM lines are also
clearly identifiable in the individual LBG spectra as is the Lyman limit. Both
galaxy redshift and apparent $R$-band magnitude (Vega) are quoted for each object.
Note that all the above spectra have been binned to $\approx16$\AA.}}
\label{exspec}
\end{figure*}

As well as $z\approx3$ galaxies, our selection also samples a number of contaminating objects. These consist of low-redshift emission line galaxies (identified by [OII]3727\AA, H$\beta$, [OIII]5007\AA\ and H$\alpha$ emission), low-redshift Luminous Red Galaxies (LRGs - identified by [OII]3727\AA\ emission, Ca H, K absorption and the 4000\AA\ break) and faint red stars (mostly M and K-type stars). We show examples of the spectra of several LBGs and contaminant low-redshift galaxies taken with the VLT VIMOS in this survey in Fig.~\ref{exspec} (note that these are not flux-calibrated spectra).

All identified objects, including stars and low-redshift galaxies, were assigned
a quality rating, q, based on the confidence of the identification. The value of
q was assigned on a scale of 0 to 1, with 1 being the most confident and 0 being
unidentified. All objects with $q<0.5$ were rejected as spurious identifications
and are not included in the spectroscopic catalogue used in the remainder of
this work. LBGs were generally classified as follows:

\begin{itemize}
\item 0.5 - Ly$\alpha$ emission or absorption line evident plus some 'noisy' ISM absorption features.
\item 0.6 - Ly$\alpha$ emission or absorption plus some ISM absorption features.
\item 0.7 - Ly$\alpha$ emission or absorption plus most ISM absorption features.
\item 0.8 - Clear Ly$\alpha$ emission or absorption plus all ISM absorption features.
\item 0.9 - Clear Ly$\alpha$ emission or absorption plus high signal-to-noise ISM features.
\end{itemize}

With this classification scheme, we have identified 392, 254, 170, 111 and 93
$z>2$ galaxies with $q=$0.5, 0.6, 0.7, 0.8 and 0.9 respectively.

\subsection{Sky Density, Completeness \& Distribution}
\label{sec:spnum}

We summarize the numbers of objects observed in Table~\ref{objtab}. Our mean
sky density for successfully identified LBGs is $0.24\text{arcmin}^{-2}$, whilst the
percentage of $z>2$ galaxies in the entire observed sample (the success rate
given in table~\ref{objtab}) is 27.5\%. The remaining observed objects are a
mix of low-redshift galaxies, stars and unidentified objects (generally very
low-signal to noise spectra). In the worst case field (J1201+0116), we have a
greater number of low-redshift galaxies than high redshift detections. We
attribute this to the relatively poor depth of the imaging observations in this
field. We also note that the PKS2126-158 field is at a
relatively low galactic latitude and thus was a higher proportion of
contamination by galactic stars. However, the field still shows a high
proportion of $z>2$ galaxies.

\begin{table*}
\caption[{\small Summary of objects identified in the VLT VIMOS observations.}]{{\small Summary of objects identified in the VLT VIMOS observations. The success rate is the number of successfully identified LBGs divided the total number objects observed. Example spectra of the high-redshift and low-redshift galaxies are shown in Fig.~\ref{exspec}. All 10 identified $z>2$ QSO spectra are provided in Fig.~\ref{qsospec}.}}
\label{objtab}
\centering
{\small
\begin{tabular}{@{}lccccccc@{}}
\hline
Field & Subfields & Slits& Galaxies & QSOs & Galaxies & Stars & Success rate\\
         &                    & & $z>2$ & $z>2.0$ & $z<2.0$   &           &                            \\
\hline
Q0042-2627    & 4 &876 & 264 ($0.29 \text{arcmin}^{-2}$) & 1 & 106  & 5  & 30.1\%\\
J0124+0044    & 4 &832 & 264 ($0.29 \text{arcmin}^{-2}$) & 0 & 54  & 18 & 31.7\% \\
HE0940-1050 & 3 &501& 169 ($0.25 \text{arcmin}^{-2}$) & 1 & 48  & 36 & 33.7\% \\
J1201+0116    & 4 &699 & 120 ($0.13 \text{arcmin}^{-2}$) & 5 & 144 & 72 & 17.2\% \\
PKS2126-158 & 4 &654 & 203 ($0.23 \text{arcmin}^{-2}$) & 3 & 49  & 126& 31.0\%\\
\hline
Total                 & 19& 3562& 1020 ($0.24 \text{arcmin}^{-2}$)& 10 & 401 & 257& 28.6\%\\
\hline
\end{tabular}
}
\end{table*}

\begin{figure}
\centering
\includegraphics[width=80.mm]{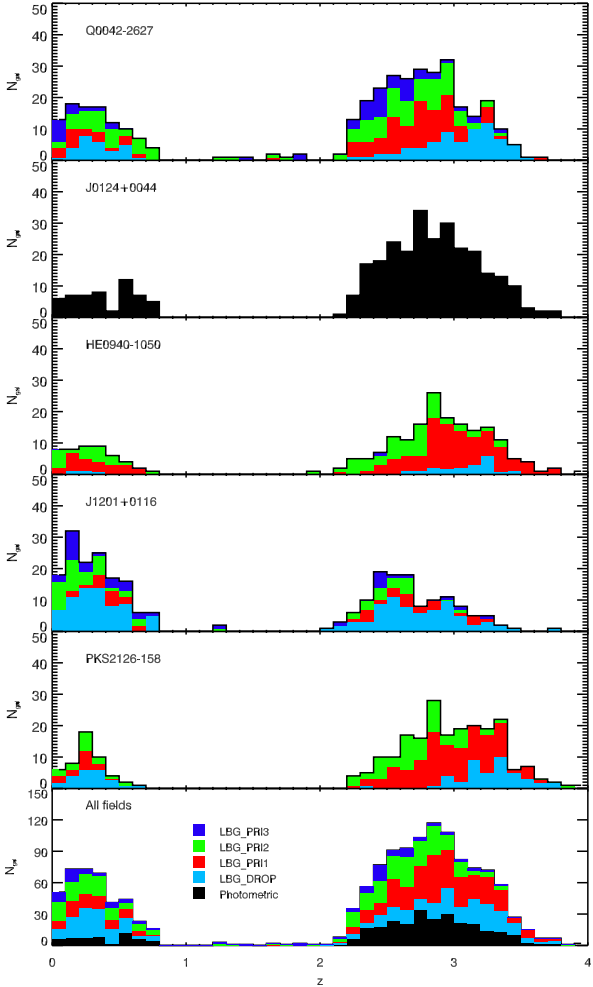}
\caption[{\small Differential redshift distribution in each of the LBG fields.}]{{\small Differential redshift distribution in each of our fields and summed over all fields. We show the number counts split by selection criteria: LBG\_DROP (cyan histograms), LBG\_PRI1 (red histograms), LBG\_PRI2 (green histograms) and LBG\_PRI3 (blue histograms). The mean redshifts for each selection are given in table~\ref{zdist}.}}
\label{lbg-nz}
\end{figure}

In Fig.~\ref{lbg-nz} and Table~\ref{zdist} we summarize the redshift distributions of each of our sample selections in our observed fields. The overall redshift distribution across all fields is shown in the bottom panel of Fig.~\ref{lbg-nz}, with the black histogram showing the redshift distribution from $UBVI$ selected objects from J0124+0044 and the red, green, blue and cyan histograms showing the LBG\_PRI1, LBG\_PRI2, LBG\_PRI3 and LBG\_DROP selections respectively. The overall mean redshift for our confirmed LBG sample is $\overset{\_}{z}=2.85\pm0.34$. It is evident from the redshift distributions that the separate selection sets give slightly differing (but overlapping) segments in redshift space. As may be expected, the LBG\_DROP selection is the most biased towards the higher end of our redshift distribution, with an overall mean redshift across all our samples of $\overset{\_}{z}=2.99$. The LBG\_PRI1 selection provides a redshift range of $2.90\pm0.32$, whilst the LBG\_PRI2 and LBG\_PRI3 give comparable redshift distributions of $2.67\pm0.26$ and $2.67\pm0.31$ respectively. We also show the redshift distributions for each individual field in the top five panels of Fig.~\ref{lbg-nz}, with the LBG\_PRI1, LBG\_PRI2, LBG\_PRI3 and LBG\_DROP identically to that in the 'all fields' plot. In each field we again see that the LBG\_PRI3 and LBG\_PRI2 selections preferentially select the lowest redshift ranges followed by LBG\_PRI1 and LBG\_DROP showing the highest redshift range (although this is less pronounced in the J1201+0116 field in which the imaging depths were least faint).

\begin{table*}
\caption[{\small Redshift ranges of $z>2$ galaxies identified from each of our photometric selections.}]
{\small Redshift ranges of $z>2$ galaxies identified from each of our photometric selections.}
\label{zdist}
\centering
{\small
\begin{tabular}{@{}lccccc@{}}
\hline
Field & LBG\_PRI1 & LBG\_PRI2 & LBG\_PRI3 & LBG\_DROP \\
\hline
Q0042-2627  & $2.74\pm0.28$ & $2.66\pm0.26$ & $2.67\pm0.30$ & $3.04\pm0.28$  \\
J0124+0044  &\multicolumn{4}{c}{$2.86\pm0.34$}  \\
HE0940-1050 & $3.02\pm0.33$ & $2.67\pm0.29$ & $2.85\pm0.39$ & $3.10\pm0.21$  \\
J1201+0116  & $2.71\pm0.29$ & $2.45\pm0.41$ & $2.61\pm0.29$ & $2.74\pm0.33$  \\
PKS2126-158 & $2.98\pm0.29$ & $2.72\pm0.27$ & n/a           & $3.30\pm0.29$  \\
\hline
All fields  & $2.90\pm0.32$ & $2.66\pm0.28$ & $2.67\pm0.30$ & $2.99\pm0.36$  \\
\hline
\end{tabular}
}
\end{table*}

\begin{figure*}
\centering
\includegraphics[width=160.mm]{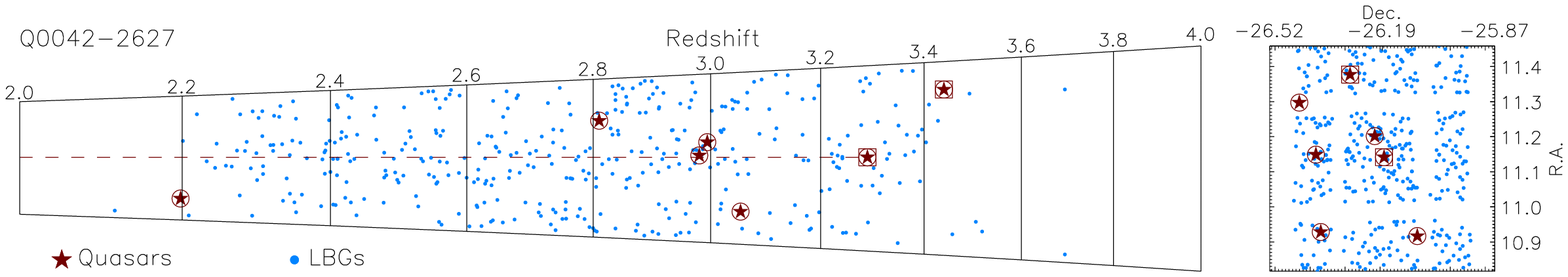}\\
\includegraphics[width=160.mm]{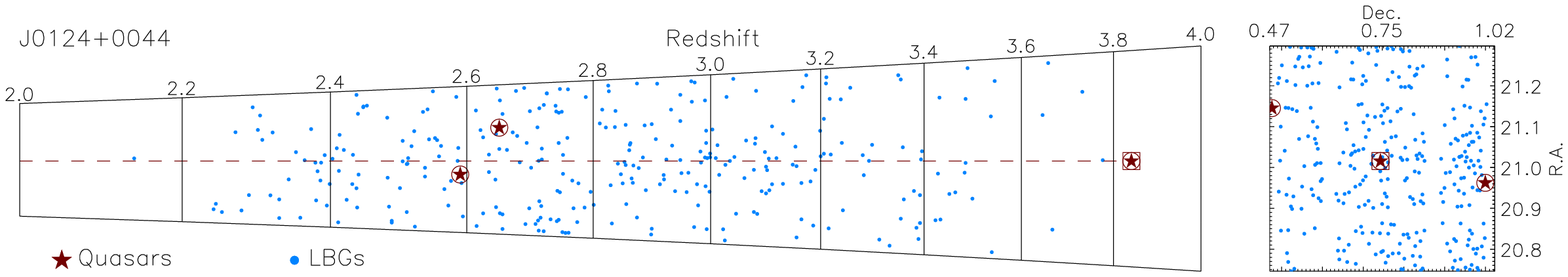}\\
\includegraphics[width=160.mm]{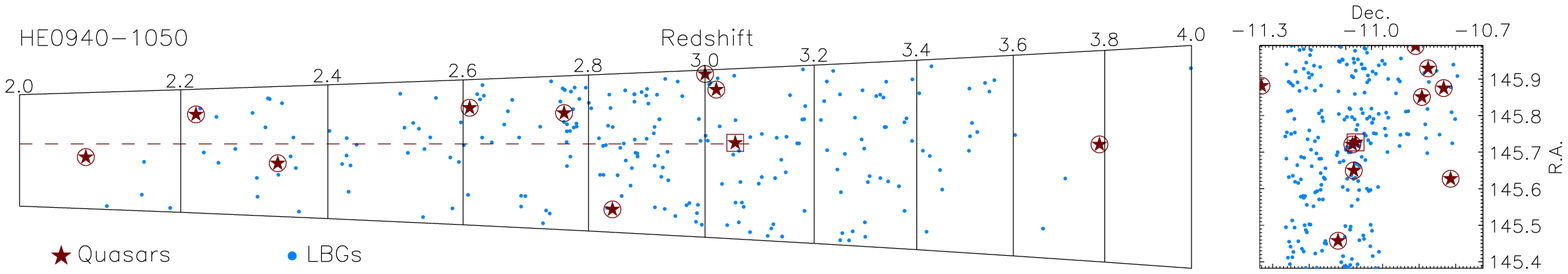}\\
\includegraphics[width=160.mm]{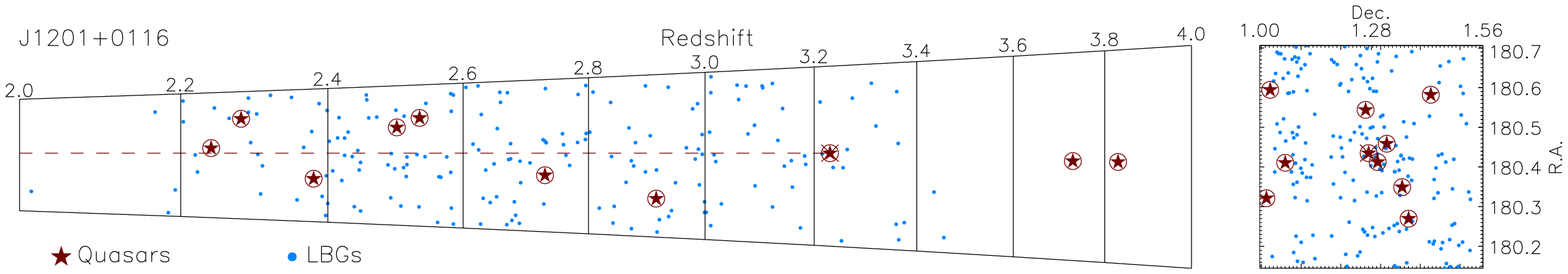}\\
\includegraphics[width=160.mm]{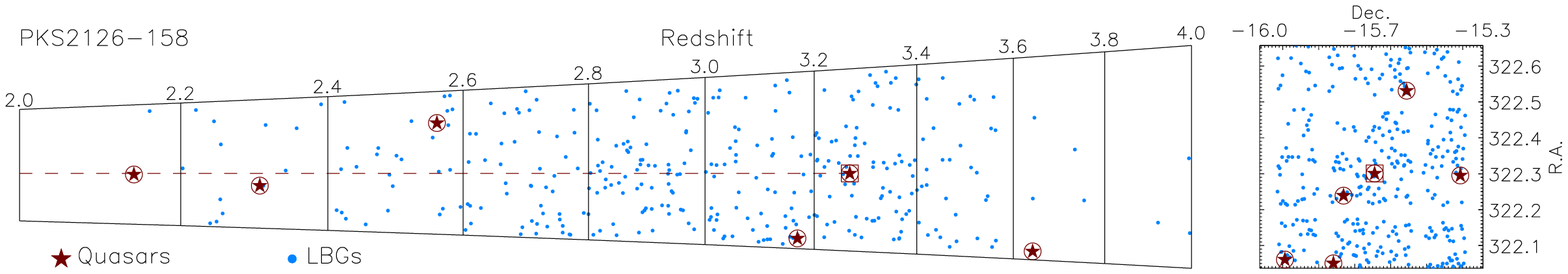}
\caption[{\small Distribution in R.A., Declination and redshift for each of our five fields.}]{{\small Distribution in R.A., Declination and redshift for each of our five fields. Spectroscopically confirmed LBGs are marked by blue filled circles and known QSOs by dark red stars. We also identify those QSOs with low-resolution spectra available (red circles, i.e. VLT VIMOS and AAT AAOmega), medium-resolution spectra (red crosses, i.e. SDSS - SDSS J1201+0116 only) and high-resolution spectra (red squares, i.e. VLT UVES, Keck HIRES).}}
 \label{wedges}
\end{figure*}

We illustrate the distribution of our spectroscopic LBG sample in each of our 5 fields in Fig.~\ref{wedges}. The fields are ordered by R.A. top to bottom and all identified $z>2$ galaxies (filled blue circles) are shown along with all known $z>2$ QSOs identified from the NASA Extragalactic Database. We also plot the positions of QSOs identified in our VIMOS observations and AAOmega QSO survey, which is described further in \citet{2010arXiv1006.4385C}.

\begin{figure}
\centering
\includegraphics[width=80.mm]{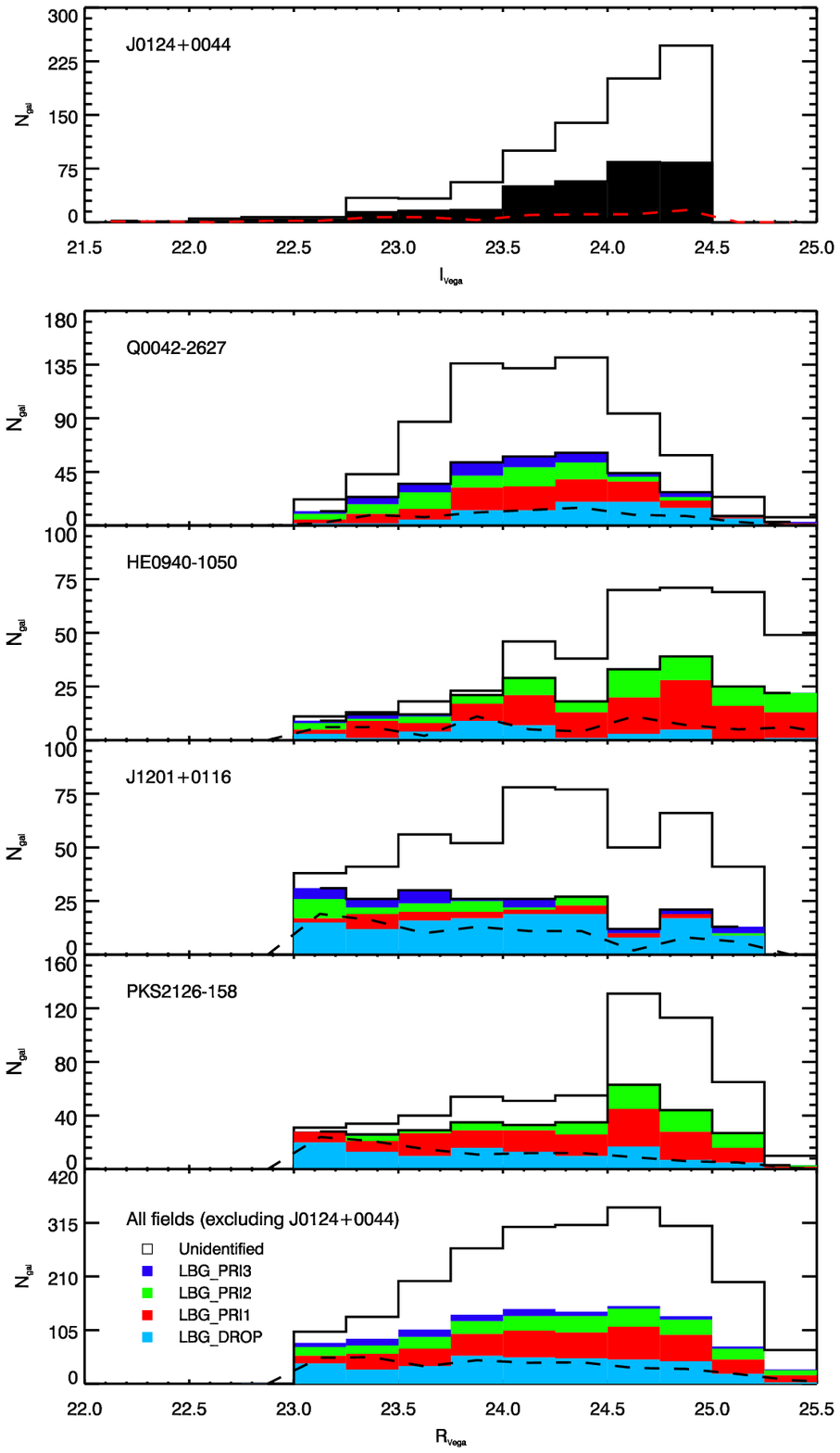}
\caption[{\small Number counts as a function of R$_{\mathrm{Vega}}$ (I$_{\mathrm{Vega}}$)
magnitude for all fields (J0124+0044).}]{{\small Number counts as a function of
R$_{\mathrm{Vega}}$ magnitude for all fields, except for J0124+0044 in which I$_{\mathrm{Vega}}$ is used. The shaded histograms show the numbers of successfully identified objects with the colour coding the same as in Fig.~\ref{lbg-nz}: the cyan histogram shows counts of LBG\_DROP objects, the red shows LBG\_PRI1 objects, the green shows LBG\_PRI2 objects and the blue shows LBG\_PRI1 objects. The unshaded histogram shows the total number of candidates observed with VLT-VIMOS in each field (i.e. the gap between the shaded regions and solid line shows the number of unidentified objects as a function of magnitude). Contamination levels from stars and low-redshift galaxies for each field are given by the dashed line in each panel.}}
\label{lbg-nm}
\end{figure}

In Fig.~\ref{lbg-nm} we plot the number of identified LBGs in magnitude bins for each of our fields. The filled histograms show the cumulative numbers of successfully identified objects (including interlopers as well as $z>2$ galaxies) split by their selection criteria. LBG\_DROP selected objects are shown by the cyan histogram, LBG\_PRI1 by the red histogram, LBG\_PRI2 by the green histogram and LBG\_PRI3 by the blue histogram. The distribution of all spectroscopically observed objects is given by the solid line histogram in each case. As the J0124+0044 objects were not selected using the same selection criteria, these are simply left as a single group shown by the filled black histogram. In all fields, we see that we are successfully identifying objects down to the magnitude limit of R$=25.5$ (I$=25$ in the case of J0124+0044), although a significant number of objects remain unidentified in each field at the fainter magnitudes as spectral features become more difficult to discern in the spectra. We note also that the shapes of the overall magnitude distributions are biased more towards brighter objects in the Q0042-2627 and J1201+0116 fields in which a greater number of LBG\_PRI3 objects are included (and also the imaging depths achieved in these fields are shallower than in the other fields).

\begin{figure}
\centering
\includegraphics[width=80.mm]{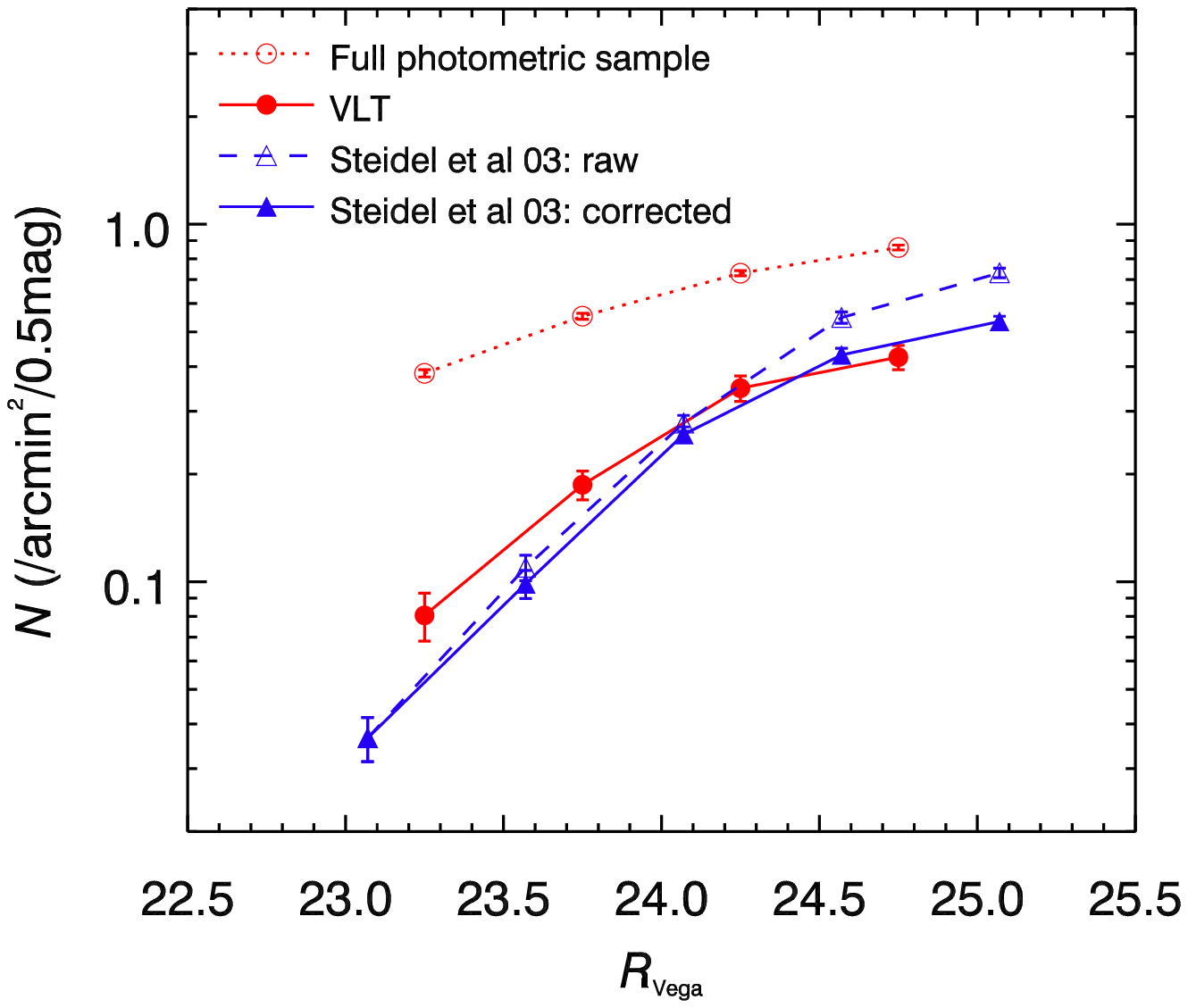}
\caption[{\small Predicted sky densities of the LBG sample.}]{\small Sky densities of the LBG sample as a function of $R_{\mathrm{Vega}}$ magnitude. The red open circles give the total densities of objects in our LBG\_PRI1, LBG\_PRI2 and LBG\_DROP photometric selections. The `VLT' densities (filled red circles) are estimated using the total photometric densities multiplied by the fraction of successfully identified LBGs from the VLT spectroscopic observations and are corrected for incompleteness in the spectroscopic sample at faint magnitudes. Raw (open blue triangles) and corrected (filled blue triangles) number counts are also shown from \citet{steidel03}. Note that we transform the \citet{steidel03} AB system ${\cal R}$ magnitudes by $-0.14$ to convert to $R_{\mathrm{Vega}}$ \citep{1993AJ....105.2017S}.}
\label{nc_lbg}
\end{figure}

In Fig.~\ref{nc_lbg}, we show the number counts of our photometrically selected LBGs (open red circles) and the estimated number counts of LBGs (filled red circles) derived from the candidate number counts and the success rate as a function of magnitude (i.e. the number of confirmed LBGs divided by the number of observed candidates). At faint magnitudes we correct the counts for incompleteness in the spectroscopic observations, however we have not made any correction for incompleteness in the original photometry. The number counts of \citet{steidel03} are also plotted, showing their candidate number counts (open blue triangles) and number counts corrected for contamination (filled blue triangles). The two data-sets show good agreement over the magnitude ranges sampled.

%We see that the VLT number counts are in agreement at $R_\textrm{Vega}<24$, whilst they are lower than those of \citet{steidel03} at magnitudes of $R>24$. This is not a physical phenomenon, but rather a consequence of the relatively low number counts in the U and $B$-band imaging observations from a number of our imaging fields (in particular Q0042-2627 and J1201+0116).

\subsection{Velocity Offsets and Composite spectra}

\begin{figure}
\centering
\includegraphics[width=80.mm]{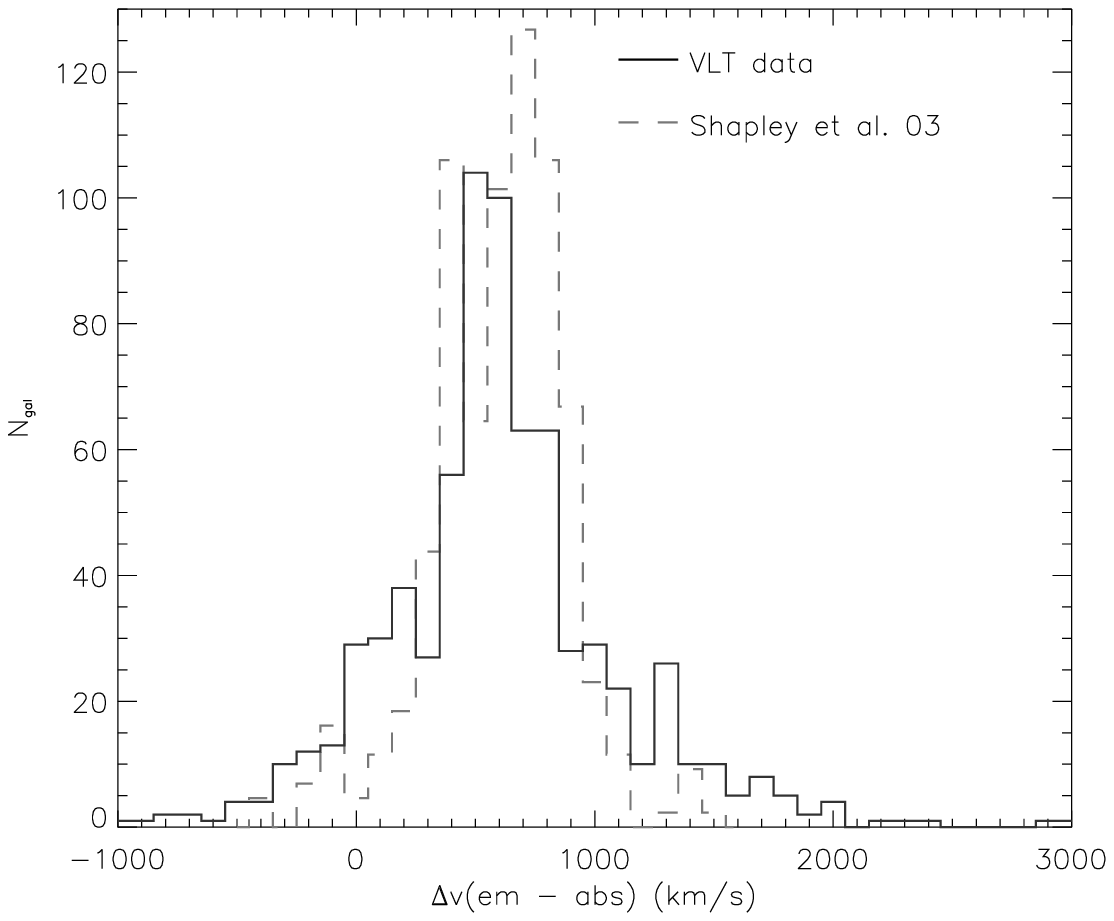}
\caption[{\small Distribution of the velocity offsets between ISM absorption
lines and the Ly$\alpha$ emission line in individual galaxies.}]{{\small
Distribution of the velocity offsets between ISM absorption lines and the
Ly$\alpha$ emission line in individual galaxies from our redshift survey (solid
histogram). We measure a mean velocity offset between Ly$\alpha$ emission and
the ISM lines of $\overset{\_}{\Delta V}=625\pm510\kps$. The result of
\citet{shapley03}, which has a mean of $650\kps$ is shown by the dashed
histogram.}}
\label{vdiff}
\end{figure}

The galaxy spectra contain a wealth of information as illustrated by the work of
\citet{shapley03}. We now look at how our spectra compare to previous work in
terms of the velocity offsets between the different spectral
features. For the galaxies that exhibit both measurable Ly$\alpha$ emission and
ISM absorption lines, we calculate the velocity offsets between these lines,
$\Delta v=v_{em}-v_{abs}$. The distribution of $\Delta v$ for our galaxy sample
is shown in Fig.~\ref{vdiff}. The distribution of velocity offsets exhibits a
strong peak with a mean of $\left<\Delta v\right>=625$ with a dispersion of $510\kps$. This compares
to a value measured by \citet{shapley03} of $650\kps$.

We have produced composite spectra in several Ly$\alpha$ equivalent width bins in order to produce spectra with increased signal-to-noise compared to the individual galaxy spectra. The \lya\ profile can be very complex, consisting of both emission and absorption features and this combination often leads to asymmetric profiles with a significant amount of absorption blue-wards of the emission line \citep{shapley03,2010ApJ...711..693K}. For the purposes of producing composite spectra of the LBGs, we take a relatively simple approach to the measurement of the equivalent widths of our galaxy sample. For a given spectrum, we measure an equivalent width for the emission line if clearly identifiable and if not we make a measurement of the absorption profile. To do this, we fit a polynomial to the continuum and a Gaussian fit to the \lya\ line profile and estimate the equivalent width from these fits.

The individual LBG spectra were normalized prior to constructing the composite, using the median of the rest-frame UV continuum in the range $1300\mbox{\AA}\lesssim\lambda_{rest}\lesssim1500\mbox{\AA}$. After this normalization, we rescale the LBG spectra to the rest-frame and re-binned the spectra before combining the samples to produce the final composite spectra. We note that all the spectra were calibrated using the VIMOS master response curves prior to this process. 

\begin{figure}
\centering
\includegraphics[width=80mm]{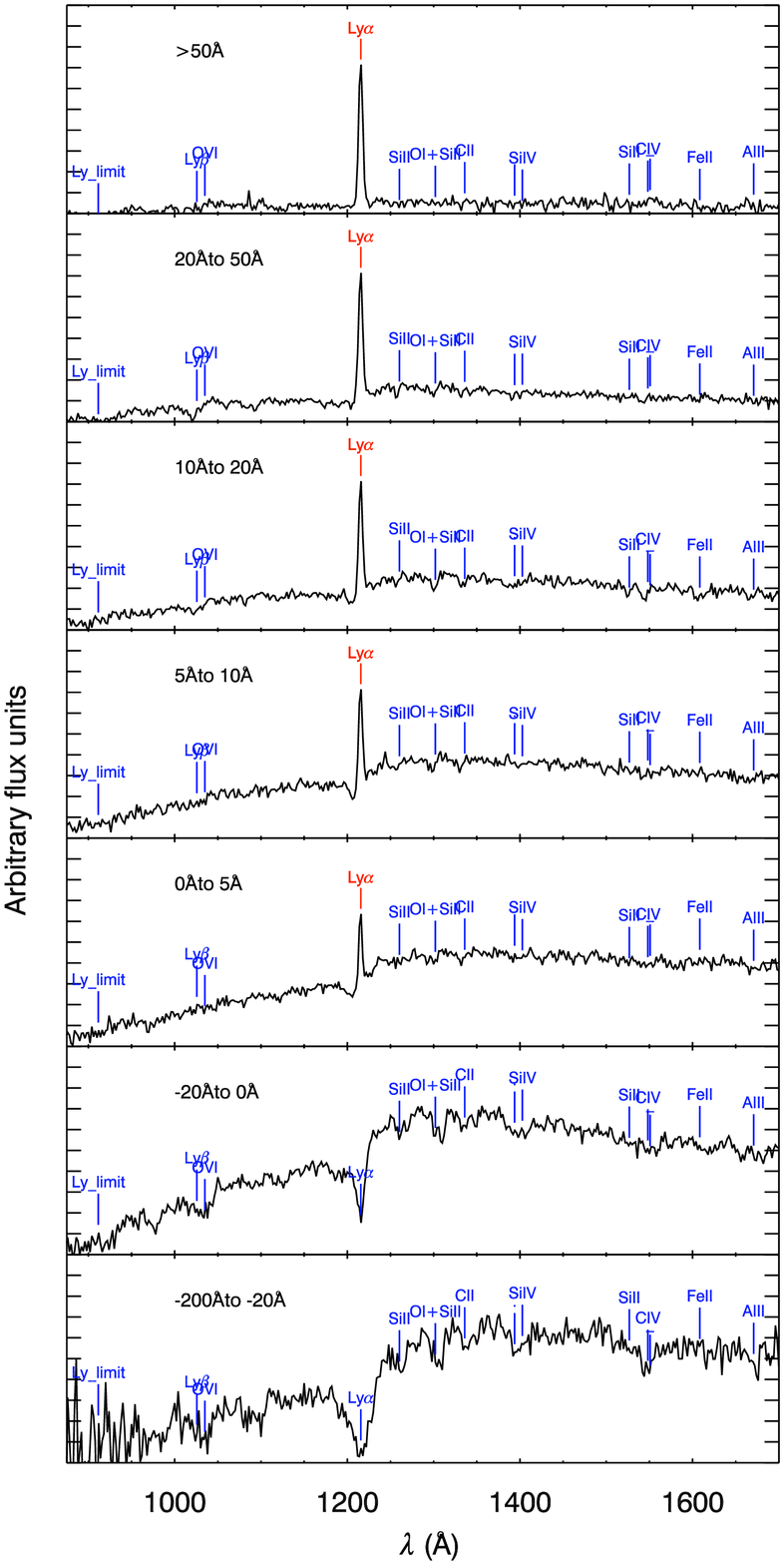}
\caption[{\small Composite spectra collated from our VLT VIMOS sample.}]{{\small
Composite spectra collated from our VLT VIMOS sample. Each spectrum shows the
composite of a sub-sample of the LBGs, grouped by Ly$\alpha$ equivalent width
measurements. The key UV spectral features discussed in the text (i.e.
Ly$\alpha$ and Ly$\beta$ emission/absorption, ISM absorption lines) are all
evident in these composite spectra.}}
\label{composite}
\end{figure}

The composite spectra are shown in Fig.~\ref{composite} and are split into (from bottom to top) equivalent width ranges of W$<-20$\AA\ (50 galaxies), -20\AA$<$W$<$0\AA\ (134 galaxies), 0\AA$<$W$<$5\AA\ (166 galaxies), 5\AA$<$W$<$10\AA\ (218 galaxies), 10\AA$<$W$<$20\AA\ (181 galaxies), 20\AA$<$W$<$50\AA\ (112 galaxies) and W$>$50\AA\ (60 galaxies). Between them, the composites incorporate a total of 921 of the galaxy sample, excluding any objects with $q<0.5$ or with
significant contamination, for example from zeroth order overlap. The key emission and absorption features are marked and we can immediately identify both absorption and weak emission for the ISM lines: SiII, OI+SiII, CII, SiIV and CIV. All the features have been marked at $z=0$. The offset between the line centres of the Ly$\alpha$ emission and the ISM absorption lines is evident in these composite spectra, a result of the asymmetry of the Ly$\alpha$, potentially combined with an intrinsic difference between the velocities of the sources of the Ly$\alpha$ emission and the ISM absorption features.

\subsection{VLT AGN and QSO observations}

\begin{figure*}
\centering
\includegraphics[width=160.mm]{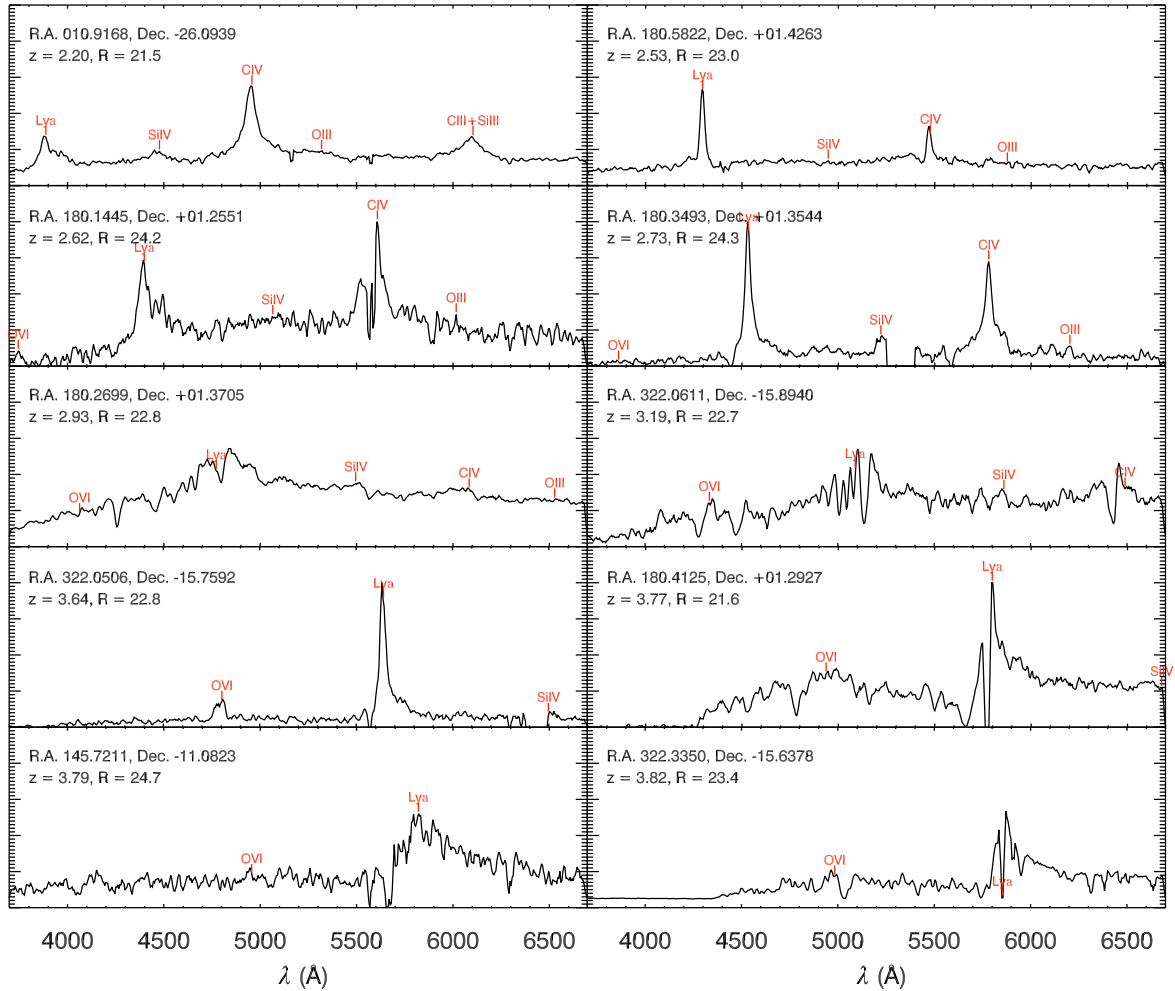}
\caption[{\small The $z>2$ QSOs observed as part of the VLT VIMOS LBG
survey.}]{{\small Spectra of the $z>2$ QSOs observed as part of the VLT VIMOS LBG survey.
Redshifts and $R$-band magnitudes are given for each QSO and significant broad
emission features are marked.}}
\label{qsospec}
\end{figure*}

As discussed earlier, we also targeted a small number of $z\approx3$ QSO
candidates selected from our $UBR$ photometry. In combination with this, due
to the similarity in the shape of the spectra of LBGs and QSOs, the LBG
selections also produced a handful of faint QSOs and AGN. We present the spectra
of these in Fig.~\ref{qsospec}, whilst the numbers of QSOs in each field are
given in table~\ref{objtab}. The positions of the observed QSOs are also shown
in Fig.~\ref{wedges}.

\section{Clustering}
\label{sec:clustering}

In this section we present the clustering analysis of the $z\approx 3$ galaxy
sample, incorporating estimates of the angular auto-correlation function for our
complete LBG candidates catalogue and the redshift space auto-correlation
function of our spectroscopically confirmed sample. Developing from these
estimates, we use a combined sample of the VLT VIMOS LBG data-set and the
\citet{steidel03} data-set to evaluate the 2-D correlation function and place
constraints on the infall parameter, $\beta$, and the bias paremeter, $b$. 
Finally, we relate the clustering properties of the $z\approx 3$ sample to those of lower-redshift samples.

\subsection{Angular Auto-correlation Function}
\label{sec:autocorr}

We now evaluate the clustering properties of our candidate and spectroscopically
confirmed LBGs. Using all five of our imaging fields, we begin by calculating
the angular correlation function of the LBG candidates. We use all LBG
candidates selected using the LBG\_PRI1, LBG\_PRI2, LBG\_PRI3 and LBG\_DROP
selections plus the candidates from the J0124+0044 field. The total number of
objects is thus 18,489 across an area of 1.8deg$^2$. First we create an artificial galaxy catalogue consisting of a randomly generated spatial distribution of points within the fields. The angular auto-correlation function is then given by the Landy-Szalay estimator \citep{1993ApJ...412...64L}:

\begin{equation}
\label{eq:landszal}
w(\theta) = \frac{\left<DD\right> - 2\left<DR\right> + \left<RR\right>}{\left<RR\right>}
\end{equation}

\noindent where $DD$ is the number of galaxy-galaxy pairs at a given separation, $\theta$, $DR$ is the number of galaxy-random pairs and $RR$ is the number of random-random pairs. The random catalogues were produced within identical fields of view to the data and with sky densities of $100\times$ the real object sky densities, in order to make the noise contribution from the random catalogue negligible. We estimated the statistical errors on the $w(\theta)$ measurement using the jack-knife estimator.

Measurements of $w(\theta)$ in small fields are subject to a bias known as the integral constraint \citep[e.g.][]{1977ApJ...217..385G,1980lssu.book.....P,1993MNRAS.263..360R}. This is given by:

\begin{equation}
\sigma^2=\frac{1}{\Omega^2}\int\int w(\theta)\rm{d}\Omega_1\rm{d}\Omega_2
\end{equation}

\noindent where the `true' $w(\theta)$ is then: 
\begin{equation}
w(\theta) = \left<w_{meas}(\theta)\right> + \sigma^2
\end{equation}

\noindent where $\left<w_{meas}(\theta)\right>$ is the measured correlation function, averaged across the observed fields, and $w(\theta)$ is the correct correlation function. As in \citet{2002MNRAS.337.1282R}, we evaluate the integral constraint using the numbers of random-random pairs in our fields:

\begin{equation}
\sigma^2=A\frac{\sum N_{RR}(\theta)\theta^{-\delta}}{\sum N_{RR}(\theta)}
\end{equation} 

\begin{figure}
\centering
\includegraphics[width=80.mm]{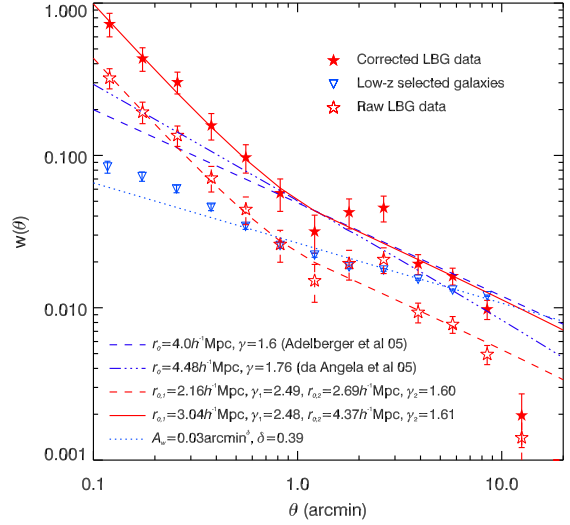}
\caption[{\small The angular correlation function, $w(\theta)$, from the five survey imaging fields.}]{\small The angular correlation function, $w(\theta)$, from our imaging fields. The open stars show the correlation function for the photometrically selected sample, whilst the filled stars show the same correlation function corrected for stellar and $z<2$ galaxy contamination as described in the text. The dashed red and solid red lines show the double power law models fitted to the raw and contamination corrected correlation functions respectively. We also show a model determined from the $r_0$, $\gamma$ measurements of \citet{daangela05b} - dash-dot blue line. The blue triangles and dotted line show the correlation function and best fitting power law model for the photometrically selected $z<2$ galaxy population. The blue dash line gives the result of \citet{2005ApJ...619..697A}, with $r_0=4.0\hmpc$ and $\gamma=1.57$.}
\label{lbg:wtheta}
\end{figure}

The results of the $w(\theta)$ calculation for the full photometrically selected LBG sample are shown in Fig.~\ref{lbg:wtheta} (open red stars). 

Additionally we show the correlation function, estimated in the same way, for
the remaining $23<R<25.5$ galaxy population (i.e. all galaxies in the given
magnitude range not selected by the LBG colour selection - blue triangles). This
gives an estimate of the clustering for the $z<2$ galaxy population in the LBG
fields. Based on the spectroscopic results, we estimate that $60\%$ of the
photometric selection consists of $z>2$ galaxies whilst the remaining $40\%$
consists of contaminant $z<2$ galaxies and galactic stars. In order to  determine 
more accurately the clustering of our selected $z>2$ galaxy population, we
therefore correct the $w(\theta)$ measurement for the effects of contamination.
The correction is given by:

\begin{equation}
\label{eq:wth_meas}
w_{\mathrm{meas}}(\theta) = w_{z<2}(\theta)f_{z<2}^2 + w_{\mathrm{LBG}}(\theta)f_{\mathrm{LBG}}^2
\end{equation}
\noindent where $w_{\mathrm{meas}}$ is the total measured correlation function,
$w_{z<2}(\theta)$ is the correlation function of the contaminant galaxies,
$f_{z<2}$ is the fraction of contaminant galaxies, $w_{\mathrm{LBG}}(\theta)$ is the
correlation function of the $z>2$ galaxies and $f_{\mathrm{LBG}}$ is the fraction of
$z>2$ galaxies. We therefore use the measured correlation function (i.e. open
red stars in Fig.~\ref{lbg:wtheta}) and the measured $z<2$ correlation
function (i.e. blue triangles in Fig.~\ref{lbg:wtheta}) along with the
spectroscopically measured fractions of $z>2$ and $z<2$ galaxies to estimate the
$z>2$ galaxy correlation function (i.e. $w_{\mathrm{LBG}}$). The result is shown by the
filled red stars in Fig.~\ref{lbg:wtheta}. At all scales we find a higher
measurement of the $z>2$ correlation function after applying this correction. We
note that the $w_{\mathrm{LBG}}(\theta)$ measurement shows signs of a change in slope at
$\theta\sim0.6-1\arcmin$, suggestive of the combination of one and two halo
terms used in Halo Occupation Distribution modeling (HOD, e.g.
\citealt{abazajian05,zheng05,2008MNRAS.387.1045W,zheng09}).

We now quantify the clustering amplitude of the raw and corrected $w(\theta)$
measurements using a simple power-law fit, with constants $A_w$ and $\delta$
such that:

\begin{equation}
\label{eq:wth_fit}
w(\theta) = A_w\theta^{-\delta}
\end{equation}

Fitting to the data to the large scale clustering
($0.8\arcmin<\theta<10\arcmin$) for the uncorrected $w(\theta)$ we obtain best
fit parameters of $A_w=1.08\pm0.27\times10^{-3}\text{deg}^{\delta}$ and
$\delta=0.76^{+0.07}_{-0.17}$. Using the same angular range with the corrected
$w(\theta)$ gives parameters of
$A_w=1.85^{+0.41}_{-0.21}\times10^{-3}\text{deg}^{\delta}$ and
$\delta=0.82^{+0.11}_{-0.12}$. We also perform a fit to the
$z<2$ correlation function. In this case, the clustering is fit by a power law
with $A_w=2.31^{+0.58}_{-0.58}\times10^{-3}\text{deg}^{\delta}$ and
$\delta=0.57^{+0.01}_{-0.01}$ (dotted blue line in Fig.~\ref{lbg:wtheta}).

We now estimate the real-space correlation function, $\xi(r)$, from our
measurement of $w(\theta)$ using Limber's formula \citep{phillipps78} with our
measured redshift distribution (Fig.~\ref{lbg-nz}). This is performed for both
the raw $w(\theta)$ and the contamination-corrected $w(\theta)$ with a double
power-law form of $\xi(r)$ given by:

\begin{equation}
\xi_1 = \left(\frac{r_{0,1}}{r}\right)^{-\gamma_1} (r<r_b)
\end{equation}
\begin{equation}
\xi_2 = \left(\frac{r_{0,2}}{r}\right)^{-\gamma_2} (r \ge r_b)
\end{equation}

\noindent where $r_b$ is the break at which the power-law is split between the
two power-laws, $r_{0}$ is the clustering length and $\gamma$ is the slope (which is
given by $\gamma=1+\delta$). We perform $\chi^2$ fitting over the $r_0$-$\gamma$ parameter space
to both the uncorrected and corrected $w(\theta)$ results. Firstly for the
uncorrected result, we find $r_{0,2}=3.14^{+0.17}_{-0.36}\hmpc$ and
$\gamma_2=1.81^{+0.09}_{-0.14}$.  For the corrected $w(\theta)$, we determine a
clustering length above the break of $r_{0,2}=4.37^{+0.43}_{-0.55}\hmpc$,
with a slope of $\gamma_2=1.61\pm0.15$. The full results are given in
table~\ref{tab:wtheta} and the best-fitting $w(\theta)$ models are plotted in
Fig.~\ref{lbg:wtheta}. We note that for continuity in the double power-law function, the break is found to be at $r_b\approx1.5\hmpc$.

%{\bf We find that the clustering result obtained by da Angela et al is a
%reasonable fit at large scales ($\theta>0.02$deg) but $w(\theta)$  steepens at
%smaller scales. Recall here that at z=3 the comoving radial distance is
%$4450\hmpc$ and so 0.02 deg corresponds to $1.5\hmpc$ in our assumed cosmology. We
%show the result of combining the above correlation function parameters with
%steeper power-law with $r_0=4.08\hmpc$ and $\gamma=2.3$ for $r_b<3\hmpc$.
%After projecting via Limber's formula, this double power-law for $\xi(r)$ gives
%an improved fit to $w(\theta)$ for $\theta<0.02$deg. }

\begin{table*}
\caption{\small Clustering results based on the raw
$w(\theta)$ and the $w(\theta)$ corrected for stellar and low-redshift galaxy
contamination.}
\label{tab:wtheta}
\centering
{\small
\begin{tabular}{@{}lcccccc@{}}
\hline
 & $A_w$ & $\delta$ & $r_{0,1}$ & $\gamma_1$ &$r_{0,2}$ &$\gamma_2$\\
 & ($\times10^{-3}$deg$^{\delta}$) & &($\hmpc$) & &($\hmpc$) & \\
\hline
Uncorrected data        & $1.08^{+0.27}_{-0.27}$ & $0.76^{+0.07}_{-0.17}$ & $2.16^{+0.24}_{-0.30}$ &$-2.49^{+0.09}_{-0.12}$ & $2.69^{+0.20}_{-0.26}$ & $-1.60\pm0.11$\\
Contamination-corrected & $1.85^{+0.41}_{-0.21}$ & $0.82^{+0.11}_{-0.12}$ & $3.04^{+0.33}_{-0.34}$ &$-2.48^{+0.10}_{-0.11}$ &$4.37^{+0.43}_{-0.55}$ &$-1.61\pm0.15$ \\
\hline
\end{tabular}
}
\end{table*}

Comparing our result to previous results, \citet{daangela05b} obtained a clustering length of $r_0=4.48^{+0.09}_{-0.14}\hmpc$ with a slope of
$\gamma=1.76^{+0.08}_{-0.09}$ and \citet{adelberger03} obtained $r_0=3.96\pm0.15\hmpc$ and $\gamma=1.55\pm0.29$, both using a single
power-law function fit ($\xi(r)=(r/r_0)^{-\gamma}$) to the same $z\approx3$ LBG data \citep{steidel03}. Our sample appears to have a comparable clustering
strength, which is slightly higher when corrected for stellar/low-redshift galaxy contamination. A further comparison can be made with the work of
\citet{foucaud03}, who measured an amplitude of $r_0=5.9\pm0.5\hmpc$ from the $w(\theta)$ of a sample of 1294 $20.0<R_{\mathrm{AB}}<24.5$ LBG candidates in the Canada-France Deep Fields Survey \citep{2001A&A...376..756M}. \citet{hildebrandt07} measure the clustering of LBGs in the GaBoDS data and find a clustering length of $r_0=4.8\pm0.3\hmpc$ for a sample of $22.5<R_{\mathrm{Vega}}<25.5$ galaxies. Subsequently to this, \citet{2009A&A...498..725H} measured the clustering properties of LBGs selected in the $ugr$ filters from the CFHTLS data and measured a clustering length of $r_0=4.25\pm0.13\hmpc$ with a magnitude limit of $r_{AB}<25$ and using redshift estimates based on the {\small HYPERZ} photometric code \citep{2000A&A...363..476B}. Our contamination-corrected result appears consistent with most previous work, although lower than the result of \citet{foucaud03}.

\subsubsection{Slit Collisions}

After calculating the angular correlation function, we next use the
redshift information from our spectroscopic survey in order to confirm the
clustering properties of the LBGs. However, before we do this we need to
evaluate the extent to which we are limited in observing close-pairs by the
VIMOS instrument set up. With the LR\_Blue grism, each dispersed spectrum
covers a length of 570 pixels on the CCD. Further to this each slit has a length
(perpendicular to the dispersion axis) in the range of 40-120 pixels. Given the
VIMOS camera pixel scale of 0.205$\arcsec/\textrm{pixel}$, each observed object therefore
covers a minimum region of $\approx120\arcsec\times8.2\arcsec$, in which no
other object can be targeted.

\begin{figure}
\centering
\includegraphics[width=80.mm]{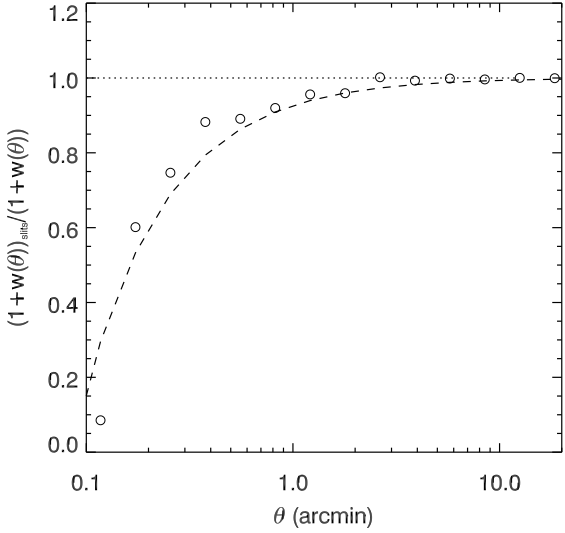}
\caption[{}]{\small Effect of 'slit collisions' on the measurement of the
angular correlation function, $w_{\theta}$. We show the ratio between the
clustering of the entire photometric sample, given by $1+w(\theta)$, and the
clustering measured from only those objects that have been spectroscopically
observed using VLT VIMOS, $1+w_{slits}(\theta)$. The observational constraints
incurred due to the constraint of preventing the dispersed spectra from
overlapping on the instrument CCD lead to a significant reduction in the
clustering measurement at $\theta<2\arcmin$. The dashed line shows our parameter
fit (equation~\ref{eq:lbgangweight}) to the measured ratio, which we use to
correct subsequent clustering measurements made using the spectroscopic galaxy
sample.}
\label{lbg:wtheta_slits}
\end{figure}

In order to evaluate this effect, we calculate the angular auto-correlation
function for only those candidate objects that were targeted in our
spectroscopic survey, $w_{slits}(\theta)$. To do so we require a tailored random
catalogue that accounts for the geometry of the VIMOS CCD layout. We therefore
create random catalogues for each sub-field using a mask based on the layout of
the four VIMOS quadrants, excluding any objects that fall within the $2\arcmin$
gaps between adjacent CCDs. The sky-density of randoms in each sub-field is set
to be $20\times$ the sky-density of data points in the corresponding parent
field. From this subset, which consists of $\approx3400$ targeted objects, we
calculate $w_{slits}(\theta)$ using the Landy-Szalay estimator
(equation~\ref{eq:landszal}). The ratio of $1+w_{slit}(\theta)$ to the original
measurement of $1+w(\theta)$ (prior to correction for contamination) is shown in
Fig.~\ref{lbg:wtheta_slits} (open circles). At $\theta>2\arcmin$ the two
correlation functions follow each other closely and give a ratio of $\approx1$.
However at separations of $\theta<2\arcmin$ we see an increasingly significant
loss of clustering showing the effect of the instrument setup. At redshifts of
$z\approx3$, the  $2\arcmin$ threshold of the effect corresponds to a comoving
separation of $r\approx2.6\hmpc$.

The dashed line in Fig.~\ref{lbg:wtheta_slits} shows a fit to the ratio between
the slit-affected clustering measurement and the original measurement. We use
this fit to provide a weighting factor dependent on angular separation,
$W_{slit}(\theta)$, which is given by:

\begin{equation}
W_{slit}(\theta) = \frac{1}{1-0.0738\theta^{-1.052}}
\label{eq:lbgangweight}
\end{equation}

Applying this weighting function to DD pairs at separations of $\theta<2\arcmin$
then allows the recovery of the original correlation function from the VIMOS
sub-sample correlation function down to separations of
$\theta\approx0.1\arcmin$. Below $\theta\approx0.1\arcmin$ however we are unable
to recreate the original candidate correlation function as no close pairs can be
observed below this scale due to the slit lengths ($8\arcsec<\theta<24\arcsec$)
used in the VIMOS masks.

\subsection{Semi-Projected Correlation Function, $w_p(\sigma)$}
\label{sec:lbgwp}

We next present the semi-projected correlation function $w_p(\sigma)$ for the 1020 $q\geq0.5$ VLT LBGs. Here,  $\sigma$ is the transverse separation given by the separation on
the sky, whilst $\pi$ will be its orthogonal, line-of-sight component. We first
estimate $w_p(\sigma)$ for the full VLT LBG sample using \citep{davispeebles83}:

\begin{equation}
w_p(\sigma) = 2\int^{\infty}_0\xi(\sigma,\pi)d\pi
\end{equation}

\begin{figure}
\centering
\includegraphics[width=80.mm]{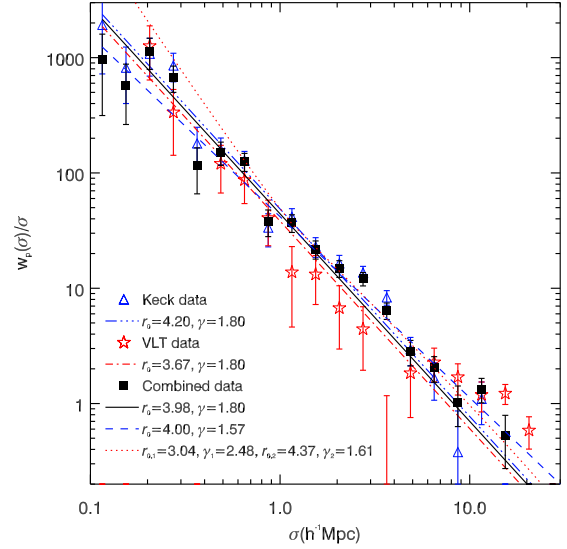}
\caption{\small Projected correlation function,
$w_p(\sigma)$ of the  full VLT, Keck \citep{steidel03} and the combined samples.
The blue dash-triple-dot line represents our best-fit (with $\gamma=1.8$) to the Keck data of
$r_0=4.20^{+0.14}_{-0.15}\hmpc$. The dot-dashed line represents the best $\gamma=1.8$ fit to
the VLT sample with $r_0=3.67^{+0.23}_{-0.24}\hmpc$. The solid line represents the best
$\gamma=1.8$ fit to the combined VLT$+$Keck sample with $r_0=3.98^{+0.23}_{-0.24}\hmpc$
and the dotted line represents the double-power-law model fitted to the VLT
$w(\theta)$. The dashed line gives the result of \citet{2005ApJ...619..697A}, with $r_0=4.0\hmpc$ and $\gamma=1.57$.}
\label{fig:lbgwpsig}
\end{figure}

\noindent We perform the integration over the line of sight range from $\pi=0$
to $100\hmpc$. This encompasses much of the bulk of the significant signal in
the correlation function and performing the calculation over a range of reasonable limits
showed the result to be robust. The VLT $w_p(\sigma)$ is shown in
Fig.~\ref{fig:lbgwpsig} with the best fit clustering model determined by a
$\chi^2$ fit to the data shown as a dotted line. For the projected correlation
function a simple power law form of $\xi(r)$ gives:

\begin{equation}
w_{p}(\sigma)/\sigma =  r_{0}^{\gamma}\sigma^{-\gamma}\left(\frac{\Gamma\left(\frac{1}{2}\right)\Gamma\left(\frac{\gamma-1}{2}\right)}{\Gamma\left(\frac{\gamma}{2}\right)}\right),
%w_p(\sigma)/\sigma = \left(\frac{\sigma}{r_0}\right)^{-\gamma}\left[\frac{\Gamma(0.5)\Gamma(0.5(\gamma-1))}{\Gamma(0.5\gamma)}\right]
\end{equation}

\noindent where $\Gamma()$ is the Gamma function. We perform the fit to the data using a
fixed value for the slope of the function of $\gamma=1.8$. With this value, we
obtain $r_0=3.67^{+0.23}_{-0.24}\hmpc$ for the full VLT sample. Comparing to the initial estimate from the $w(\theta)$ measurement in Fig.~\ref{lbg:wtheta_slits}, we find the $w_p(\sigma)$ measurement gives a somewhat lower value for $r_0$. The difference is at the $\lesssim2\sigma$ level and given the level of contamination in the photometric sample, we expect the $w_p(\sigma)$ measurement to be the more reliable.

We next compare the VLT result to the LBG Keck sample of 
\citet{steidel03}. This sample consists of 940 LBGs in the redshift
range $2.0\lesssim z\lesssim3.9$, with a mean redshift of
$\left<z\right>=2.96\pm0.29$ (compared to $2.0\lesssim z\lesssim4.0$ and
$\left<z\right>=2.87\pm0.34$ for the VLT LBG survey). The survey is
based on observations within 17 individually observed fields, with most
of these being $\approx8\arcmin\times8\arcmin$ with a few exceptions
(the largest field being $\approx15\arcmin\times15\arcmin$). The Keck
spectroscopic data covers a total area of $0.38\deg2$, with just a small
number of the fields being adjacent. The median rest-frame UV absolute
magnitude is $M_{1700}=-17.92\pm0.02$, based on the commonly used
transformations to $M_{1700}$ using the observed magnitudes ${\cal R}$
and $G$ \citep[e.g. ][]{2006ApJ...642..653S,2008ApJS..175...48R}. With
the same method (and the transformations to ${\cal R}$ and $G$ AB
magnitudes given by \citealt{1993AJ....105.2017S}), we estimate a median
rest-frame UV absolute magnitude of $M_{1700}=-18.19\pm0.03$ for our VLT
sample. The samples appear broadly compatible, with the Keck sample
having a marginally fainter average absolute magnitude, most likely due
to the greater number of fainter objects (${\cal R}\gtrsim25$) observed
with the deeper spectroscopy obtained for the Keck sample.

Combining the two spectroscopic data-sets gives a total of 1,980 LBGs over a total area of $1.56\deg2$. In Fig.~\ref{fig:lbgwpsig} we  further present the  Keck and combined  results
for $w_p(\sigma)$. The VLT results are  slightly lower than for the Keck data in
the range $1<\sigma<7\hmpc$. The result for the combined sample
is dominated by Keck pairs for $\sigma<7\hmpc$ and VLT pairs at larger scales.
The solid line represents the $r_0=4.20^{+0.14}_{-0.15}\hmpc$, $\gamma=1.8$ fit for the Keck
data. The dashed line represents $r_0=3.98^{+0.13}_{-0.12}\hmpc$,  which gives the
best $\gamma=1.8$ fit to the VLT$+$Keck combined  $w_p$ data. Also shown is the
best $\gamma=1.8$ fit to the full VLT sample with $r_0=3.67^{+0.23}_{-0.24}\hmpc$.

To calculate $w_p(\sigma)$ for the double power-law $\xi(r)$ that we fitted
above to the  VLT $w(\theta)$ we used the relation

\begin{equation}
w_{p}(\sigma) = 2\int_{\sigma}^{\infty}\frac{r\xi(r)}{\sqrt{r^{2}-\sigma^{2}}}dr
\end{equation}

\noindent The dot-dashed line in Fig.~\ref{fig:lbgwpsig} then  shows that this 
model also gives a good fit to the combined $w_p(\sigma)$.

\subsection{Redshift-Space Correlation Function}
\label{sec:lbgxis}

\begin{figure}
\centering
\includegraphics[width=80.mm]{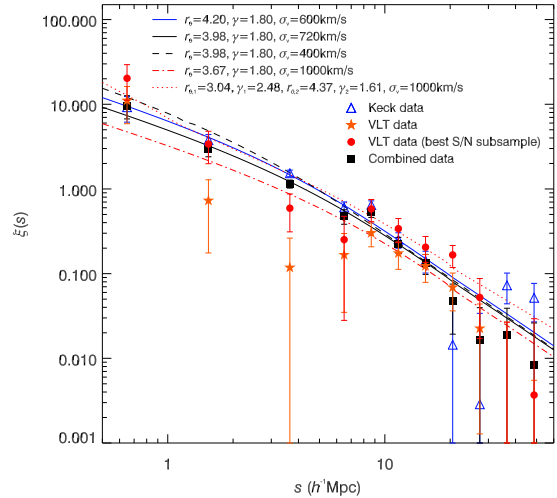}
\caption[{\small Redshift-space clustering function, $\xi(s)$, calculated from
1020 spectroscopically identified Lyman Break galaxies.}]{{\small Redshift-space
clustering function, $\xi(s)$, calculated from 1020 spectroscopically identified
LBGs in the full VLT, Keck  and combined samples. Also shown is the result from
the 529 LBGs in the high S/N VLT sample. The models generally adopt the
$\gamma=1.8$, $\xi(r)$ amplitudes fitted to $w_p(\sigma)$. Thus the combined
VLT$+$Keck model assumes $r_0=3.98\hmpc$ and expected velocity dispersions of
$<w_z^2>^{1/2}=720\kps$ (VLT) and $<w_z^2>^{1/2}=400\kps$ (Keck). Also shown is
a model with $r_0=3.67\hmpc$ from the full VLT $w_p(\sigma)$ result and
$<w_z^2>^{1/2}=1000\kps$, improving the VLT fit. A further model with
$r_0=4.2\hmpc$ from the Keck $w_p(\sigma)$ gives a good fit to the Keck $\xi(s)$
with $<w_z^2>^{1/2}=600\kps$. Finally, we show the 2-power-law VLT $w(\theta)$
model, assuming $<w_z^2>^{1/2}=1000\kps$. All models assume $\beta=0.48$ (see
Sect. \ref{sec:beta}.) }
}
\label{fig:lbgxis}
\end{figure}

The redshift-space correlation function, $\xi(s)$, is an estimator of the
clustering of a galaxy population as a function of the redshift-space distance,
$s$, which is given by $s=\sqrt{\sigma^2+\pi^2}$.  Now, using the full VLT
sample of 1,020 $q\geq0.5$ spectroscopically confirmed $z>2$ galaxies, we
estimate $\xi(s)$ using the simple  estimator $\xi(s)=DD(s)/DR(s)-1$. Again the
random catalogues were produced individually for each field to match the VIMOS
geometry and with $20\times$ the number of objects as in the associated data
catalogues. The DD pairs were then corrected for slit collisions using the
angular weighting function (equation~\ref{eq:lbgangweight}) applied to pairs
with separations of $\theta<2\arcmin$. The result is shown in
Fig.~\ref{fig:lbgxis} (filled circles) with Poisson error estimates. The
accuracy of these errors is supported by analysis of mock catalogues generated
from N-body simulations \citep{daangela05b,2000MNRAS.317L..51H}. Plotted for
comparison is the Keck result as analysed by \citet{daangela05b}. Also shown is
the combined VLT$+$Keck $\xi(s)$ result.

The VLT and Keck  samples show good agreement at separations of
$s>8\hmpc$, however the VLT sample shows a significant drop in
clustering strength at $1<s<8\hmpc$ compared to the Keck measurement.
This seems at odds with the $w(\theta)$ result, which points to the two
samples having similar clustering strengths. However, we note that the
estimate of the line-of-sight distances is sensitive to any intrinsic
peculiar velocities and also errors on the redshift estimate, which will
have a consequent effect on the measured redshift space correlation
function. In addition to this, the peculiar velocities are an important element in the cross-correlation between the galaxy population and the \lya forest, which is presented with this galaxy sample in \citet{2010arXiv1006.4385C}. We therefore now estimate the effect of our redshift errors on
this result. The error on a given LBG redshift is a combination of the
mean error on the spectral feature measurements, which is given by the
measurement error on the Ly$\alpha$ emission line from
Fig.~\ref{lbg:linefitting}, (i.e. $\approx\pm450\kps$ given average
spectral S/N=5.5 in the full VLT sample) combined with the error on the
estimation of the redshift from the measurement of the outflow features
($\approx\pm200\kps$). In addition, there will be some contribution from
intrinsic peculiar velocities. We estimate this contribution based on the work of \citep{nokinprep}. \citet{nokinprep} use the Galaxies-Intergalactic Medium Interactions Calculation \citep[GIMIC, ][]{2009MNRAS.399.1773C}, which samples a number of sub-grids of the Millennium Simulation \citet{2005Natur.435..629S}, populating these with baryons using hydrodynamic simulations. \citet{nokinprep} measure a mean intrinsic peculiar velocity based on galaxies in the GIMIC simulations in redshift slices at $z=3.06$ and find a value of $\approx140\kps$. Combining this in quadrature with the estimated measurement errors gives an overall velocity dispersion  of
$\sigma_z=\sqrt{(450\kps)^2+(200\kps)^2+(140\kps)^2}\approx510\kps$. The
expected overall VLT pairwise velocity dispersion is therefore
$<w_z^2>^{1/2}=\sqrt{2}\times510\approx720\kps$. Substituting a
Ly$\alpha$ emission-line velocity error of $\pm150\kps$
\citep[based on a measurement error of $\Delta z\approx0.002$ from ][]{steidel03} in the above expression similarly implies an expected
$<w_z^2>^{1/2}\approx400\kps$ for the Keck pairwise velocity dispersion.

On small scales, the above  random pair-wise velocity dispersion leads to the well known
`finger-of-god' effect on redshift-space maps and correlation functions. On
larger scales, bulk infall motion towards over-dense regions becomes a
significant factor and causes a flattening in the line-of-sight direction in
redshift space. We now model these two effects to see if  the $\xi(r)$ estimates 
measured from the LBG semi-projected correlation function, $w_p(\sigma)$, and the
angular correlation function, $w(\theta)$, are  consistent with the measured LBG
redshift-space correlation function, $\xi(s)$. Following \citet{hawkins03}, we
use the real-space prescription for the large scale infall effects given by
\citet{hamilton92} whereby the 2-D infall affected correlation function is given
by:

\begin{equation}
\xi^\prime(\sigma,\pi) = \xi_0(s)P_0(\mu) + \xi_2(s)P_2(\mu) + \xi_4(s)P_4(\mu)
\label{reddisti}
\end{equation} 

\noindent where $P_l(\mu)$ are Legendre polynomials, $\mu=cos(\theta)$ and $\theta$ is the
angle between $r$ and $\pi$. For a simple power-law form of $\xi(r)$ the forms
of $\xi_l(s)$ are:

\begin{equation}
\xi_0(s) = \left(1+\frac{2\beta}{3}+\frac{\beta^2}{5}\right)\xi(r)
\label{xi0}
\end{equation} 

\begin{equation}
\xi_2(s) = \left(\frac{4\beta}{3}+\frac{4\beta^2}{7}\right)\left(\frac{\gamma}{\gamma-3}\right)\xi(r)
\end{equation} 

\begin{equation}
\xi_4(s) = \frac{8\beta^2}{35}\left(\frac{\gamma(2+\gamma)}{(3-\gamma)(5-\gamma)}\right)\xi(r)
\label{xi4}
\end{equation} 

\noindent where $\gamma$ is the slope of the power-law form of the real-space correlation
function: $\xi(r)=(r/r_0)^{-\gamma}$. For the 2-power-law model case we use the equivalent expressions
derived by \citet{daangela05b}.  As in \citet{hawkins03}, the infall
affected clustering, $\xi^\prime(\sigma,\pi)$ is then convolved with the random
motion (in this case the pair-wise motion combined with the measurement
uncertainties):

\begin{equation}
\xi(\sigma,\pi)=\int_{-\infty}^{\infty}\xi^\prime(\sigma,\pi-v(1+z)/H(z))f(v)dv
\label{reddistf}
\end{equation}

\noindent where $H(z)$ is Hubble's constant at a given redshift, $z$, and $f(v)$ is the
profile of the random velocities, $v$, for which we use a Gaussian with width equal to
the pair-wise velocity dispersion, $<w_z^2>^{1/2}$.

With this form of $f(v)$,  we take the expected  pair-wise velocity dispersion,
$<w_z^2>^{1/2}=720\kps$ for the full VLT sample and $<w_z^2>^{1/2}=400\kps$ for
the Keck sample. Now taking an estimate of $\beta=0.48$ (see Section
\ref{sec:beta}), we may model the effect of these velocity components on the LBG
sample $\xi(\sigma,\pi)$, first using the  single power-law fit to the combined
sample $w_p(\sigma)$ with $r_0=3.98\hmpc$ and $\gamma=1.8$. The form of $\xi(s)$
estimated from the resultant $\xi(\sigma,\pi)$ is plotted in
Fig.~\ref{fig:lbgxis} (solid line). While the model with $<w_z^2>^{1/2}=400\kps$
gives a good fit to the Keck data, the model with $<w_z^2>^{1/2}=720\kps$ 
appears to overestimate the VLT correlation function at $s<8\hmpc$. Even
increasing the velocity dispersion to $1000\kps$ did not significantly improve
the fit.  We also analysed the LBG sub-sample defined by having spectral
$S/N>5$. We found that $\xi(s)$ for this subsample did rise and  would require
a pair-wise velocity dispersion of $\approx1000\kps$ for the model to fit the
data. This is significantly more than the predicted pair-wise velocity
dispersion of $\approx600\kps$, calculated by replacing the velocity error of
$450\kps$ for the full sample by $350\kps$ in this case, corresponding to
average S/N=8.25 in Fig.~\ref{lbg:linefitting}. The fact that the points at
$s<1\hmpc$ and those at $s>8\hmpc$ agree with the model argues against an even
larger velocity dispersion.

The other possibility is that the $r_0=3.98\hmpc$ model may be too high for the
VLT $\xi(r)$. Certainly the amplitude of $\xi(r)$ from the VLT $w_p(\sigma)$
appears lower than either that from the VLT $w(\theta)$ or the Keck
$w_p(\sigma)$. Fig.~\ref{fig:lbgxis} shows that the fit improves for the full  VLT
samples and the high S/N subsample if the correlation function amplitude
reduces to $r_0=3.67\hmpc$ as fitted to the VLT $w_p(\sigma)$, coupled with
the velocity dispersion increasing to $<w_z^2>^{1/2}=1000\kps$.

The combined VLT$+$Keck sample is very similar to the Keck sample at small
scales. Even for the Keck sample we find that an increased pairwise velocity
dispersion of $<w_z^2>^{1/2}\approx600\kps$ is needed to fit $\xi(s)$ if $r_0=4.2\hmpc$. For the Keck LBGs, the velocity error ($\pm150\kps$, \citealt{steidel03})  $+$ intrinsic outflow error ($\pm200\kps$, \citealt{adelberger03}) combines in quadrature to give $\pm250\kps$ as the error for the line measurement. Subtracting from $\pm600/\sqrt{2}\kps$ would imply $\approx340\kps$ for the pairwise intrinsic velocity dispersion. Clearly for the VLT samples the implied velocity dispersion would be even larger.

We have also used the double power-law $\xi(r)$ indicated by the VLT $w(\theta)$ 
to predict $\xi(s)$. Since  the steepening takes place at $r<3\hmpc$, this means that we
would need even higher velocity dispersions to fit $\xi(s)$. Fig.~\ref{fig:lbgxis}
shows that the double power-law model needs at least a velocity dispersion of $\approx1000\kps$
to fit the VLT$+$Keck combined sample.

We conclude that the low $\xi(s)$ we find in the full VLT sample may be  caused by a
statistical fluctuation in the LBG clustering due to a lower than average $r_0$
and a higher than average velocity dispersion. The VLT sample is designed to
improve correlation function accuracy at large scales, particularly in the
angular direction, and the somewhat  noisy result for $\xi(s)$ at the smallest scales
reflects this. Overall, we conclude that the velocity dispersions required by $\xi(s)$
are bigger than reported previously for the Keck data \citep[$400\kps$ by ][]{daangela05b} 
with the Keck and VLT samples now being fitted by $<w_z^2>^{1/2}=600-1000\kps$, close to 
what is expected from estimates of the redshift errors.

\subsection{Estimating the LBG infall parameter, $\beta(z=3)$}
\label{sec:beta}

The infall parameter, $\beta$, quantifies the extent of large scale coherent
infall towards overdense regions via the imprint of the infall motion on the
observed redshift space distortions. Given its dependence on the distribution of
matter, measuring $\beta$ can provide a useful dynamical constraint on
$\Omega_m(z)$ \citep{hamilton92,heavenstaylor95,hawkins03,daangela08,cabre09}.
It relates the real-space clustering and redshift-space clustering as outlined
in the previous section (see equations~\ref{reddisti} to \ref{xi4}).

We shall measure  $\beta(z=3)$, using the combination of our VLT LBG data and
the LBG data of \citet{steidel03}. As noted above, the VLT and Keck samples
complement each other in the wide range of  separation, $\sigma$, in the angular
direction for the VLT sample and the high sky densities of the Keck samples,
which help define the clustering better at small scales. As discussed in
section~\ref{sec:lbgxis}, the two samples possess comparable real-space
clustering strengths, with measured clustering lengths of $r_0=3.67^{+0.23}_{-0.24}\hmpc$ 
and $r_0=4.20^{+0.14}_{-0.15}\hmpc$ for the VLT and Keck LBG  samples respectively. 
The higher  estimated velocity error of the VLT sample at $\pm450\kps$ compared 
to the Keck $\pm300\kps$ will make little
difference due to the further contributions of the outflow errors and intrinsic
velocity dispersions, the dominance of the Keck data at small scales and the
smaller effect of velocity errors at large spatial scales where the VLT data is
dominant.  We shall therefore combine the two samples in the two methods we use
to measure $\beta$.

\begin{figure}
\centering
\includegraphics[width=80.mm]{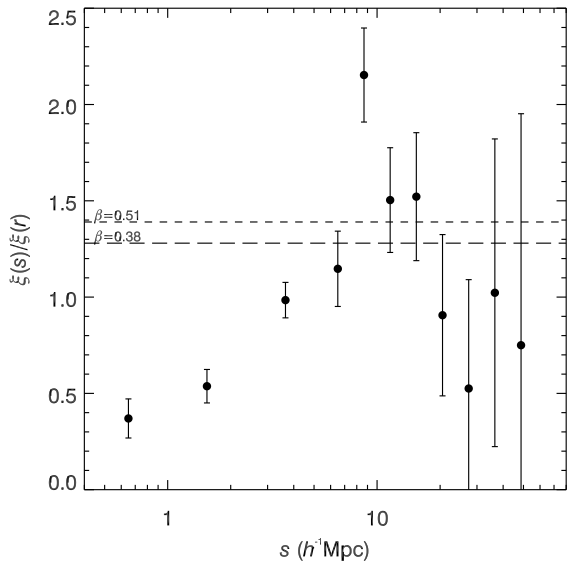}
\caption[]{The redshift space correlation function, $\xi(s)$ divided by the real
space correlation function, $\xi(r)$, with the latter assuming $r_0=3.98\hmpc$
and $\gamma=1.8$. The short and long dashed lines represent the best fit to the
data in the ranges $10<s<25\hmpc$ and $10<s<60\hmpc$, which correspond
to $\beta(z=3)=0.51^{+0.20}_{-0.23}$ and  $\beta(z=3)=0.38^{+0.19}_{-0.23}$ from
equation \ref{xi0}.
}
\label{fig:xis_xir_b}
\end{figure}

We first estimate $\beta$ by simply comparing the amplitude of $\xi(s)$ and
$\xi(r)$ and using equation \ref{xi0} at large scales. Fig. \ref{fig:xis_xir_b}
shows the $\xi(s)$ from the combined VLT and Keck samples divided by the best
fit model for $\xi(r)$ from the semi-projected correlation function,
$w_p(\sigma)$, with $r_0=3.98^{+0.13}_{-0.12}\hmpc$ and $\gamma=1.8$. Equation \ref{xi0}
applies only in the linear regime, so we do not expect it to fit at small
separations. We therefore fit at $s>10\hmpc$. Fitting in the ranges
$10<s<25\hmpc$ and $10<s<60\hmpc$ gives the two dashed lines in
Fig.~\ref{fig:xis_xir_b}, which correspond to $\beta(z=3)=0.51^{+0.20}_{-0.23}$
and  $\beta(z=3)=0.38^{+0.19}_{-0.23}$  with the difference between these two
giving a further estimate of the uncertainty in $\beta$ from this method.

\begin{figure}
\centering
\includegraphics[width=80.mm]{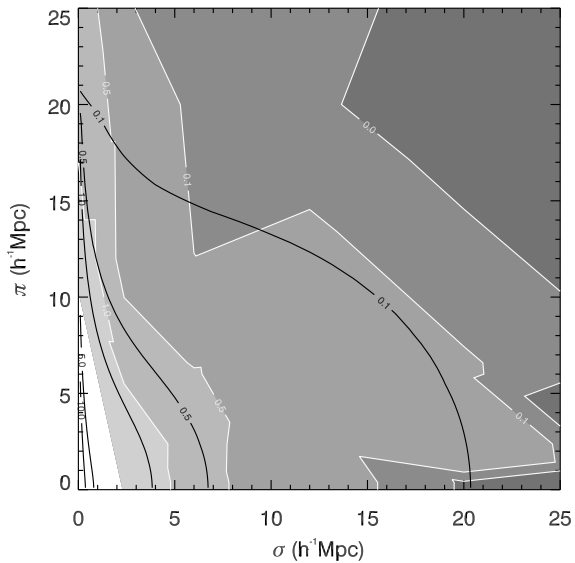}
\caption[{\small $\xi(\sigma,\pi)$ projected correlation function calculated
from the combined \citet{steidel03} and VLT VIMOS LBG samples.}]{\small
$\xi(\sigma,\pi)$ projected correlation function calculated from the
spectroscopically confirmed LBGs from the combined \citet{steidel03} and VLT
VIMOS LBG samples. The best fit model contours are marked as solid lines with 
$\beta(z=3)=0.48$ and $<w_z^2>^{1/2}=700\kps$.}
\label{fig:lbgxisipi}
\end{figure}

We next estimate $\beta$ using the shape of the 2-point correlation
function, $\xi(\sigma,\pi)$, to measure the effect of redshift space
distortions. We calculate $\xi(\sigma,\pi)$ for  the combined  sample. 
As with our determination of $\xi(s)$, we use the simple $DD/DR$
estimator taking randoms tailored to each individual field, with errors 
again calculated using the Poisson estimate.
The resultant $\xi(\sigma,\pi)$ is plotted in Fig.~\ref{fig:lbgxisipi}.
%To show more clearly  the redshift distortions, we
%mirror the result obtained purely in the positive $\sigma$ and $\pi$ directions
%into the negative directions for the purpose of this plot. Having done this, 
The elongation in the $\pi$ dimension, due to the pair-wise velocity dispersion and
redshift errors, is clearly evident at small scales.

%\begin{figure}
%\centering
%\includegraphics[width=80.mm]{xisp_Steidel_DDDR.eps}
%\caption[{\small $\xi(\sigma,\pi)$ projected correlation function calculated
%from the \citet{steidel03} LBG sample.}]{\small $\xi(\sigma,\pi)$ projected
%correlation function calculated from the spectroscopically confirmed LBGs from
%the \citet{steidel03} LBG sample.}
%\label{fig:lbgxisipi}
%\end{figure}

%\begin{figure}
%\centering
%\includegraphics[width=80.mm]{xisp_VLT_DDDR_w.eps}
%\caption[{\small $\xi(\sigma,\pi)$ projected correlation function calculated
%from the VLT VIMOS LBG sample.}]{\small $\xi(\sigma,\pi)$ projected correlation
%function calculated from the spectroscopically confirmed LBGs from the VLT VIMOS
%LBG sample.}
%\label{fig:lbgxisipi}
%\end{figure}

\begin{figure}
\centering
\vspace {-17mm}
\includegraphics[width=100.mm]{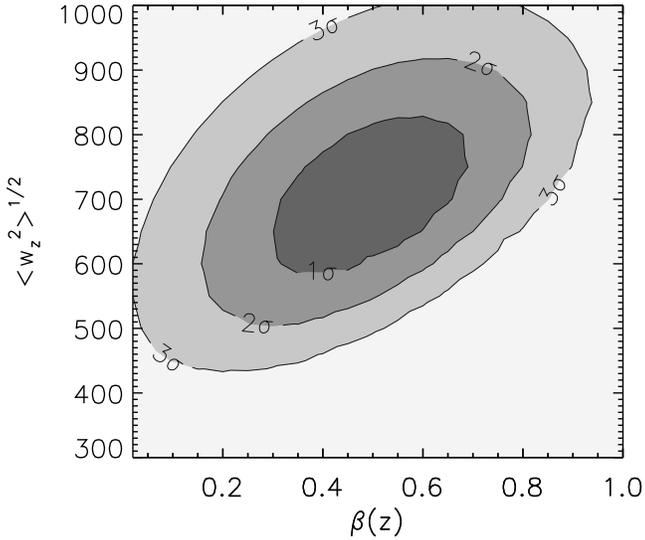}
\vspace {-15mm}
\caption[]{LBG pairwise velocity dispersion ($<w_z^2>^{1/2}$)-infall
parameter($\beta$) $\Delta \chi^2$ contours for the VLT$+$Keck sample, fitting to
$\xi(\sigma,\pi)$ with $s<40\hmpc$. The best fit values are $\beta=0.48\pm0.17$
and $<w_z^2>^{1/2}=700\pm100\kps$, assuming $r_0=3.98\hmpc$ and $\gamma=1.8$. }
\label{fig:w_b_comb}
\end{figure}

Now using this measurement of $\xi(\sigma,\pi)$, we  make an estimate of the
infall parameter, $\beta$. For this we use the single power-law model of
$\xi(r)$ with $r_0=3.98\hmpc$ and $\gamma=1.8$ based on the semi-projected
correlation function of the combined data in Fig.~\ref{fig:lbgwpsig}. With these
parameters set, we calculate the model outlined in equations~\ref{reddisti} to
\ref{reddistf} over a range of values of $<w_z^2>^{1/2}$ and $\beta$. We then
perform a simple $\Delta\chi^2$ fitting analysis and jointly estimate
$<w_z^2>^{1/2}=700\pm100\kps$ and  infall parameter of
$\beta_{\mathrm{LBG}}(z\approx3)=0.48\pm0.17$ for our combined LBG sample. The
contour plot of $\Delta\chi^2$ for the fit in the $<w_z^2>^{1/2}:\beta$ plane
is given in Fig.~\ref{fig:w_b_comb}.

We note that if we allow the  amplitude of $\xi(r)$ to be fitted as well as the
other two parameters, then the results move to $\beta=1.1\pm0.4$ and
$<w_z^2>^{1/2}=800\pm100\kps$ for a best fit $\gamma=1.8$ value of
$r_0=3.64\hmpc$. Taking the Keck sample on its own, we again find
$\beta=0.9-1.5$ and $<w_z^2>^{1/2}=650-750\kps$ if $r_0$ is not
or is allowed to float respectively. The Keck fits have to be resticted to $s<25\hmpc$
because of the small $\sigma$ range in the angular direction and if we apply the
same cut to the combined sample, values of $\beta$ again rise to $\beta=0.8-1.1$
and $<w_z^2>^{1/2}\approx800\kps$, similar to the results for the Keck sample.
Although the errors are clearly still significant, we prefer values of
$\beta\approx0.5-0.6$ given by the amplitude of $\xi(s)$  and the shape of
$\xi(\sigma,\pi)$ for the combined sample which seems best to exploit the
advantages of the Keck sample at small scales and the VLT sample at large
scales.

We have also checked the effect of assuming the double power-law model
fitted to the LBG $w(\theta)$ in Fig.~\ref{lbg:wtheta} with
$r_{0,1}=3.19\hmpc$, $\gamma_1=2.45$, $r_{0,2}=4.37\hmpc$,
$\gamma_2=1.61$ and $r_b=1\hmpc$. The best $\xi(\sigma,\pi)$ fits are
then given by $\beta=0.20\pm0.2$ and $<w_z^2>^{1/2}=750\pm150\kps$. The
reduced $\chi^2$ was 3.44 compared to 3.16 for the single power-law
model. However, allowing the $\xi(r)$ amplitude to vary gave
$\beta=0.48^{+0.24}_{-0.33}$ and  $<w_z^2>^{1/2}=725^{+175}_{-150}\kps$
with fitted amplitudes $\approx80$\% below those estimated from
$w(\theta)$. The small scale rise at $r<1\hmpc$ will not affect our fit
much because of the lack of statistical power at small separations.
Also, the models we are using are expected to be accurate only in the
linear regime at larger scales. The 80\% reduction of the amplitude to 
the large scale power-law implies an $r_0=4.05\hmpc$ which is close to
the $r_0=3.98\hmpc$ value assumed for our single power-law fits above,
leading to similar fitted values for $\beta$ and $<w_z^2>^{1/2}$ in
these two cases. The lower $\beta$ from the actual 2 power-law model is
simply a result  of the high $\xi(r)$ amplitude implied by $w(\theta)$
forcing $\beta$ down in the $\xi(\sigma,\pi)$ fit according to equation
\ref{xi0}.

Comparing our result of $\beta=0.48\pm0.17$ to previous estimates of
$\beta(z\sim3)$, we generally find  somewhat higher values than 
\citet{daangela05b}, who estimate a value of $\beta=0.15^{+0.20}_{-0.15}$. This
is partly because we have assumed $\Omega_m(z=0)=0.3$ and fitted for the
velocity dispersion $<w_z^2>^{1/2}$ whereas \citet{daangela05b} assumed
$<w_z^2>^{1/2}=400\kps$ and fitted for $\Omega_m(z=0)$. If we assume
$<w_z^2>^{1/2}=400\kps$ for the VLT $+$ Keck samples, our estimate of $\beta$
reduces to $\beta=0.18$ for the combined sample. The assumption of
$<w_z^2>^{1/2}=400\kps$ seems to be the main factor that drove  $\beta$ to lower
values, also helped by the different model for $\xi(r)$ assumed by
\citet{daangela05b}, a 2-power law model with $\gamma_1=1.3$ and $\gamma_2=3.29$
with $r_b=9\hmpc$ motivated by fitting the form of $\xi(s)$. The contours in the
$<w_z^2>^{1/2}:\beta$ plane in Fig. \ref{fig:w_b_comb} show that $\beta$ and
$<w_z^2>^{1/2}$ are degenerate - higher $\beta$ implies more flattening in the
$\pi$ direction which can be counteracted by fitting a higher $<w_z^2>^{1/2}$ to
produce elongation in $\pi$. A flatter small scale slope for $\xi(r)$ also
allows a smaller $<w_z^2>^{1/2}$  to be fitted which can then allow lower values
of $\beta$ to be fitted. We have also fitted our combined data with a further 
2-power-law form for $\xi(r)$, now with $r_{0,1}=3.98\hmpc$, $\gamma_1=1.8$,
$r_{0,2}=5.99\hmpc$, $\gamma_2=2.6$ and $r_b=15\hmpc$ but we find that the
results for $<w_z^2>^{1/2}$ and $\beta$ from the combined sample are similar to
those for the single power-law model.

As well as the higher value of $\beta$, we note that we are also fitting higher
velocity dispersion values to the  combined sample. Again the degeneracy
between $<w_z^2>^{1/2}$ and $\beta$ may be the cause. However, the need for high
velocity dispersions was also noted in the small scale fits to $\xi(s)$
particularly for the VLT sample but also for the Keck sample. Even
$<w_z^2>^{1/2}=600\kps$ for the Keck sample implies an intrinsic velocity
dispersion of $<w_z^2>^{1/2}\approx440\kps$ taking into account velocity and
outflow errors on the redshift, much higher than $<w_z^2>^{1/2}=200\kps$
expected from the simulations. If our velocity errors were underestimated then
this could be a cause but they would have to be underestimated in  both the Keck
and VLT datasets. Larger velocity errors are also contradicted by the
consistent widths of the emission-absorption difference histograms in
Fig.~\ref{vdiff}. For example, assuming $\pm450\kps$ for the VLT 
emission velocity error is consistent with $\pm200\kps$ for the outflow error and 
$\pm130\kps$ for the absorption line error.

We conclude that for $\Omega_m(z=0)=0.3$, the combined survey is best fitted by
$<w_z^2>^{1/2}=700\pm100\kps$ with $\beta=0.48\pm0.17$ for a
single power-law model with $\gamma=1.8$ and $r_0=3.98\hmpc$. Based on the
$\beta=0.49\pm0.09$ value, $r_0=5.05\hmpc$ and $\gamma=1.8$ values found for
2dFGRS \citep{hawkins03} linear theory predicts $\beta(z=3)=0.22$  in the
$\Omega_m=1$ case and $\beta=0.37$ in the $\Omega_m(z=0)=0.3$ case, with
$r_0=3.98\hmpc$ for the latter and transformed appropriately  for $\Omega_m=1$.
Our measurements appear to produce  values of $\beta$ that are marginally more
acceptable with $\Omega_m(0)=0.3$ than $\Omega_m(0)=1$ but neither case is
rejected at high significance; $\beta=0.22$ with  $<w_z^2>^{1/2}=600\kps$ is
rejected only at $1.5\sigma$ in Fig.~\ref{fig:w_b_comb}. More importantly, these
measurements provide a useful check of the impact of small- and large-scale
dynamics on our measurement of the clustering of our $z\approx3$ galaxies. The
estimates of $<w_z^2>^{1/2}$ will also be useful in  interpreting the effect of
star-formation feedback from our LBGs on the IGM as measured by the Lyman-alpha
forest in background QSOs \citep{2010arXiv1006.4385C}.

\subsection{Estimating the LBG bias parameter, $b(z=3)$}

We can now estimate the bias, $b$, of the VLT$+$Keck LBG sample from our $\beta$
measurements. The bias gives the relationship between the galaxy clustering and
the underlying dark matter clustering:

\begin{equation}
\overset{\_}{\xi}_g = b^2\overset{\_}{\xi}_{DM} \label{eq:bias}
\end{equation}

\noindent where $\overset{\_}{\xi}_{DM}$ is the volume averaged clustering of
the dark matter distribution and $\overset{\_}{\xi}_{g}$ is the volume averaged
clustering of a given galaxy distribution. In a spatially flat universe, the relationship
between the bias, $b$, and the infall parameter, $\beta$, can be approximated by
\citep{lahav91}:

\begin{equation}
\beta = \frac{\Omega_m^{0.6}}{b} \label{eq:biasbeta}
\end{equation} 

Using this relation with our estimate of $\beta=0.48\pm0.17$ and assuming that
$\Omega_m(z=0)=0.3$ and then given that $\Omega_m(z=3)=0.98$, this implies
$b(z=3)=2.06^{+1.12}_{-0.53}$.

We now compare this to an estimate of the bias from our earlier clustering analysis using equation~\ref{eq:bias}. To do this we calculate the dark matter clustering using the {\small CAMB} software incorporating the {\small HALOFIT} model of non-linearities \citep{smith03}. From this we determine a second estimate of the bias using equation~\ref{eq:bias} and calculating the volume averaged clustering function \citep{1980lssu.book.....P} within a radius, x, for our galaxy sample and the dark matter:

\begin{equation}
\overset{\_}{\xi}(x) = \frac{3}{x^3}\int_{0}^{x} r^2\xi(r) dr
\label{eq:xi20}
\end{equation}

\noindent where $\xi(r)$ is the 2-point clustering function as a function of separation, $r$. We use an integration limit of $x=20\hmpc$, ensuring a significant signal, whilst still being dominated by linear scales. Taking the volume averaged non-linear matter clustering, with the volume averaged clustering of our galaxy sample (with $r_0=3.98\hmpc$ and $\gamma=1.8$, from the VLT$+$Keck $w_p(\sigma)$ measurement) and determining the bias using equation~\ref{eq:bias}, we find $b=2.22\pm0.16$, consistent with the estimate from the bulk flow measurement of $\beta=0.48\pm0.17$ which implies $b=2.06^{+1.12}_{-0.53}$. Both values are somewhat lower than the measurement of the bias of a sample of LBGs from the Canada-France Deep Survey by \citet{foucaud03} who measured a value of $b=3.5\pm0.3$.

\begin{figure}
\centering
%\vspace {-1mm}
\includegraphics[width=80.mm]{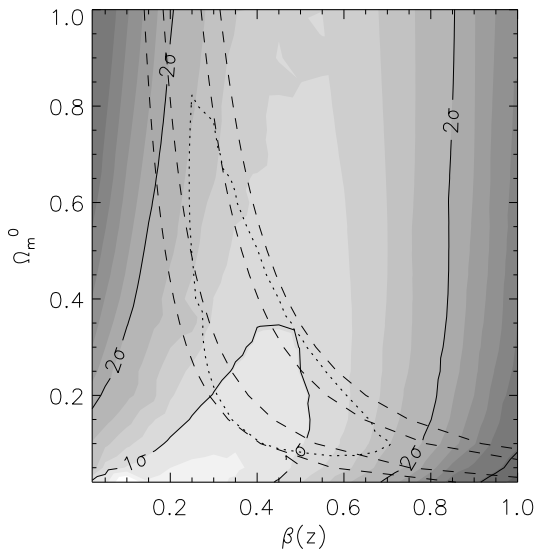}
\vspace {-4mm}
\caption[]{The shaded regions are the $\Omega_m^0$-$\beta(z)$ contours for the VLT$+$Keck 
sample, fitting to $\xi(\sigma,\pi)$ with $s<40\hmpc$. The dashed lines are the 1- and 2-$\sigma$
contours from comparing the $z\approx3$ LBG and the 2dFGRS $z\approx0.1$ clustering amplitudes 
and also using the 2dFGRS $\beta(z\approx0.1)=0.49\pm0.09$ result. The dotted line is the $1-\sigma$
joint contour from applying both of these constraints.
}
\label{fig:o_b_comb}
\end{figure}

We now estimate the mass of typical host haloes for the $z\approx3$ LBG
sample using the \citet{2001MNRAS.323....1S} prescription for the
relation between halo mass and bias, determining a host halo mass of
$M_{DM}=\times10^{11.1\pm0.1}h^{-1}\rm{M_\odot}$. Comparing this to
other LBG samples, \citet{foucaud03}, \citet{hildebrandt07} and
\citet{2008ApJ...679..269Y} measure halo masses of bright $z\approx3$
LBG samples of $M_{DM}\sim10^{12}h^{-1}\rm{M_\odot}$. This difference in
mass estimates reflects the deeper magnitude limits of our survey
compared to a number of the above results and also a slightly lower
redshift range that contribute to our LBG selection sampling a lower
mass range. Work using the \citet{steidel03,2004ApJ...604..534S} data,
which is closer to our own in redshift and depth, report halo masses of
$M_{DM}\sim10^{11.5}h^{-1}\rm{M_\odot}$
\citep{2005ApJ...619..697A,2008ApJ...679.1192C}, which is closer to the
estimate presented here, although our result is still somewhat low.

\subsection{Further test of the standard cosmology}

Following the analysis of \citet{daangela05b} we can make a further test of the
standard cosmology by directly comparing the independent values of
the bias from the z-space distortion and the LBG clustering amplitude. Whereas
in the above case we assumed the DM clustering for the standard model, here we
simply assume the 2dFGRS clustering scale length which we approximate as
$r_0=5.0\hmpc$ and $\gamma=1.8$. We also assume their value of
$\beta(z\approx0.1)=0.49\pm0.09$ from  redshift space distortions. In similar
fashion to \cite{daangela05b} we can then for any $\Omega_m^0$, find the mass
clustering amplitude at $z=3$ and then  we can find the LBG bias, $b(z=3)$, by
comparing this to the amplitude of LBG clustering given by $r_0=3.98\hmpc$ and
$\gamma=1.8$. This can then be converted to $\beta(z=3)$ by using the value for
$\Omega_m(z=3)$ implied by the assumed $\Omega_m^0$ and therefore the
$\beta(z):\Omega_m^0$ relation can be drawn. The 1- and 2-$\sigma$ upper and
lower limits on this relation are shown in Fig. \ref{fig:o_b_comb}. These are
overlaid on the $\Delta\chi^2$ contours (greyscale) from a similar 
redshift-space distortion analysis as seen in Fig. \ref{fig:w_b_comb} but now
allowing $\Omega_m^0$ and $\beta(z=3)$ to vary  while keeping
$<w_z^2>^{1/2}=700\kps$ constant. In this case we have also allowed the LBG
clustering amplitude to be fitted within a 50\% range; this is to ensure that
the dynamical constraint is as independent as possible of the  other constraint
which is  directly taken from the LBG clustering amplitude. We see that although 
the best fit from redshift-space distortions has now moved to lower $\Omega_m^0$ 
and lower $\beta(z=3)$, there is still a good overlap between the $\pm1-\sigma$ regions
of both constraints. The $1-\sigma$ joint contours from both constraints are shown by 
the dotted line with the best joint-fit being $\Omega_m^0=0.2$ and $\beta(z=3)=0.45$.
Thus there is certainly no inconsistency with the standard $\Lambda$CDM model 
although, as  before,  the $\Omega_m^0=1$ model is still rejected at 
less than the $2\sigma$ level. With the values of $\Omega_m^0$ in a 
reasonable range, there appears no inconsistency with the evolution of gravitational 
growth rates as predicted by Einstein gravity, extending the results presented by 
\citet{2008Natur.451..541G} to $z\approx3$.

\subsection{Clustering Evolution}

The space density and clustering evolution of LBGs have frequently been
used to infer their descendant galaxy populations at the present day.
Initially, their relatively high clustering amplitudes were taken to
mean that they would evolve on standard halo models into luminous red
galaxies in the richest galaxy clusters at $z=0$
\citep{steidel96,1998Natur.392..359G,2005ApJ...619..697A}. On the other
hand, \citet{1996Natur.383..236M,metcalfe01} noted that the comoving
density of LBGs was close to that of local spirals. Indeed, they showed
that a simple, Bruzual \& Charlot (1996), pure luminosity evolution
model with e-folding time, $\tau=9$Gyr, plus a small amount of dust,
could explain the LBG  luminosity function  at $z\approx3$. Recently,
more detailed merger tree models have been used to interpret LBG space
densities and clustering. For example, \citet{2008ApJ...679.1192C} have
concluded on this basis  that the descendants are  varied, with LBGs
evolving to become both  blue and red $L^*$ and sub-$L^*$ galaxies. 

\begin{figure}
\centering
\includegraphics[width=80.mm]{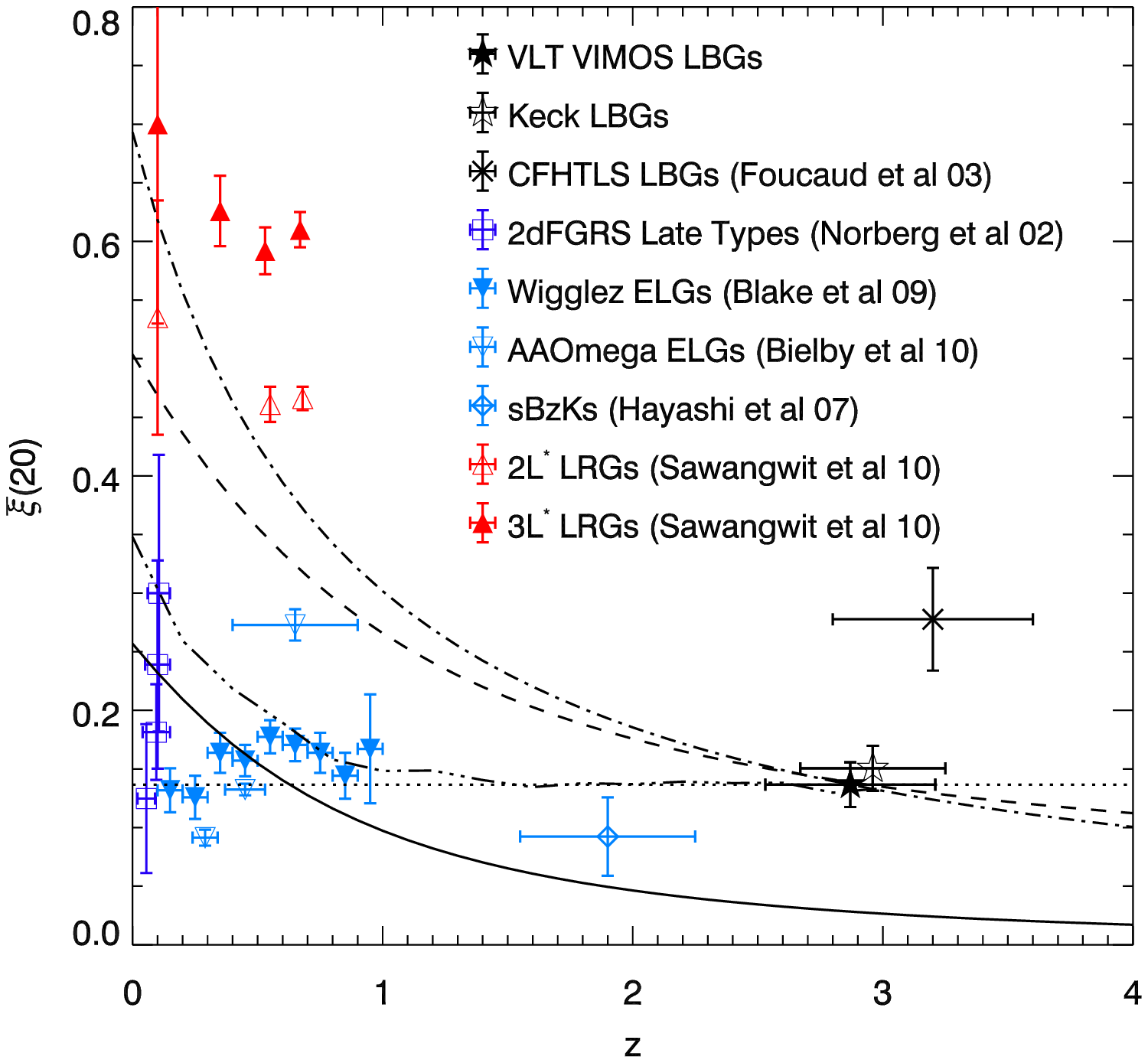}
\caption[{\small The evolution of the volume-averaged correlation function to
$z\approx3$}]{{\small The volume-averaged correlation function,
$\overset{\_}{\xi}(20)$, is plotted for our LBG sample alongside
$\overset{\_}{\xi}(20)$ measurements for several other galaxy populations,
including LRGs at $z<1$ \citep{2009arXiv0912.0511S}, star-forming galaxies at $z<1$
\citep{norberg02,2010MNRAS.403.1261B,2010MNRAS.406..803B} and $z\sim2$ and other LBG populations
\citep{adelberger03,foucaud03}. Further to the observational data, the solid
line shows the estimated evolution of the underlying dark-matter clustering
using the CAMB software \citep{lewis00}, whilst the horizontal dotted line,
dashed line and dot-dash line show the clustering evolution given no evolution
in comoving coordinates, the long-lived model and the stable model. The dot-dot-dot-dash line shows the clustering evolution based on the modeling of the merger history of dark matter haloes.}}
\label{clustevo}
\end{figure}

We now qualitatively compare the clustering strength of our LBG samples
to that of lower redshift galaxies. We first determine the
volume-averaged correlation function at $20\hmpc$ using the single
power-law form of the clustering of both our own and the Keck LBG sample
as prescribed in equation~\ref{eq:xi20}. The $\overset{\_}{\xi}(20)$
measured for the VLT LBG sample is shown in
Fig.~\ref{clustevo}, compared to a number of measures of the clustering
of other galaxy samples across a range of redshifts. The VLT$+$Keck
result ($r_0=3.98\hmpc$, $\gamma=1.8$, $z=2.87$) is shown by the filled
star. We also show the measure for the Keck LBG sample alone (open star)
and the \citet{foucaud03} LBG sample (cross). The apparent
$B$-band magnitude range of the VLT$+$Keck sample is $B=25.69\pm0.76$.
Using the overall redshift range of the sample ($z=2.87\pm0.34$) and
K$+$e corrections determined using the \cite{bruzualcharlot03} stellar
population evolution, this equates to an absolute $B$-band magnitude of
$M_B\approx-21.5\pm1.1$. 

For comparison with our data, we have also plotted the estimated
volume-averaged correlation function values for a number of low and high
redshift galaxy samples. The open and filled red triangles show the LRG
samples of \citet{2009arXiv0912.0511S}, giving the clustering for a $2L^*$ and
$3L^*$ sample respectively (and having absolute i-band magnitudes of
$M_{i(AB)}=-22.4\pm0.5$ and $M_{i(AB)}=-22.6\pm0.4$). The open squares
show the clustering of late-type galaxies from the 2dFGRS as given by
\citet{norberg02} with the individual points giving the clustering of
galaxies in the absolute magnitude ranges of $-18>M_{bj}>-19$,
$-19>M_{bj}>-20$, $-20>M_{bj}>-21$ and $-20.5>M_{bj}>-21.5$ (in order of
lowest to highest clustering data-points). In addition we plot the blue
spiral galaxies of \citet{2010MNRAS.403.1261B} with the open upside-down
triangles and \citet{2010MNRAS.406..803B} with filled upside-down triangles, plus
the sBzKs (open blue diamond) of \citet{hayashi07}.

As an illustration of how we may expect the clustering of the samples to
evolve with time, we first consider a model based on the simulated
merger history of dark matter haloes (dot-dot-dot-dash line) calculated
from the simulations of \citet{2010MNRAS.407.1449G}, whilst the method used to follow the merger trees is described in \citet{2010MNRAS.409..184P}. The simulation was performed using parameter values of $\Omega_m=0.26$, $\Omega_\Lambda=0.74$,
$\sigma_8=0.80$ and $n_s=1.0$ and consisted of a box size of
$L_{box}=123\hmpc$ containing $512^3$ particles with a particle mass of
$10^9h^{-1}\rm{M_\odot}$. The normalization to the LBG data was performed by finding the halo
mass ($10^{11.12\pm0.08}h^{-1}\rm{M_\odot}$) for which the halo $\overset{\_}{\xi}_{20}$ matches
the $\overset{\_}{\xi}_{20}$ measurement for the VLT LBG sample at
its mean redshift. We see that the model predicts little change in the
clustering amplitude at $z=1$ and then stronger evolution to a higher
clustering amplitude at $z=0$. The amplitude of the clustering at z=0
appears consistent with that of late-type galaxies in the 2dFGRS survey
\citep{norberg02}. The predicted descendant number density at $z=0$ based on the halo merger tree model is $\mbox{log}_{10}(n/(h^3\mbox{Mpc}^{-3}))=-3.49_{-0.51}^{+0.59}$ and is also
consistent with number density of the \citet{norberg02} $-20.5>M_{bj}>-21.5$ late-type population, which is equal to $\mbox{log}_{10}(n/(h^3\mbox{Mpc}^{-3}))=-3.64_{-0.01}^{+0.01}$. These
models are able to estimate the transition scale between the 1-halo and
2-halo terms in the correlation function of 
$0.71^{+1.80}_{-0.51}\hmpc$, consistent with the transition scale of
$r_b\approx1.5\pm0.3\hmpc$ in  our measured LBG $w(\theta)$. Overall,
these conclusions are not dissimilar to those of \citet{2008ApJ...679.1192C}.
However, \citet{2008ApJ...679.1192C} predicted higher clustering amplitudes,
$r_0\approx5-6$h$^{-1}$Mpc or $\xibar(20)=0.21-0.29$, at $z\approx1$ and
$r_0\approx6-7$h$^{-1}$Mpc or $\xibar(20)=0.29-0.38$, at $z\approx0$ for the LBG
descendants. Given these differences between the merger-tree 
models of \citet{2010MNRAS.409..184P} and \citet{2008ApJ...679.1192C}, we conclude the results appear 
somewhat  model dependent.

We next compare the $\overset{\_}{\xi}(20)$ results to simpler clustering models. This 
approach is partly motivated by the interpretation of \citet{1996Natur.383..236M,metcalfe01} whose passive luminosity evolution (PLE) models connected the LBG population at $z\approx3$ to the late-type 
population at $z\approx0$. Such models assume that the comoving density of the
LBG/late-types remains constant with time and the clustering models considered here
also make this assumption. Although the models do not take into account 
halo mergers, it has been shown that in the case of Luminous Red Galaxies, 
such models can still provide useful phenomenological fits to LRG clustering
out to significant redshifts \citep{2008MNRAS.387.1045W,2009arXiv0912.0511S}.
Therefore we first plot in Fig.~\ref{clustevo} three simple clustering evolution
models: the long-lived model (dashed blue lines), stable clustering (dot-dashed
cyan lines), and no evolution of the comoving-space clustering (short-dashed line).
All the models have been normalised to the VLT LBG clustering amplitudes.

The long-lived model is equivalent to assuming that the galaxies have ages of
order the Hubble time. The clustering evolution is then governed by their motion
within the gravitational potential and assuming no merging \citep{fry96,croom05}. 
The bias evolution is thus governed by:

\begin{equation}
b(z) = 1+\frac{b(0)-1}{D(z)}\label{eq:longlive}
\end{equation}

\noindent where $D(z)$ is the linear growth rate and is determined using
the fitting formulae of \citet{carroll92}. We evaluate
$\overset{\_}{\xi}(20)$ using the bias evolution in conjunction with the
dark matter clustering evolution, again determined using the {\small
CAMB} software incorporating the {\small HALOFIT} model. This is then
normalized to the measured LBG clustering at the appropriate redshift.

The stable clustering model represents the evolution of virialized
structures and is characterised by \citep{peacock99book}:

\begin{equation}
\overset{\_}{\xi}(r,z) \propto r^{-\gamma} \propto (1+z)^{\gamma-3}
\end{equation}

\noindent where r is the comoving distance.

Finally, the no-evolution model simply assumes that there is no
evolution of the clustering in comoving coordinates. From Eq~\ref{eq:longlive},
this model  can be thought of as a long-lived model in the limit of very 
high bias, ($b(0)>>1$) since then $b(z)\approx b(0)/D(z)$.

Evaluating the clustering evolution of the LBGs, first using the stable
clustering prescription, we would expect the clustering of the
$z\approx3$ galaxies to evolve to a level comparable to that of
low-redshift LRG galaxy samples \citep{2009arXiv0912.0511S}, giving a highly
clustered modern day population. However, as argued by \citet{2008ApJ...679.1192C}, the number density of luminous, early-type galaxies may not
match that of LBGs at $z\approx3$ as required by this virialised
clustering model. Alternatively, on the basis of the long-lived model,
the LBG descendants could either be lower luminosity red galaxies or
higher luminosity blue galaxies. The space density of such  galaxies is
probably more consistent with that of the LBG population. This assumes
the $\Lambda$CDM cosmology and its specific value of $\sigma_8=0.80$. For
a lower mass clustering amplitude the long-lived model would have higher
bias and the $z=0$ predicted amplitude would reduce to more resemble the
no-evolution model. In this case, the descendants of high redshift LBGs
could even be the relatively poorly clustered, star-forming galaxies of
\citet{2010MNRAS.406..803B}. Thus the long-lived models tend to make LBGs the
progenitors of  bluer, or lower-luminosity red galaxies at the present
day, similar to the conclusion from the merger-tree model of \citet{2008ApJ...679.1192C}. The no-evolution (or long-lived, high bias) model would suggest LBGs are the progenitors of  bluer galaxies with lower
clustering amplitudes, more similar to the conclusions of the
merger-tree models of \citet{2010MNRAS.409..184P} or the simple pure luminosity 
evolution models of \citet{1996Natur.383..236M,metcalfe01}.

\section{Conclusions}
\label{sec:conclusions}

In this paper we have described the VLT VIMOS survey of $z\approx3$ galaxies in
a number of fields around bright $z>3$ QSOs. In total this survey has so far
produced a total of 1020 LBGs at redshifts of $2<z<3.5$ over a total area of
$1.18\deg2$. This concludes the data acquisition for the initial phase of the
VLT VIMOS LBG Survey. At the time of writing, these are the most up to date
observations, however the survey has a number of observations only recently
acquired, comprising another 25 VIMOS pointings. Upon completion, the
survey will comprise a total of 45 VIMOS pointings, building significantly on this
initial data-set and providing a catalogue of $\approx2,000$ $z>2$ galaxies over a
sky area of $2.11\textrm{deg}^{2}$. The wide angular coverage of VLT VIMOS makes the new
LBG study very complementary to the previous Keck study which has higher space
densities over smaller areas and hence increased power at the smallest LBG 
separations but little information in the angular direction beyond $10\hmpc$. 
We therefore have frequently used the two surveys in combination in the studies 
of LBG clustering we have presented here.

Based on the fraction of objects observed for this initial VLT LBG survey, we find that
our estimated number densities are consistent with previous studies of LBGs in
this redshift range. Overall we obtain a mean redshift of
$\overset{\_}{z}=2.85\pm0.34$. From the data obtained we have shown evidence for
the existence of galactic outflows with comparable offsets between emission and
absorption lines as in previous studies (e.g. \citealt{2000ApJ...528...96P, 2002ApJ...569..742P} and
\citealt{shapley03})

We have further measured  the clustering properties of the VLT VIMOS LBG sample.
Based on the angular auto-correlation function of the photometric LBG
candidates, the real-space LBG correlation function, $\xi(r)$, is estimated to take
the form of a double power-law, with a break at $r_b\approx1.5\hmpc$. This is
parametrised by a clustering length and slope below the break of
$r_{0,1}=3.19\pm0.55\hmpc$, $\gamma_1=2.45\pm0.15$ and above the break of
$r_{0,2}=4.37^{+0.43}_{-0.55}\hmpc$, $\gamma_2=1.61\pm0.15$.

Assuming $\gamma=1.8$, the semi-projected  LBG correlation function
$w_p(\sigma)$ gives $r_0=3.67^{+0.23}_{-0.24}\hmpc$ for the VLT LBGs,
slightly lower than $r_0=4.2^{+0.14}_{-0.15}\hmpc$ for the Keck LBGs,
and the combined VLT$+$Keck sample gives
$r_0=3.98^{+0.13}_{-0.12}\hmpc$. At $r_b>1\hmpc$, the $\xi(r)$ estimates
from $w(\theta)$ and $w_p(\sigma)$ are therefore quite consistent. At
$r_b<1\hmpc$ the steeper power-law from the angular correlation function
rises above the single power-law that best fits $w_p$, but the
difference is only marginally statistically significant. These
measurements of LBG clustering are broadly  consistent with previous
measurements of the clustering of LBGs at $z\approx3$ made by
\citet{adelberger03} and \citet{daangela05b} but  lower than those made
by \citet{foucaud03}

We then  measured the redshift-space LBG auto-correlation function,
$\xi(s)$. As expected, this presents a flatter slope at  scales
$s<8\hmpc$ due to the effect of  velocity errors, outflows and intrinsic
velocity dispersions. Both the VLT and Keck samples require total
pairwise velocity dispersions in the range $<w_z^2>^{1/2}=600-1000\kps$
to fit $\xi(s)$, higher than the $<w_z^2>^{1/2}=400\kps$ previously
assumed \citep{daangela05b}. The VLT and Keck samples' $\xi(s)$ results
both imply an intrinsic pairwise velocity dispersion of $\pm400\kps$ for
a $\xi(r)$ model with $r_0=3.98\hmpc$ and $\gamma=1.8$. A higher 
$<w_z^2>^{1/2}$ will imply a higher  infall parameter, $\beta(z=3)$, due
to the degeneracy between these parameters. The high value of the
velocity dispersion will also  have an impact on our search for the
effects of star-formation feedback on the QSO Lyman-$\alpha$ forest
\citep{2010arXiv1006.4385C} because any sharp  decrease in absorption
near an LBG will tend to be smoothed away  by this dispersion acting 
as an effective redshift error.

We combine our LBG sample with that of \citet{steidel03} with the aim
of measuring the infall parameter, $\beta(z=3)$. Using a  single power-law with
$r_0=3.98\hmpc$ and $\gamma=1.8$ as our model for the real space $\xi(r)$, our
fits to our measurement of the LBG $\xi(\sigma,\pi)$ from the combined data-set
produce a best fitting infall parameter of $\beta=0.48\pm0.17$. We find that
this value is consistent with the  $\beta=0.37$ value expected in the standard
$\Lambda$CDM cosmology. For this cosmology the value of the LBG bias implied
from the galaxy dynamics is $b=2.06^{+1.12}_{-0.53}$, again consistent with the value of
$b=2.22\pm0.16$ measured from the amplitude of the LBG $\xi(r)$, assuming the
standard cosmology. 

We have also made the cosmological test suggested by \citet{2002MNRAS.329..336H} and
\citet{daangela05b} and shown that the  values of $\Omega_m^0$ and $\beta(z=3)$
derived from LBG redshift-space distortion are consistent with those derived by
comparing the amplitude of LBG clustering at $z=3$ from the combination of the
measured 2dFGRS clustering amplitude and $\beta$ at $z=0.1$, using linear theory. Our
measurement of $\beta(z=3)$  is therefore consistent with what is expected from
the gravitational growth rate predicted by Einstein gravity in the standard
cosmological model (see \citealt{2008Natur.451..541G}).

Finally, we have used the clustering amplitude measured for the LBGs to test
simple models of clustering evolution. In particular, we find that if the LBGs are
long-lived then they could be the progenitors of low redshift $L^*$ spirals or 
early-type galaxies by the present day.

The VLT LBG Survey is an ongoing project and we hope to double the
survey area and LBG numbers by completion of the project. In combination
with this work we are performing a survey of $z\approx3$ QSOs in our LBG
survey fields using the AAOmega instrument at the AAT. Bringing these
two data-sets together will present a significant data resource for the
study of the relationship between galaxies and the IGM at $z\approx3$.

\section*{Acknowledgments}

This work was based on data obtained with the NOAO Mayall 4m Telescope
at Kitt Peak National Observatory, USA (programme ID: 06A-0133), the NOAO
Blanco 4m Telescope at Cerro Tololo Inter-American Observatory, Chile
(programme IDs: 03B-0162, 04B-0022) and the ESO VLT, Paranal, Chile
(programme IDs: 075.A-0683, 077.A-0612, 079.A-0442). RMB acknowledges the
support of a STFC PhD Studentship grant and funding from the Agence
Nationale de la Recherche (ANR). This work was partially supported by
the Consejo Nacional de Investigaciones Cient\'ificas y T\'cnicas and
Secretaria de Ciencia y T\'cnica de la Universidad Nacional de
C\'ordoba, and the European Union Alfa II Programme, through LENAC, the
Latin American-European Network for Astrophysics and Cosmology. DM, LI
and NP are supported by FONDAP CFA 15010003, and BASAL CATA PFB-06. This
research has made use of the NASA/IPAC Extragalactic Database (NED)
which is operated by the Jet Propulsion Laboratory, California Institute
of Technology, under contract with NASA.

\bibliographystyle{mn2e}
\bibliography{$HOME/Dropbox/rmb}

\label{lastpage}

\end{document}